\documentstyle[12pt,twoside,epsfig,color,amssymb,subfig]{report}
\def\tv{\setlength{\unitlength}{8pt}\begin{picture}(1,1)\mbox{\scriptsize T}\put(-1,-0.08){\line(2,1){1.3}}\end{picture}}    % T-violating

\def\cpv{\setlength{\unitlength}{8pt}\begin{picture}(2,1)\mbox{\scriptsize CP}\put(-1.6,-0.1){\line(2,1){1.7}}\end{picture}}    % CP-violating

\def\cptv{\setlength{\unitlength}{8pt}\begin{picture}(2,1)\mbox{\scriptsize CPT}\put(-2.4,-0.1){\line(3,1){2.5}}\end{picture}}    % CPT-violating

\definecolor{mygreen}{rgb}{0,.8,0}
\definecolor{brown}{rgb}{.3,0,0}
\definecolor{VIOLET}{rgb}{.8,0,.9}
\definecolor{YELLOW}{rgb}{0,.5,.5}

\newcommand{\1}{\'{\i}}
\newcommand{\be}{\begin{equation}}
\newcommand{\bea}{\begin{eqnarray}}
\newcommand{\ee}{\end{equation}}
\newcommand{\eea}{\end{eqnarray}}

\newcommand{\kk}{{\bar K}^0}
\renewcommand{\b}{B^0}
\newcommand{\bb}{\bar {B^0}}
\newcommand{\p}{P^0}
\newcommand{\pp}{{\bar P}^0}
\renewcommand{\H}{\mbox{\boldmath$H$}}
\newcommand{\GAMMA}{\mbox{\boldmath$\Gamma$}}
\newcommand{\M}{\mbox{\boldmath$M$}}
\newcommand{\X}{\mbox{\boldmath$X$}}

\renewcommand{\P}{P^{^{(WW)}}}
\newcommand{\eq}[1]{Eq.\,(\ref{#1})}
\newcommand{\eqs}[2]{Eqs.\,(\ref{#1}) and (\ref{#2})}
\newcommand{\then}{\Longrightarrow}
\newcommand{\ok}{^{\sqrt{}}}
\newcommand{\CP}{\mbox{CP}}
\newcommand{\CPT}{\mbox{CPT}}
\newcommand{\T}{\mbox{T}}
\newcommand{\sm}{Standard Model }
\newcommand{\ti}{\Theta \mbox{ and } \Theta^\prime}
\newcommand{\cp}{CP_{\Theta \Theta '}}
\newcommand{\2}{\longrightarrow}
\newcommand{\h}{{\cal H}}
\newcommand{\ww}{Weisskopf-Wigner }
\newcommand{\px}{-\vec x,\, t}
\newcommand{\tx}{\vec x,\, -t}
\newcommand{\C}{{\cal C}}
\newcommand{\parity}{{\cal P}}
\renewcommand{\and}{\mbox{ and }}
\newcommand{\nota}[1]{}
\newcommand{\pepe}[1]{}
\newcommand{\hecho}[1]{}
\newcommand{\parami}[1]{}
\newcommand{\addons}[1]{}
\newcommand{\comentariodelta}[1]{}
\newcommand{\mec}{|\epsilon|^2}
\newcommand{\Dt}{\Delta t}

\newcommand{\sq}{\frac{1}{\sqrt{2}}}
\newcommand{\w}{\omega}
\newcommand{\Dm}{\Delta m}
\newcommand{\DG}{\Delta \Gamma}
\newcommand{\G}{\Gamma}
\newcommand{\BB}{{|B^0B^0\rangle}}
\newcommand{\BbB}{{|\bar B^0B^0\rangle}}
\newcommand{\BBb}{{|B^0 {\bar B}^0\rangle}}
\newcommand{\BbBb}{{|\bar B^0\bar B^0\rangle}}
\newcommand{\abs}[1]{\left| #1 \right|}
\newcommand{\re}{\frac{Re(\epsilon)}{1+|\epsilon|^2}}
\newcommand{\nn}{\nonumber}
\newcommand{\fig}[1]{Fig.~(\ref{#1})}
\newcommand{\DA}{\Delta A_{sl}}

\newtheorem{theorem}{Proposition}[chapter]

\begin{document}
\pagenumbering{roman}

%%%%%%%%%%%%%%%%%%%%%%%%%%%%%%%%%%%%%%%%%%%%%%%%%%%%%%%%%%%
%\parami{
\thispagestyle{empty} \vskip1cm
\begin{center}
{\Large {\bf CP, T and CPT analyses in EPR-correlated $\b\bb$
decays}} \vskip2cm {\large PhD Thesis} \vskip3cm {\it author} \vskip
.1cm {\bf Ezequiel \'Alvarez} \vskip .4cm {\it director} \vskip .1cm
{\bf Jos\'e Bernab\'eu Alberola} \vskip 3.8cm
\resizebox{11cm}{3cm}{\includegraphics{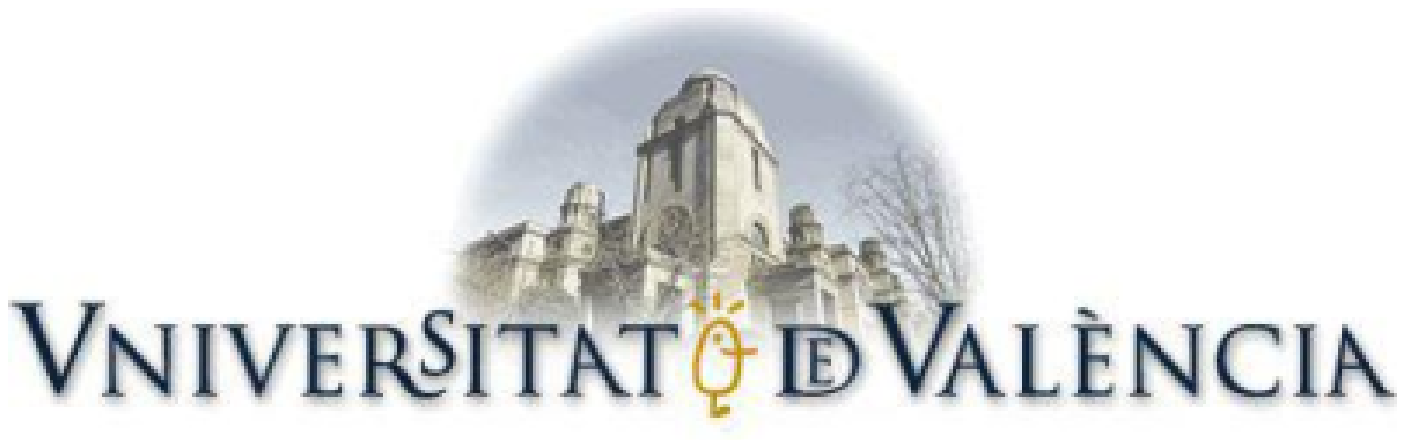}}
\end{center}
\newpage
%%%%%%%%%%%%%%%%%%%%%%%%%%%%%%%%%%%%%%%%%%%%%%%%%%%%%%%%%%%
~ \thispagestyle{empty} \newpage
%} %termina el "parami" de la tapa
\setcounter{page}{3} 
~ \thispagestyle{empty} \newpage
~ \thispagestyle{empty} \newpage

%%%%%%%%%%%%%%%%%%%%%%%%%%%%%%%%%%%%%%%%%%%%%%%%%%%%%%%%%%%
\vskip 5cm
\begin{center}
{\it {\bf A Carla, mi} raison d'\^{e}tre}
\end{center}
\thispagestyle{empty}
\newpage

~ \thispagestyle{empty} 
\newpage

%%%%%%%%%%%%%%%%%%%%%%%%%%%%%%%%%%%%%%%%%%%%%%%%%%%%%%%%%%
\thispagestyle{empty}
\vskip 2cm
JOS\'E BERNAB\'EU ALBEROLA, Catedr\'atico de F\1sica Te\'orica de la Universidad de Valencia,

\vskip 1cm

\begin{flushright}
CERTIFICA: \hskip 0.4cm
\parbox[t]{8cm}{
\noindent Que la presente memoria "CP, T AND CPT ANALYSES IN EPR-CORRELATED $\b\bb$ DECAYS" ha sido realizada bajo su direcci\'on en el Departamento de F\1sica Te\'orica de la Universidad de Valencia por el Li\-cen\-cia\-do D.~EZEQUIEL \'ALVAREZ y constituye su Tesis Doctoral.

~ 

\noindent Y para que as\1 conste, presenta la referida memoria en

~

\noindent Burjassot, a 1 de marzo de 2006

\vskip 4cm

Firmado: Jos\'e Bernab\'eu Alberola
}
\end{flushright}
\newpage

~ \thispagestyle{empty}
\newpage

%%%%%%%%%%%%%%%%%%%%%%%%%%%%%%%%%%%%%%%%%%%%%%%%%%%%%%%

%\setcounter{page}{5} 
\tableofcontents ~\newpage
~\thispagestyle{empty}
 ~\newpage

%%%%%%%%%%%%%%%%%%%%%%%%%%%%%%%%%%%%%%%%%%%%%%%%%%%%%%%
\pagenumbering{arabic}
\setcounter{page}{1}
\addcontentsline{toc}{chapter}{Abstract}
\begin{center}
{\large{\bf Abstract}}
\end{center}

In this work we study the T, CP and CPT symmetries in
the B-meson system.  Our analysis and results are addressed to the case
of correlated mesons in the B-factories.

In the first set of theory and results we investigate the
consequences of these discrete symmetries in the B-mixing and
interference between mixing and decay.  With the help of the CP-tag
we compute all possible intensities and asymmetries which concern
flavour-specific and CP Golden Plate decays.  Our proposed
observables are a new self-consistency check for the Standard Model,
as well as a new exploration for the traces of the discrete
symmetries in the B-system.

In the second set of results we study CPT violation in the initial
state of the B-factories through the loss of indistinguishability of
$\b$ and $\bb$.  We show that, if the consequence of Bose statistic
is relaxed, then the equal-sign dilepton events are considerably
modified.  We analyze the demise of flavour tagging and we also
prove that the equal-sign charge asymmetry, $A_{sl}$, is an optimal
observable where to look for this new CPT violation effect.  The
detailed study of this asymmetry allows us to predict different
behaviours according to the possible values of the CPT violating
parameter $\w$.  We conclude that the best observable where to find
traces of this novel kind of CPT violation is the analysis of
$A_{sl}$ at small $\Dt$'s.  We use existing data on $A_{sl}$ to put
the first limits on $\w$.

\cleardoublepage

\addcontentsline{toc}{chapter}{Resumen [espa\~nol]}
\begin{center}
{\large{\bf Resumen [espa\~nol]}}
\end{center}

En este trabajo estudiamos las simetr\1as T, CP y CPT en el sistema
de mesones B.  Nuestros an\'alisis y resultados est\'an dirigidos
para el caso de mesones correlacionados en las f\'abricas de mesones
B.

En el primer conjunto de teor\1a y resultados investigamos las
consecuencias de estas simetr\1as discretas en el mixing de B's y en
la interferencia entre mixing y decaimiento.  Con la ayuda del
r\'otulo por CP (CP-tag) calculamos todas las posibles intensidades
y asimetr\1as que conciernen a sabor-espec\1fico y decaimiento
CP Golden Plate.  Nuestros observables propuestos son una nueva
verificaci\'on de auto-consistencia para el Modelo Est\'andar, as\1
como una nueva exploraci\'on para los rastros de las simetr\1as
discretas en el sistema de mesones B.

En el segundo conjunto de resultados estudiamos violaci\'on de CPT
en el estado inicial de las f\'abricas de B a trav\'es de la
p\'erdida de indistinguibilidad de $\b$ y $\bb$.  Mostramos que, si
se relaja el requisito de estad\1stica de Bose, entonces los eventos
dilept\'onicos de igual signo son considerablemente modificados.
Analizamos el 'fin' del r\'otulo por sabor (demise of flavour-tag) y
tambi\'en probamos que la asimetr\1a de carga de eventos del
mismo-signo, $A_{sl}$, es un observable \'optimo donde buscar esta
nueva clase de violaci\'on de CPT.  El estudio detallado de esta
asimetr\1a nos permite predecir diferentes comportamientos de
acuerdo con los posibles valores del par\'ametro de violaci\'on de
CPT, $\w$.  Concluimos que el mejor observable para hallar rastros
de esta flamante clase de violaci\'on de CPT es el an\'alisis de
$A_{sl}$ a tiempos cortos.  Tambi\'en usamos medidas existentes de
$A_{sl}$ para poner los primeros l\1mites en $\w$.

\cleardoublepage\chapter{Introduction}

\section{General Introduction \hecho{90}}
{\it Symmetry} may be one of the most interesting and outstanding
archetypes of mankind. From the earliest homo-sapiens-sapiens
legacy's art motivation, running by the Egyptians, Greek, Romans,
Arabs and every-other Civilization, and until the present day, we
can detect it and feel it in a considerable amount of creations or
inventions of man's mind.  Its power is so strong that some times
may even corrupt the line between {\it archetype} and {\it
instinct}.
%: did we invent symmetry or comes deep insight in everybody of us?

Archetype or instinct, symmetry has proven to be an essential tool
for the development of science.  From the very first days of Natural
Philosophy, Pythagoras VIth century BC, symmetry has furnished
insight into the laws of physics and the nature of the Cosmos.  This
insight has found always a constant evolution with the pass of time and the
depth of knowledge, and it is worth to briefly point out some landmarks.

In the late XVIIth century I.\ Newton and G.W.\ Leibniz created the
infinitesimal calculus and, in particular, the latter developed the
analytic notation, which replaced geometry as the essential tool to
study physical systems.  At first sight this could have looked as a
step backward in the art of taking profit of symmetries to
understand physical systems, but in the following century J.-L.\
Lagrange and W.\ Hamilton proved this was not the case.  They
invented the Lagrangian formalism, in which all the symmetries of
the system are explicitly incorporated and {\it transformed} into
conservation laws, and hence improving considerably, with respect to
geometry, the depth of the insight that symmetries can furnish in
the physical theories and in the understanding of the Universe.
Moreover, in this formalism many times the symmetries of the system
itself can determine the Lagrangian, and therefore all the physical
theory behind it.  It is clear that all these achievements would not
have been possible without a mathematical baggage which could, first
{\it define} symmetry, and then {\it incorporate it} analytically
into the theory:  this was the invention of Group Theory, in the
XVIIIth and XIXth century by the great mathematicians J.-L.\
Lagrange and E.\ Galois.  At present day, symmetry is one of the
chief concepts of modern physics and mathematics. The two
outstanding theoretical development of the XXth century, Relativity
and Quantum Theory, involve notions of symmetry in a fundamental and
irreplaceable way.  We should not be surprised if, in the future,
the laws of Nature end up being written {\it uniquely} in terms of
symmetry notions.

It is clear that the study and analysis of symmetries is essential
for the understanding and development of physics.  In their study,
the first major division occurs between {\it continuous} and {\it
discrete} symmetries.  In this work we propose to study, within the
frame of the B-meson system in particle physics, the discrete
symmetries C, P, T and their relevant combinations CP, T, and CPT.

In particle physics, charge conjugation (C) is a mathematical
operation that changes all the charge's sign of a particle, for
instance, changing the sign of the electrical charge. Charge
conjugation implies that every charged particle has an oppositely
charged antimatter counterpart, or antiparticle. The antiparticle of
an electrically neutral particle may be identical to the particle,
as in the case of the neutral pi meson, or it may be distinct, as
with the anti-neutron due to baryon number. Parity (P), or space
inversion, is the reflection in the origin of the space coordinates
of a particle or particle system; i.e., the three space dimensions
$x$, $y$, and $z$ become, respectively, $-x$, $-y$, and $-z$. Time
reversal (T) is the mathematical operation of replacing the
expression for time with its negative in formulas or equations so
that all the motions are reversed. A resultant formula or equation
that remains unchanged by this operation is said to be time-reversal
invariant, which implies that the same laws of physics apply equally
well in both situations. A motion picture of two billiard balls
colliding, for instance, can be run forward or backward with no clue
of which of both is the original sequence.

For years it was assumed that charge conjugation, parity and time
reversal were exact symmetries of elementary processes, namely those
involving electromagnetic, strong, and weak interactions.  However,
a series of discoveries from the mid-1950s caused physicists to
alter significantly their assumptions about the invariance of C, P,
and T. An apparent lack of the conservation of parity in the decay
of charged K mesons into two or three pi mesons prompted C.N.\ Yang
and T.-D.\ Lee to examine the experimental foundation of parity
itself. In 1956 they showed that there was no evidence supporting
parity invariance in weak interactions. Experiments conducted the
next year by Madame C.-S.~Wu verified decisively that parity was
violated in the weak interaction beta decay. Moreover, they revealed
that charge conjugation symmetry also was broken during this decay
process. The discovery that the weak interaction conserves neither
charge conjugation nor parity separately, however, led to a
quantitative theory establishing combined CP as a symmetry of
nature. Physicists reasoned that if CP were invariant, time reversal
T would have to remain so as well due to the CPT theorem. But
further experiments, carried out in 1964, demonstrated that the
long-life electrically neutral K meson, which was thought to break
down into three pi mesons, decayed a fraction of the time into only
two such particles, thereby violating CP symmetry. CP violation
implied nonconservation of T, provided that the long-held CPT
theorem was valid. In this theorem, regarded as one of the basic
principles of quantum field theory, charge conjugation, parity, and
time reversal are applied together. As a combination, these
symmetries constitute an exact symmetry of all types of fundamental
interactions.  In any case, experiments are testing this CPT
invariance -- which up to now it has not been seen to be violated.

CP and T-violation have important theoretical consequences. The
violation of CP symmetry enables physicists to make an absolute
distinction between matter and antimatter. The distinction between
matter and antimatter may have profound implications for cosmology.
One of the unsolved theoretical questions in physics is why the
present Universe is made chiefly of matter. With a series of
debatable but plausible assumptions, it can \nota{esta bein lo de
fraction of seconds?}be demonstrated that the observed
matter-antimatter ratio may have been produced by the occurrence of
CP-violation in the first fractions of a second after the Big Bang.
But, contrary to our expectations, the CP-violation measured in
particle physics insofar is not enough to generate baryogenesis.

~

~

\section{Intoducci\'on General [espa\~nol]}
La {\it simetr\1a} es, tal vez, uno de los arquetipos m\'as
asombrosos e interesantes de la raza humana.  Desde las primeras
motivaciones del legado art\1stico del homo-sapiens-sapiens, pasando
por los egipcios, griegos, romanos, \'arabes y cada una de las
civilizaciones, hasta el d\1a de hoy, la podemos percibir y sentir
en una considerable cantidad de invenciones y creaciones de la mente
humana.  Su poder es tan fuerte que hasta a veces puede llegar a
corromper la frontera entre {\it arquetipo} e {\it instinto}.

Arquetipo o instinto, la simetr\1a ha demostrado ser una herramienta
e\-sen\-cial para el desarrollo de la ciencia.  Desde los primeros
momentos de la Filosof\1a Natural, Pit\'agoras siglo VI a.C., la
simetr\1a nos ha proporcionado importantes nociones, conceptos y
se\~nales sobre las leyes de la f\1sica y la naturaleza del Cosmos.
Estos conceptos han hallado siempre una constante evoluci\'on con el
paso del tiempo y la profundidad del conocimiento, y es importante
destacar algunos hitos en esta relaci\'on.

Hacia finales del siglo XVII I.~Newton y G.W.~Leibniz crearon el
c\'alculo infinitesimal, en particular, este \'ultimo desarroll\'o
la notaci\'on anal\1tica, que reemplaz\'o a la geometr\1a en su
papel de herramienta esencial para el estudio de sistemas f\'isicos.
A primera vista, esto pudo haber parecido un paso atr\'as en el arte
de aprovechar las simetr\1as para comprender los sistemas f\1sicos,
pero en el siglo siguiente J.-L.~Lagrange y W.~Hamilton mostraron
que esto no era as\1.  Inventaron el formalismo Lagrangiano, en el
cual todas las simetr\1as del sistema son expl\1citamente
incorporadas y {\it transformadas} en leyes de conservaci\'on.
Mejorando as\1 considerablemente, con respecto a geometr\1a, la
profundidad de las nociones que las simetr\1as pueden aportar a las
teor\1as f\1sicas en la comprensi\'on del Universo.  Es m\'as, en
este formalismo muchas veces las simetr\1as mismas del sistema
pueden determinar el Lagrangiano, y por ende toda la teor\1a f\1sica
detr\'as de \'el. Es claro, a este punto, que todos estos logros no
hubiesen sido jam\'as posible sin un bagaje matem\'atico que
pudiese, primero {\it definir} simetr\1a, y luego {\it incorporarla}
anal\1ticamente a la teor\1a: esto fue la invenci\'on de Teor\1a de
Grupos en los siglos XVIII y XIX por los geniales matem\'aticos
J.-L.~Lagrange y E.~Galois.  Al d\1a de hoy, la simetr\1a es uno de
los conceptos protagonistas de la f\1sica y matem\'atica moderna.
 Los dos desarrollos te\'oricos brillantes del siglo XX, la Teor\1a
de la Relatividad y la Teor\1a Cu\'antica, incorporan nociones de
simetr\1a en un modo fundamental e irreemplazable.  No ser\1a una
sorpresa si, en un futuro, las Leyes de la Naturaleza terminan
escribi\'endose {\it \'unicamente} en t\'ermino de nociones de
simetr\1a.

Es claro que el estudio y an\'alisis de las simetr\1as es esencial
para la comprensi\'on y el desarrollo de la f\1sica.  En su estudio,
la primera gran divisi\'on ocurre entre simetr\1as {\it continuas} y
{\it discretas}.  En este trabajo proponemos estudiar, dentro del
marco del sistema de mesones B en la f\1sica de part\1culas, las
simetr\1as discretas C, P, T y sus combinaciones relevantes CP, T y
CPT.

En f\1sica de part\1culas, conjugaci\'on de carga (C) es la
operaci\'on mate\-m\'atica que cambia los signos de todas las cargas
de una part\1cula, por ejemplo, cambia el signo de la carga
el\'ectrica.  Conjugaci\'on de carga implica que para cada
part\1cula cargada existe en contraparte una antipart\1cula con la
carga opuesta.  La antipart\1cula de una part\1cula el\'ectricamente
neutra puede ser id\'entica a la part\1cula, como es el caso del
pi\'on neutro, o puede ser distinta, como pasa con el anti-neutr\'on
debido al n\'umero bari\'onico.  Paridad (P), o inversi\'on
espacial, es el reflejo en el origen del espacio de coordenadas de
un sistema de part\1culas; i.e., las tres dimensiones espaciales
$x$, $y$ y $z$ se convierten en $-x$, $-y$ y $-z$, respectivamente.
Inversi\'on temporal (T) es la operaci\'on matem\'atica que
reemplaza la expresi\'on del tiempo por su negativo en las
f\'ormulas o ecuaciones de modo tal que describan un evento en el
cual todos los movimientos son revertidos. La f\'ormula o ecuaci\'on
resultante que resta sin modificaciones luego de esta operaci\'on se
dice de ser invariante bajo inversi\'on temporal, lo cual implica
que las mismas leyes de la f\1sica se aplican en ambas situaciones,
que el segundo evento es indistinguible del original.  Una pel\1cula
de dos bolas de billard que colisionan, por ejemplo, puede ser
pasada hacia adelante o hacia atr\'as sin ninguna pista sobre cu\'al
es la secuencia original en que ocurrieron los eventos.

Durante a\~nos ha sido supuesto que conjugaci\'on de carga, paridad
e inversi\'on temporal eran simetr\1as exactas de los procesos
elementales, ll\'amese aquellos que involucran interacciones
electromagn\'eticas, fuertes y d\'ebiles.  Sin embargo, una serie de
descubrimientos de mediado de los 50's causaron que los f\1sicos
alterasen significativamente sus presunciones respecto a la
invarianza de C, P y T.  La aparente falta de conservaci\'on de
paridad en el decaimiento de los mesones cargados K en dos o tres
mesones pi llevaron a C.N.~Yang y T.D.~Lee a examinar los
fundamentos experimentales de la conservaci\'on de paridad.  En 1956
demostraron que no exist\1a evidencia experimental de invarianza
bajo paridad para las interaccciones d\'ebiles.  Los experimentos
llevados a cabo al a\~no siguiente por Madame C.-S.~Wu verificaron
definitivamente que paridad era violada en el decaimiento d\'ebil
beta.  Es m\'as, tambi\'en revelaron que la simetr\1a de
conjugaci\'on de carga era tambi\'en violada en este proceso.  El
descubrimiento de que las interacciones d\'ebiles no conservan ni
paridad ni conjugaci\'on de carga separadamente condujeron a una
teor\1a cuantitativa que establec\1a la combinaci\'on CP como una
simetr\1a de la Naturaleza. De este modo los f\1sicos razonaban que
si CP era inva\-rian\-te entonces T deber\1a serlo tambi\'en debido
al teorema CPT.  Sin embargo los experimentos siguientes, llevados a
cabo en 1964, demostraron que los mesones K el\'ectricamente neutros
de vida media larga, que deb\1an decaer en tres piones, deca\1an una
fracci\'on de las veces en s\'olo dos de estas part\1culas, violando
as\1 la simetr\1a CP.  Provisto que valiese el teorema fundamental
de CPT, violaci\'on de CP implicaba tambi\'en violaci\'on de T.  En
este teorema, con\-si\-de\-ra\-do uno de los pilares de teor\1a
cu\'antica de campos, conjugaci\'on de carga, paridad e inversi\'on
temporal son aplicadas todas juntas y, combinadas, estas simetr\1as
constituyen una simetr\1a exacta de todos los tipos de interacciones
fundamentales.  Cabe notar que constantemente se realizan
experimentos para verificar la validez de la simetr\1a CPT -- que
hasta el d\1a de hoy siempre se ha visto respetada.

Las violaciones de CP y de T tienen importantes consecuencias
te\'oricas.  La violaci\'on de la simetr\1a CP permite a los
f\1sicos realizar una distinci\'on absoluta entre materia y
antimateria.  Esta distinci\'on puede tener implicaciones profundas
en el campo de la cosmolog\1a: una de las inc\'ognitas te\'oricas en
f\1sica es por qu\'e este Universo esta formado principalmente por
materia.  Con una serie de debatibles, pero plausibles,
presunciones, se puede demostrar que la relaci\'on entre materia y
antimateria que se observa pudo haber sido producida por el efecto
de violaci\'on de CP durante las primeras fracciones de segundo
luego del Big Bang. Sin embargo, contrario a nuestras previsiones,
la violaci\'on de CP medida en f\1sica de part\1culas hasta ahora no
es suficiente para generar bariog\'enesis.

\section{Let the games begin}
\subsection{General considerations \hecho{89}}
In 1973, nine years after the first measurement of CP-violation by
Christenson, Cronin, Fitch and Turlay \cite{cc64}, the two physicist
Kobayashi and Maskawa published a paper \cite{km73} which explained
CP-violation within the electro-weak $SU(2)\times U(1)$ theory
\cite{w67} only if a third generation of fermions would exist.  In
fact, the third generation was found years later and additional
experiments confirmed what is now called the Standard Model.
Kobayashi and Maskawa's theory, which is an extension of Cabibbo's
universality \cite{cabibbo}, became the \sm explanation for
CP-violation.  Through the Cabibbo-Kobayashi-Maskawa (CKM) matrix
all source of CP-violation is reduced to one complex-phase coming from the
fact of having three generations.  This complex-phase has accounted
for the CP-violation observed in the Kaons and also for the CP-violation measured
in the B-sector since 2000 \cite{Bsector}.  The consistency of the
model, measured already in different sectors of particle physics, is up to now in very
good agreement with the experiments.

A geometrical description of the \sm CP-violation is done through
the well known unitarity triangles.  These triangles summarize all
the information contained in the CKM-model.  In the experimental and
theoretical fields physicists prove the \sm over-constraining these
triangles.   Any deviation would be a sign of physics
beyond the \sm .

Although the CKM-model has explained up to now all the experiments
in particle physics, a very important observation in Cosmology
remains unexplained by the \sm CP-violation: the matter-antimatter
asymmetry observed in today's Universe.   Although the
matter-antimatter asymmetry was accepted as one of the fundamental
parameters of the big bang model through $\eta = \eta_{baryons}
/\eta_{photons}$, it was not until 1967, three years after CP
violation was discovered, that Sakharov pointed out in his seminal
paper \cite{sakharov} that a baryon asymmetry can actually arise
dynamically during the evolution of the Universe from an initial
state with baryon number equal zero if the following three
conditions hold:
\begin{itemize}
\item baryon number (B) violation,
\item C and CP violation,
\item departure from thermal equilibrium (i.e., an "arrow of time").
\end{itemize}
Sakharov's work and successive works have studied how much CP
violation was needed in order to get the matter asymmetric Universe
that we see today.  These works, together with particle physics and
cosmology studies show that  the CP-violation given by the CKM-model
is probably not enough to describe today's Universe.  More sources of
CP-violation are required to close a circle in Particle Physics and
Cosmology.

These experimental and theoretical facts have pushed physicists to
study CP-violation, and to place in it a possible window where to
find physics beyond the \sm . The search for such a physics goes
through a wide range of possibilities, among the most important ones
we could recall the existence of new quarks, the existence of
Supersymmetry and the existence of more space-time dimensions.  All
of them include new phases and thus new sources for CP-violation.
The refutation or acceptance of any new theory requires a deep
understanding of the present theory.

The recent measures of CP and T-violation and the forthcoming experiments
in the B-sector, make B-physics one of the best places where to study
the CP, T and CPT symmetries.  The data to be collected by future
experiments as Babar, Belle, BTeV and LHC-b are
expected to give light on the problem.  The comparison of the
out-coming data with the theoretical predictions may clarify whether
the \sm with its unique complex phase can explain all the observed
CP-violation or, on the contrary, New Physics is needed.

CPT symmetry, contrary to CP and T, has not been measured to be
violated \cite{cpt measures}.  Indeed, the CPT theorem
\cite{pauliCPT} states that a quantum field theory that respects the
four hypothesis of $(i)$ locality, $(ii)$ Lorentz invariance,
$(iii)$ causality and $(iv)$ vacuum as the lowest energy state, it
will be CPT-invariant.  In any case, the experimental and
theoretical search for CPT violation is a hot topic \cite{cpt
measures,hot}, and its measurement would be a turning point in all
physical theories and in the comprehension of the Universe.
Throughout this work we propose not only several new observables for
measuring CPT violation in the neutral meson system in a
conventional manner, but also new observables to measure a novel
kind of CPT violation \cite{prl} which accounts for the loss of
particle-antiparticle identity.

\subsection{Structure of this work \hecho{89}}

In this thesis we study, analyze and propose observables that are
concerned with the violation of the discrete symmetries T and CP,
and the possible violation of CPT.  These analysis are performed in
the B-meson system, where the high performance of the present experimental
facilities (Babar, Belle, BTev, etc.) and the promising future machines (SuperKek,
LHC-b) make of it a very fertile field where to explore.  The
results of this work are divided mainly into two groups: the first
corresponds to the analysis of T, CP and CPT violation in the mixing
and decay of the correlated neutral B-meson experiment; whereas the
second it is a study of a distinct expression of CPT violation that
would appear as a modification of the initial state of the B-factories.

In the first piece we study T, CP and CPT observables through the
correlated decays of the neutral $B$-mesons into CP eigenstates and flavour specific channels in a
general framework.  We analyze and compare the time-dependent
intensities associated with the decays on both sides which are
flavour-flavour, flavour-CP, CP-flavour and CP-CP eigenstates, in
order to look for new characteristic effects and regularities.  This
analysis (in particular the CP-flavour and CP-CP decays) requires an
appropriate introduction and definition of the CP-tag, which we
perform in this work. Needless to say, the observation of CP-CP
decays, like $(J/\psi K_S, J/\psi K_S)$, is highly demanding in
statistics: upgraded facilities, like the Super $B$-Factories
\cite{10}, would be needed for a complete study of them.  The complete set of
observables allows therefore to close the circle for correlated
decays into flavour and CP-eigenstates in the spirit of the
theoretical study of Ref. \cite{mcb2} where the CP-tag was first
introduced.  The present results include the effect of the mixing and
decay interference CP and T-violating parameter $\sin(2\beta)$, the
mixing CP and T-violating parameter $Re(\epsilon)$, the lifetime
difference $\Delta \Gamma$, and the CPT violating parameter
$\delta$.  The results we predict in this work, involving CP-flavour
and double CP-CP decays, will over-constrain and test the \sm .

In the second part of the results we analyze possible CPT violation
appearing, in this case, as a relaxation of the demand from Bose
statistics due to distinguishability of $\b-\bb$ in the B-factories.
This novel form of CPT violation is corresponded with a perturbative
modification of the initial state of the $\b\bb$-pair after the
$\Upsilon(4S)$ decay. The parameter which accounts for this
perturbation, $\w$, is unrelated to the parameter that measures CPT
violation through the time evolution of the B mesons, $\delta$.  The
modification of the initial state introduces several changes, in
particular we show in this work how $\w\neq 0$ accounts for what
would be the demise of flavour tagging \cite{plb}.  Besides of being
an important conceptual change, this feature is measurable; we
propose an observable for it which requires $|\w|^2$ precision.   In
a different set of time-dependent observables, we also propose to
measure the equal-sign dilepton charge asymmetry, $A_{sl}$, as a
function of time and to seek for important behaviour changes in it
which are linear in $\w$.  Briefly, we prove that $\w\neq0$ implies
a time dependent $A_{sl}$ asymmetry with an enhanced peak for short
$\Dt$.  We show that this is a very sensitive observable to explore
the existence of CPT violation through $\w$.

This work is divided as follows.  Next Chapter contains an
exhaustive study of CP violation in the \sm.  In Chapter \ref{meson
system} we describe the correlated meson system and all the features
in it. Inspired in the \ww formalism, Section \ref{apendice}
contains theoretical results about the impossibility of extending
the definition of probability if the physical states were
non-orthogonal.  Chapter \ref{cp} describes the first piece of
results of this thesis: after defining the CP-tag, analyzes the T,
CP and CPT violation observables in all possible CP and flavour
correlated neutral B-meson decays.  Chapter \ref{w} is the study of
the consequences of CPT violation through the modification of the
B-factories' initial state.
 And at last, Chapter \ref{conclusiones} contains the conclusions of the
 thesis.

%%%%%%%%%%%%%%%%%%%%%%%%%%%%%%%%%%%%%%%%%%%%%%%%%%%%%%%%%%%%%%%%%%%%%%%%%%%%%%%%%
%%%%%%%%%%%%%%%%%%%%%%%%%%%%%%%%%%%%%%%%%%%%%%%%%%%%%%%%%%%%%%%%%%%%%%%%%%%%%%%%%
%%%%%%%%%%%%%%%%%%%%%%%%%%%%%%%%%%%%%%%%%%%%%%%%%%%%%%%%%%%%%%%%%%%%%%%%%%%%%%%%%
\cleardoublepage\chapter{CP violation in the \sm } \label{ch2}
\section{Introduction \hecho{90}}
The first experimental evidence for CP-violation was obtained in
1964 when Christenson, Cronin, Fitch and Turley \cite{cc64} observed
the long-life neutral Kaon decaying to two pions.  At that time this
came to such a surprise that its origin could not be assigned either
to weak interactions or elsewhere.  Years later when the
electromagnetic and weak forces were unified through the $SU(2)
\times U(1)$ electroweak Hamiltonian and the origin of the masses of
the fermions could be explained through the Higgs mechanism, the
situation changed.

It was then in 1973 when Kobayashi and Maskawa \cite{km73} showed
that the $SU(2) \times U(1)$ theory could allow for CP-violation if
the number of families of quarks is three or more.  Although at that time
there was experimental evidence for only two families, a few years
later the third family was discovered.  This discovery would set the
\sm as the $candidate$ theory to explain CP-violation.  According to
the Standard Model, CP-violation is all originated in what is going to be
called the Cabibbo-Kobayashi-Maskawa matrix, or CKM matrix for
short.

In this chapter we will first introduce the C, P and T operators, and
their relevant combinations CP and CPT, through their definition in the free
field Lagrangian.  Then we will analyze their action on the
interacting \sm Lagrangian, and at last study the CKM matrix and
show how it accounts for all the CP-violation observed up to now in
particle physics.  It is worth noticing at this point, though, that
in Cosmology CP-violation has important consequences, and the
CKM-model is not enough to explain the observed matter-antimatter
asymmetry in the Universe. If there would be found some extra
CP-violation besides the Standard Model then we could be able to
understand some explanation for this matter-dominated Universe.  In
this sense CP-violation is one of the possible windows where
physicists look for physics beyond the \sm .

\section{Transformation of fields and currents \hecho{85}}
\nota{Revisar?}

 \label{transformations} The operations P, C, T, and
their relevant combinations CP and CPT, are defined for free fields.
The transformation requirements which define C, P and T at the level
of the free Lagrangian $L_0$, do not determine the action of these
operations completely, since an arbitrary phase can be included in
the definition of each transformed field with no effect in the
equations.  We analyze first the basic discrete operations P, C and
T, and then the combined transformations CP and CPT.

\subsubsection{Parity}

The parity transformation changes the sign of the space coordinates describing the system, so that
\bea
\vec{x} &\stackrel{P}{\rightarrow}& -\vec{x} \nn\\
\vec{p} &\stackrel{P}{\rightarrow}& -\vec{p} \\
\vec{J} &\stackrel{P}{\rightarrow}& \vec{J}. \nn\\
\nonumber \eea The transformation of the fields upon parity is shown
in the first column in Table \ref{basic}.  This general
transformation includes arbitrary phases $\eta_i$ in the definition
of transformed fields. The free Lagrangian $L_0$, is parity
invariant for any phase of the transformed field. On the contrary,
the transformation of the interaction terms involves relative
parities for different fields and hence its invariance will depend
on the value of the phases.  P is a good symmetry, if it is
possible to choose all the arbitrary phases $\eta_i$ in such a way
that $L(\vec{x},t) \stackrel{P}{\rightarrow} L(-\vec{x},t)$, so that
the action, $S$, does not change under the transformation. But if P
is broken, no choice of the phases will leave $S$ invariant.
\begin{table}[t]
\begin{center}
{\footnotesize
\begin{tabular}{|c|c|c|c|c|}
\hline
\multicolumn{2}{|c|}{Field} &   P   &   C   &   T   \\
\hline
%c-number & $c$ & $c$ & $c$ & $c^*$ \\
scalar & $\Phi (x)$ & $e^{i\eta_\phi} \Phi (\px)$ & $e^{i\xi_\phi} \Phi^\dagger (x)$ & $e^{i\zeta_\phi} \Phi(\tx) $ \\
pseudo-scalar   &  $P(x) $  & $-e^{i\eta_P } P(\px)  $ & $e^{i\xi_P } P^\dagger(x) $ & $e^{i\zeta_P} P(\tx)  $ \\
fermion   &  $\psi(x) $  & $e^{i\eta_\psi} \gamma^0 \psi(\px) $ & $e^{i\xi_\psi } \C \bar \psi ^T (x)   $ & $- e^{i\zeta_\psi} \gamma_5 \C \psi(\tx)   $ \\
~   & $\bar \psi(x)  $  & $e^{-i\eta_\psi } \bar \psi(\px) \gamma^0  $ & $-e^{-i\xi_\psi}\psi(x)^T \C^{-1}   $ & $-e^{-i\zeta_\psi} \bar \psi(\tx) \C\gamma_5   $ \\
vector  &  $V^\mu (x)$  & $e^{i\eta_V} V_\mu (\px)  $ & $e^{i\xi_V} V^{\mu \dagger}(x)$ & $-e^{i\zeta_V} V_\mu (\tx)$ \\
axial-vector  &  $A^\mu (x)$  & $-e^{i\eta_A} A_\mu (\px)  $ & $e^{i\xi_A} A^{\mu \dagger}(x)$ & $-e^{i\zeta_A} A_\mu (\tx)$ \\
\hline
%   &  $ $  & $e^{i\eta_\ }  $ & $e^{i\xi_\ }  $ & $e^{i\zeta_\ }   $ \\
%   &  $ $  & $e^{i\eta_\ }  $ & $e^{i\xi_\ }  $ & $e^{i\zeta_\ }   $ \\

\end{tabular}
}% ends small
\end{center}
\caption{Transformation of the free fields upon the discrete
operations P, C and T.  The phases $\eta_i,\ \xi_i$ and $\zeta_i$
cannot be determined from the requirement of invariance of the free
Lagrangian.  Here $\cal C$ is a unitary $4 \times 4$ matrix
satisfying the condition ${\cal C}^{-1} \gamma_{\mu} {\cal
C}=-\gamma_{\mu}^T$ (usually taken to be ${\cal C}=i \gamma^2
\gamma^0$).} \label{basic}
\end{table}

\subsubsection{Charge Conjugation}
The essence of Charge Conjugation operation is to change the sign of
all internal charges. On the free fields, it exchanges the
annihilation (creation) operators for particles and antiparticles.
Notice, however, that since C is not a good symmetry of Nature, this
does not mean that physical particle and antiparticle are exchanged.
The action of C on the free fields is found in the second column of
Table \ref{basic}.

\subsubsection{Time Reversal}
The operation of time reversal corresponds to change the sign of $t$
in the equations of motion.

The operator T, on the other hand of P and C, results to be
$antiunitary$ and therefore cannot be an observable (this means that
it cannot give rise to conserved quantum number analogous to
parity).  The action of T on the free fields is shown in the third
column of Table \ref{basic}.

\begin{table}[t]
\begin{center}
\begin{tabular}{|c|c|c|}
\hline
Field & CP & CPT \\
\hline
$\Phi(x)$ & $e^{i\phi_\phi} \Phi^\dagger(\px)$ & $e^{i\chi_\phi} \Phi^\dagger(-x)$ \\
$P(x)$ & $-e^{i\phi_P} P^\dagger(\px)$ & $e^{i\chi_P} P^\dagger(-x)$ \\
$\psi(x)$ & $e^{i\phi_\psi} \gamma^0 \C \bar \psi^T(\px)$ & $-e^{i\chi_\psi} \gamma_5 \psi^*(-x)$ \\
$\bar \psi(x)$ & $-e^{-i\phi_\psi}\psi^T(\px) \C^{-1} \gamma^0$ & $e^{-i\chi_\psi} \bar\psi(-x)^*\gamma_5 $ \\
$V^\mu(x)$ & $-e^{i\phi_V} V_\mu^\dagger(\px)$ & $-e^{i\chi_V} V^{\mu\dagger}(-x) $ \\
$A^\mu(x)$ & $e^{i\phi_A} A_\mu^\dagger(\px)$ & $-e^{i\chi_A} A^{\mu\dagger}(-x) $ \\
\hline
\hline
operator & CP & CPT \\
\hline
$c$ & $c$ & $c^*$ \\
$\bar\psi_1  \psi_2$ & $e^{i(\phi_2-\phi_1)} \, \bar\psi_2  \psi_1$ & $e^{i(\chi_2-\chi_1)} \left(\bar\psi_1   \psi_2\right)^\dagger $ \\
$\bar\psi_1 \gamma_5  \psi_2$ & $-e^{i(\phi_2-\phi_1)}\,  \bar\psi_2 \gamma_5 \psi_1$ & $e^{i(\chi_2-\chi_1)} \left(\bar\psi_1 \gamma_5 \psi_2\right)^\dagger $ \\
$\bar\psi_1 \gamma^\mu \psi_2$ & $-e^{i(\phi_2-\phi_1)}\,   \bar\psi_2 \gamma_\mu \psi_1$ & $-e^{i(\chi_2-\chi_1)} \left(\bar\psi_1 \gamma^\mu  \psi_2\right)^\dagger $ \\
$\bar\psi_1 \gamma_5\gamma^\mu \psi_2$ & $-e^{i(\phi_2-\phi_1)}\,  \bar\psi_2 \gamma_5\gamma_\mu  \psi_1$ & $-e^{i(\chi_2-\chi_1)} \left(\bar\psi_1 \gamma_5\gamma^\mu \psi_2\right)^\dagger $ \\
$\bar\psi_1 \sigma^{\mu\nu} \psi_2$ & $-e^{i(\phi_2-\phi_1)}\,  \bar\psi_2 \sigma_{\mu\nu}  \psi_1$ & $e^{i(\chi_2-\chi_1)} \left(\bar\psi_1 \sigma^{\mu\nu}   \psi_2\right)^\dagger $ \\
\hline
\end{tabular}
\end{center}
\caption{Transformation of the free fields and the Dirac bilinear operators when acting with the combined operators CP and CPT.  The phases $\phi_i$ and $\chi_i$ cannot be determined from the transformation of the free fields.}
\label{combined}
\end{table}

\subsubsection{CP and CPT}

The CP and CPT transformations are combined operations of the basic
transformations C, P and T.  The interest of these operations comes
out when applied to the complete interaction Lagrangian, where
besides the free fields, due to Lorentz invariance, we find the
spinors describing fermion fields as bilinear operators.  Therefore
the CP and CPT transformation rules for free fields and for bilinear
operators are shown in Table \ref{combined}.  The $\phi_i$ and
$\chi_i$ are undetermined phases until their value is needed in
order to leave invariant --if possible-- the interaction Lagrangian.

Notice in the Dirac bilinear operators, comparing the CP and the CPT
columns, how the operation T, when added to CP, completes the
transformation to retrieve --up to a phase-- the hermitian conjugate
of the original operator.  This constitutes one of the keys to
understand how, with an appropriated choice of phases, the CPT
operator may always leave invariant an hermitian scalar interaction.

\section{CP operation in the Standard Model Lagrangian \hecho{90}}

We have seen that the discrete operations C, P and T are well
defined in the free Lagrangian and leave it invariant.  The next
step is to study the interacting \sm Lagrangian in order to analyze
how these operators, in particular CP, act on it.

The Lagrangian density of the Electroweak Standard
Model~\cite{GSW61} has a local gauge invariance under $SU(2) \times
U(1)$. It can be symbolically written as~\cite{Jar89}
 \be
 L=L(f,G)+L(f,H)+L(G,H)+L(G)-V(H),
 \label{eq:sch.L}
 \ee
 where $f$ represents the fermions, $G$ the gauge bosons and $H$ the scalar
doublet in the theory.

The hadronic sector consists of $N$ quark families, organized in
$SU(2)$ doublets, for the left-handed components, and $SU(2)$
singlets for the right-handed ones. That is, there are $N$
multiplets
 \be \left ( \begin{array}{c} q_j \\ q_j' \end{array}
\right )_L,\hskip.5cm {q_j}_R,\hskip.5cm {q_j' }_R, \hskip.5cm
j=1,...,N. \ee

As the theory contains N families of fermions with the same flavour
charges, the generalized CP operation for quarks involves, instead
of arbitrary phases for each field, a unitary $N \times N$ matrix
acting on flavour space for the up sector, and an independent
one for the down sector. The invariance condition for strong and
electromagnetic interactions requires these matrices to be unitary,
leaving them otherwise unfixed. We are interested in the way CP
transformation acts on each piece of the Lagrangian \eq{eq:sch.L}.

The first term in the Lagrangian, representing interaction bet\-ween
fer\-mions and gauge bosons, reads, for the hadronic part, \bea
\lefteqn{L(f,G)\!=\!\!\sum_{j=1}^{N} \left \{ \overline{(q_j,
q_j')}_L i \gamma^{\mu}\! \left [
 \partial_{\mu} I \!-\! i g_2 \frac{\vec{\sigma}}{2}\! \cdot\! \vec{W}_{\mu} \!-\! i g_1 \frac{
1}{6} B_{\mu} I \right ]\! \left (\! \begin{array}{c} {q_j}_L \\ {q_j'}_L \end{array}\! \right
) \right. \!+} \nonumber \\
& & \! \! \! + \!\left. {\bar{q_j}}_R i \gamma^{\mu}\!\! \left [\partial_{\mu}\! - i g_1 \frac{
2}{3} B_{\mu} \right ]\! {q_j}_R \!+\! {\bar{q_j}}_R' i \gamma^{\mu}\!\! \left [ \partial_{\mu}
 - i g_1 \!\left (\!-\frac{1}{3}\!\right )\! B_{\mu} \right ] \! {q_j'}_R \! \right\}.
\label{eq:LfG} \eea
Left-handed and right-handed quarks interact in
a different way, so that Parity is formally violated by this term.
Charge conjugation is also broken, due to the simultaneous presence
of axial and vector currents. Nevertheless, it is possible to choose
the transformation phases for the different fields so that
 \bea
 L_{(f,G)}(\vec{x},t) \stackrel{CP}{\rightarrow}
 L_{(f,G)}(-\vec{x},t),
 \nn\eea
 if up and down sectors transform the same.
Thus, in a theory with no scalar sector, where there is no
connection between both sectors, the Lagrangian would be CP
invariant~\cite{GR97}.

However, in order to have CP-conservation in the whole Lagrangian, one
needs to find a phase choice which leaves all the terms in $L$
invariant at the same time. It is possible to show that $L(G)$,
$L(G,H)$ and $V(H)$ preserve, separately, P and C. So, it is left to
analyze the transformation properties of $L(f,H)$, that represents
the interaction between fermions and scalars. The hadronic part of
this term is \bea L(f,H)\!&=&\!\!\sum_{j,k=1}^{N} \left \{ Y_{jk}
\overline{(q_j, q_j')}_L \left (\begin{array}{c
} \Phi^{(0)*} \\ -\Phi^{(-)} \end{array}\right ) {q_k}_R \ + \right. \nonumber \\
& & \left. + \: Y_{jk}' \overline{(q_j, q_j')}_L \left
(\begin{array}{c} \Phi^{(+)} \\ \Phi^{(0 )} \end{array} \right )
{q_k'}_R + h.c. \right \}, \eea
where $Y_{jk}$ and $Y_{jk}'$ are the Yukawa couplings, complex numbers that the Standard Model do not determine. %completely undetermined in the Standard Model.

After spontaneous symmetry breaking, in the unitary gauge %refs SSB y gauge?
the field $\Phi^{(0)}$ becomes real, while $\Phi^{(+)}$ disappears.
So, out of the four degrees of freedom in the scalar doublet, three
are absorbed by the longitudinal components of the $W$ and $Z$
bosons, which acquire mass in this way. What is left from $L(f,H)$
is \be L(f,H) \stackrel{SSB}{\rightarrow} -\sum_{j,k=1}^{N} \left
\{m_{jk} \bar{q_j}_L {q_k}_R + m_{jk}' \bar{q_j'}_L {q_k'}_R + h.c.
\right \}\left [1 + \frac{1}{v} H \right ], \label{L.Yukawa} \ee
where
 \bea
 m_{jk}\equiv [M]_{jk} &=& -\frac{v}{\sqrt{2}} Y_{jk}\nn\\
 m_{jk}'\equiv [M']_{jk} &=& -\frac{v}{\sqrt{2}} Y_{jk}'
 \eea
 are the
complex mass matrices, and $H$ is the real field of the Higgs boson.
Although $M$ and $M'$ need to be neither real nor hermitian, due to
the structure of gauge interactions any unitary transformation on
the right-handed quarks is unobservable, it is possible to
restrict ourselves to hermitian mass matrices without loss of
generality~\cite{FJ85}.

The most general CP transformation for up and down weak quark fields
is defined up to the $N \times N$ unitary matrices $\Phi$, $\Phi'$
as
 \bea q (\vec{x},t) & \stackrel{\rm CP}{\rightarrow}
& \Phi \gamma_0 {\cal C} \bar{q}^T(-\vec{x},t), \nonumber \\
q' (\vec{x},t) & \stackrel{\rm CP}{\rightarrow} & \Phi' \gamma_0
{\cal C} \bar{q'}^T(-\vec{x},t). \label{eq:transweak} \eea
 The
question is whether there is a way of choosing $\Phi$ and $\Phi'$
leaving $L$ invariant. Imposing CP-invariance to the mass term, we
get the following condition on the CP matrices: \bea
M^* &=& \Phi^+ M \Phi, \nn\\
{M'}^* &=&{\Phi'}^+ M' \Phi' \ . \label{eq:mass.inv} \eea Being
hermitian, $M$ and $M'$ can be diagonalized by unitary matrices \bea
 M &=& U^+ D \, U, \nn\\
M' &=& {U'}^+ D' U' \ ,\nn
\eea
with $D={\rm diag} (m_u, m_c, m_t)$ and $D'={\rm diag} (m_d, m_s, m_b)$.

It is then immediate to find matrices $\Phi$, $\Phi'$ which satisfy the conditions \eq{eq:mass.inv}, namely:
\bea
\Phi &=& U^+ e^{2 i \Theta} U^* ,  \nn\\
\Phi'&=&{U'}^+ e^{2 i \Theta^{'}} {U'}^* \ ,
\label{eq:CP.weak}
\eea
where $\Theta$ and $\Theta^{'}$ are real diagonal matrices given by
$\Theta\equiv {\rm diag}(\theta_u,\theta_c,\theta_t)$ and $\Theta'\equiv {\rm diag}(\theta_d',\theta_s',\theta_b')$.

By construction, such a transformation leaves the mass term
unchanged. However, when we look at its action on the charged
current term, we get
$$
{\bar{q_i}}_L \gamma_{\mu} {q_i'}_L \stackrel{\rm CP}{\rightarrow} -{\bar{q_k'}}_L \Phi_{ji}^\dagger {\Phi'}_{ik}  \gamma^{\mu} {q_j}_L .
$$
However, $\Phi_{ji}^\dagger {\Phi'}_{ik}  = [\Phi^\dagger \cdot
\Phi' ]_{jk}$ and the Lagrangian will only be invariant if
 \bea
 \Phi'=\Phi.
 \label{argeval}
 \eea
\section{The CP invariance condition \hecho{90}}
\label{condition}

We have seen that the problem of CP invariance of the Electroweak
Lagrangian reduces to the joint study of the transformation
properties of the mass and the charged current terms. This is so
because they involve common quark fields, so that  in order to have
invariance of $L$, one needs to find a transformation which respects
both terms.

Such a study can be performed in any basis. Up to this point, all
the analysis has been done in a weak basis, in which the charged
currents are diagonal. We may as well move to the physical basis
which is where one usually works, and wonder what the invariance
requirements look like.

Definite mass quark fields are given by
 \bea
 u &\equiv & q_{phys}=U q, \nn\\
 d &\equiv & q_{phys}'=U' q'.
 \eea
 In this basis the two relevant terms of the Lagrangian read
 \bea
 L&=&\left ( \begin{array}{c c} \overline{q}_{phys} &
 \overline{q}_{phys}' \end{array} \right ) \left ( \begin{array}{c c}
 D & 0 \\ 0 & D' \end{array} \right ) \left ( \begin{array}{c}
 q_{phys} \\ q_{phys}' \end{array} \right )+\nn\\
 && g {{\overline{q}}_{phys}}_L \gamma_{\mu}V {q_{phys}'}_L {W^+}^{\mu}
 \label{relevant}
 \eea
  being $V \equiv U {U'}^+$, the quark mixing
matrix also known as the Cabibbo-Kobayashi-Maskawa (CKM) matrix.

Since the phase of physical quark fields can be arbitrarily chosen,
the most general $V$ for $N$ quark families can be written in terms
of $N(N-1)/2$ moduli and $(N-1)(N-2)/2$ phases~\cite{BP81}. This
means that for three families, the mixing matrix is completely
determined by four real parameters: three rotation angles and one
phase.

The mixing matrix $V$ can be parameterized in multiple ways,
according to the definition of parameters and the choice of quark
phases for, under a redefinition of the type
 \bea
 u_i \rightarrow e^{i \varphi_i} u_i, \nn\\
 d_i \rightarrow e^{i \varphi_i'} d_i,
 \label{eq:rephase}
 \eea
  the matrix elements change as $V_{ij}
\rightarrow e^{i (\varphi_j'-\varphi_i)} V_{ij}$.

The change of basis affects also the CP transformation matrices  $\Phi$ and $\Phi'$ which transform as
 \bea
 \Phi & \rightarrow & \tilde{\Phi} \equiv
 U \Phi \, U^T=e^{2 i \Theta}, \nonumber \\
 \Phi' & \rightarrow & \tilde{\Phi}' \equiv U' \Phi' {U'}^T=e^{2 i
 \Theta^{'}}. \label{eq:CP.phys} \eea
 The transformation law for
physical fields has then the same form of \eq{eq:transweak} with the
transformed matrices $\tilde{\Phi}$ and  $\tilde{\Phi}'$. Then
$\Theta$ and $\Theta'$ correspond to the CP transformation  phases
for the physical quark fields.

These phases are not invariant under a rephasing of the
physical fields as the one in \eq{eq:rephase}.
Such a rephasing corresponds to a rotation of the quark fields
under a diagonal unitary matrix, so that also the CP transformation
phases in $\Theta$ and $\Theta'$ transform as
$\theta_i^{(')} \rightarrow \theta_i^{(')}+\varphi_i^{(')}$.

Regarding \eq{relevant}, in the physical basis the mass term is
diagonal, and thus invariant under a CP transformation (see Table
\ref{combined}).  On the contrary the charged current interaction is
not invariant, but transforms as
 \be V \stackrel{\rm
 CP}{\rightarrow} e^{2 i \Theta} V^* e^{-2 i \Theta'},
 \ee
 so that,
for the Lagrangian to be symmetric, it has to be possible to find
$\Theta$ and $\Theta'$ so that
 \be
 V^*=e^{-2 i \Theta} V e^{2 i \Theta'},
 \label{condicion}
 \ee
  or either
\addtocounter{equation}{-1}
 \be
 V^*_{ij}=V_{ij} e^{2i(\theta'_j-\theta_i)} .
 \label{condicion2}
 \ee
Therefore, by means of a change of basis, we have moved all the CP
problem to the charged current term.

Two seconds of reflection show that if, as is the case of three
generations, the CKM matrix $V$ has at least one complex phase which
is independent of the rephasing of the quarks, then \eq{condicion}
will generally not be satisfied.  And therefore this is the \sm
explanation for CP-violation in particle physics.

~

{\it NOTE:}  The invariance condition for $V$, corresponding to the
physical basis, is equivalent to the relation we found by requiring
invariance in the weak basis, $\Phi=\Phi'$.  To show this we may
consider a matrix $\Phi$ in the weak basis which satisfies both
equalities in \eq{eq:mass.inv}. When we transform it into the
physical basis, it yields
 \bea
 \tilde{\Phi} &=& U \Phi U^T, \nn\\
 \tilde{\Phi}'&=& U' \Phi {U'}^T,
 \eea
 so that
\be
\tilde{\Phi}^+ V \tilde{\Phi}'=U^* {U'}^T = V^*,
\ee
q.e.d.

\section{The CP operators \hecho{90}}
As we have learned in the previous Section, the \sm with three
generations does not allow, in principle, to determine in a unique
way a CP operator which leaves the Lagrangian invariant.  Moreover,
as experimental results have shown, the CP-operation {\it is not} a
symmetry of the Universe and hence, in fact, there is not a CP
operator which commutes with Nature's Lagrangian.

However, although not totally determined, we can still define a
family of CP-operators.   We can label the variety
of CP operators that come from different choices of the $\Theta$ and
$\Theta'$ matrices as $\cp$. In fact, the CP transformation in
\eq{eq:transweak} is written in the physical basis as
 \be \left.
 \begin{array}{rcl}
 q'_{phys,\, j}(\vec{x},t) &\stackrel{\rm \cp}{\longrightarrow}& e^{2i\theta'_j} \gamma_0 {\cal C} \bar{q'}_{phys\,j}^T(-\vec{x},t) \ \ , \label{trafo1}\\
 \bar{q'}_{phys\,j}(\vec{x},t) &\stackrel{\rm \cp}{\longrightarrow}& -{q^\prime}_{phys\,j}^T(-\vec{x},t){\cal C}^{-1} \gamma_0 e^{-2i\theta'_j} , \label{trafo2}
 \end{array}
 \right. \label{cpoperator}
 \ee
  where $(j=d,s,b)$ and similar
 relations hold for the up sector but with no primes in the $q$'s
 nor in the $\theta$'s.  Observe then, how in \eq{cpoperator} is
 naturally defined each $\cp$ operator:  its label through the
 diagonal matrices $\ti$ determines the transformation phases of
 each quark and therefore the '$\cp$-$operation$'.

This definition allows us to express formally the analysis of CP in
the \sm and, as it will be shown in this work, to extract firm
sentences and conclusions of what is known as the {\it CP tag}.

The picture of CP in the \sm Lagrangian can now be written as
follows: There is no choice of $\ti$ such that a $\cp$ operator
commutes with the Lagrangian,
 \be
 \not\exists \ \ti \ \mbox{such that} \ [ \cp \ ,L  ] =0 .
 \ee
 Whereas if the charged current piece ($L^{cc}$) is
extracted from the Lagrangian then the remaining Lagrangian commutes
with $any$ $\cp$, \be [ \cp \ ,\  L - L^{cc} ] =0, \ \ \forall
(\Theta, \Theta^\prime) . \ee Moreover, the crucial point for the
definition of the CP-tag is that {\it although there is no $\cp$
which commutes with all the $L^{cc}$, for some pieces of it there
exist precise $\cp$ which do commute.}  And thus there exist some
decays which conserve $\cp$, hence providing a CP-tag.  This idea is
displayed in Chapter \ref{cp}.

\comentariodelta{

%%%%%%%%%%%%GGGGGGGGGGGGGGGGGGGGG mierda!!!

%

The here given description of the Electro-Weak Lagrangian shows that
the two pieces which have potential problems with CP invariance are
the charged current term in \eq{eq:LfG} and the mass terms in
\eq{L.Yukawa}.  We next study the CP invariance condition by
imposing invariance in each of these two pieces.

\section{CP invariance condition \hecho{90}}

%de aca

} %fin de comentario delta!!!
\section{The CKM matrix: hierarchy of flavour mixing \hecho{90}}

In Section \ref{condition} it was shown how in the physical basis
all the CP-violation gets reduced to occur in the charged current
term, moreover, in the CKM matrix $V$ of flavour mixing.  In this
section we advocate to study in depth the CKM matrix $V$, its
parameterizations and its experimental values which define a
hierarchy of flavour mixing in the different families.
\subsection{Parametrization of the CKM matrix \hecho{90}}
\subsubsection{Standard Parametrization}
In the case of three generations, three Euler-type angles and one
{\it complex phase} are needed to parameterize the CKM matrix. This
complex phase allows us to accommodate CP-violation in the Standard
Model, as was pointed out by Kobayashi and Maskawa in 1973
\cite{km73}. In the ``standard parametrization'' \cite{SMpar}, the
three-generation CKM matrix takes the following form:
\begin{equation}\label{standard}
\hat V_{\rm CKM}=\left(\begin{array}{ccc}
c_{12}c_{13}&s_{12}c_{13}&s_{13}e^{-i\delta_{13}}\\ -s_{12}c_{23}
-c_{12}s_{23}s_{13}e^{i\delta_{13}}&c_{12}c_{23}-
s_{12}s_{23}s_{13}e^{i\delta_{13}}&
s_{23}c_{13}\\ s_{12}s_{23}-c_{12}c_{23}s_{13}e^{i\delta_{13}}&-c_{12}s_{23}
-s_{12}c_{23}s_{13}e^{i\delta_{13}}&c_{23}c_{13}
\end{array}\right),
\end{equation}
where $c_{ij}\equiv\cos\theta_{ij}$ and $s_{ij}\equiv\sin\theta_{ij}$.
Performing appropriate redefinitions of the quark-field phases, the real
angles $\theta_{12}$, $\theta_{23}$ and $\theta_{13}$ can all be made to
lie in the first quadrant. The advantage of this parametrization is that
the generation labels $i,j=1,2,3$ are introduced in such a way that
the mixing between two chosen generations vanishes if the corresponding
mixing angle $\theta_{ij}$ is set to zero. In particular, for
$\theta_{23}=\theta_{13}=0$, the third generation decouples, and we arrive
at a situation characterized by the Cabibbo $2\times 2$ matrix.
\begin{figure}[t]
\makebox[\textwidth][c]{
\framebox[1\textwidth]{
\includegraphics[width=.57\textwidth]{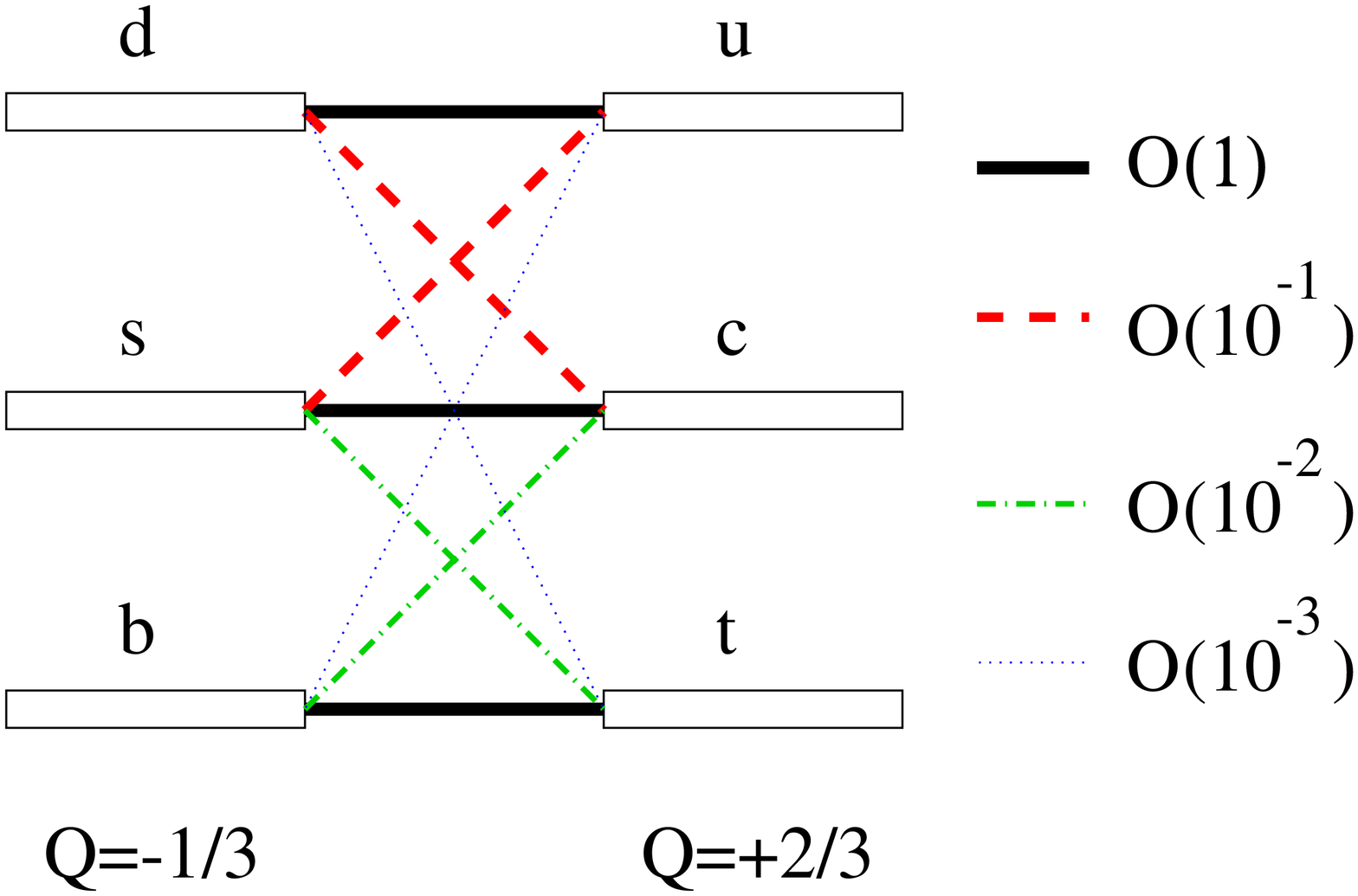}
}}
\caption[]{Hierarchy of the quark transitions mediated through charged
currents.}\label{fig:term}
\end{figure}
\subsubsection{Wolfenstein Parametrization}
In \fig{fig:term}, we have illustrated the hierarchy of the
strengths of the quark transitions mediated through charged-current
interactions: transitions within the same generation are governed by CKM
elements of ${\cal O}(1)$, those between the first and the second
generation are suppressed by CKM factors of ${\cal O}(10^{-1})$, those
between the second and the third generation are suppressed by
${\cal O}(10^{-2})$, and the transitions between the first and the third
generation are even suppressed by CKM factors of ${\cal O}(10^{-3})$.
In the standard parametrization \eq{standard}, this hierarchy
is reflected by
\bea
s_{12}&=&0.222 \nn\\
s_{23}&=&{\cal O}(10^{-2})\\
s_{13}&=&{\cal O}(10^{-3}).\nn
\eea
If we introduce a set of new parameters $\lambda$, $A$, $\rho$ and $\eta$
by imposing the relations \cite{schu,blo}
 \bea
 s_{12}&\equiv&\lambda=0.222,\nn\\
 s_{23}&\equiv& A\lambda^2,\label{set-rel}\\
 s_{13}e^{-i\delta_{13}}&\equiv& A\lambda^3(\rho-i\eta),\nn
 \eea
 and go back to the standard parametrization \eq{standard}, we obtain an {\it exact} parametrization of the CKM matrix as a function of $\lambda$
(and $A$, $\rho$, $\eta$). Now we can expand straightforwardly each CKM
element in the small parameter $\lambda$. Neglecting terms of
${\cal O}(\lambda^4)$, we arrive at the famous ``Wolfenstein
parametrization'' of the CKM matrix \cite{wolf}:
 \begin{equation}\label{W-par}
 \hat V_{\mbox{{\scriptsize CKM}}} =\left(\begin{array}{ccc}
 1-\frac{1}{2}\lambda^2 & \lambda & A\lambda^3(\rho-i\eta) \\
 -\lambda & 1-\frac{1}{2}\lambda^2 & A\lambda^2\\
 A\lambda^3(1-\rho-i\eta) & -A\lambda^2 & 1
 \end{array}\right)+{\cal O}(\lambda^4).
 \end{equation}
 Since this parametrization makes the hierarchy of the CKM matrix explicit,
it is very useful for phenomenological applications.

\subsection{Requirements for CP Violation \hecho{90}}
As we have seen in the previous Section, at least three
generations are required to accommodate CP-violation in the Standard
Model. However, still more conditions have to be satisfied for
observable CP-violating effects.  These conditions can be obtained
by working in the weak basis and imposing $$\Phi \equiv \Phi'$$ to assure
CP invariance in the charged current terms (see \eq{argeval}).  In
this case the necessary and sufficient condition for CP invariance
is found in \eq{eq:mass.inv}.  Using the matrices
\bea
H&=&M\, M^\dagger \nn \\
H'&=&M' \, M'^\dagger \nn \eea
 these conditions read
  \bea
\begin{array}{rcl}
\Phi^\dagger H \Phi &=& H^* =H^T \\
\Phi^\dagger H' \Phi &=& H'^* =H'^T .
\end{array}
\label{asado} \eea These are necessary and sufficient conditions for
CP invariance with any number of generations.  In the case of three
generations \eq{asado} is equivalent to \cite{BBG} \bea Tr\, [ H,H'
]^3 = 0 . \label{traza} \eea To see the physical meaning of this
condition we can work in the weak basis where $H$ is diagonal,
$H=diag\{m_u^2,m_c^2,m_t^2\}$, to get
 \bea (m_t^2 -m_c^2)(m_c^2-m_u^2)(m_t^2-m_u^2)\, Im[H'_{12} H'_{23} H'_{31}] =0 .
 \label{queseyo}
 \eea
  It is clear that if we would have chosen to work in the basis where
$H'$ is diagonal, then we would have obtained a condition analogous
to \eq{queseyo} but with the down sector masses and the imaginary
part of the $H_{ij}$ elements instead.  Therefore we see that a
condition for CP invariance is to have no degeneracy in the masses
of the up and down sectors. Moreover, it can be seen \cite{jarlskog}
that \eq{queseyo} contains, beside of the relationship between the
masses, the condition of reality of the CKM matrix.  Adding up all
together we write the necessary and sufficient condition for CP
invariance as
\begin{eqnarray}\label{CP-req}
\lefteqn{(m_t^2-m_c^2)(m_t^2-m_u^2)(m_c^2-m_u^2)}\nonumber\\
&&\times(m_b^2-m_s^2)(m_b^2-m_d^2)(m_s^2-m_d^2)\times
J_{\rm CP}\,\not=\,0,
\end{eqnarray}
where
\begin{equation}
J_{\rm CP}=|\mbox{Im}(V_{i\alpha}V_{j\beta}V_{i\beta}^\ast
V_{j\alpha}^\ast)|\quad(i\not=j,\,\alpha\not=\beta)\,.
\end{equation}

The factors in \eq{CP-req} involving the quark masses are related
to the fact that the CP-violating phase of the CKM matrix could be
eliminated through an appropriate unitary transformation of quark
fields if any two quarks with the same charge had the same mass.
Consequently, the origin of CP-violation is not only closely related
to the number of fermion generations, but also to the hierarchy of
quark masses and cannot be understood in a deeper way unless we have
insights into these very fundamental issues, usually referred to as
the ``flavour problem''.

The second ingredient of \eq{CP-req}, the ``Jarlskog Parameter''
$J_{\rm CP}$ \cite{jarlskog}, is sometimes interpreted as a
measure\footnote{The problem is that there is no unique
CP-conserving reference in the Standard Model, so that the
"relative" value of $J_{CP}$ can be either small or large.  If the
flavour mixing amplitude is $\lambda$ (as in $K$-physics) or
$\lambda^2$ (as in $B_s$-physics), CP-violation is small; if is
$\lambda^3$ (as in $B_d$-physics), CP-violation is large because the
CP-conserving probability is also $\lambda^6$.} of the strength of
CP-violation in the Standard Model. It does not depend on the chosen
quark-field parametrization, i.e.\ is invariant under
\eq{eq:rephase}, and the unitarity of the CKM matrix implies that
all combinations $|\mbox{Im}(V_{i\alpha}V_{j\beta}V_{i\beta}^\ast
V_{j\alpha}^\ast)|$ are equal. Using the Standard and Wolfenstein
parameterizations, we obtain
\begin{equation}
J_{\rm CP}=s_{12}s_{13}s_{23}c_{12}c_{23}c_{13}^2\sin\delta_{13}=
\lambda^6A^2\eta={\cal O}(10^{-5}),
\end{equation}
where we have taken into account the present experimental information
on the Wolfenstein parameters in the quantitative estimate. Consequently, CP-violation is a small
effect in the Standard Model. Typically, new complex couplings are present
in scenarios for new physics, yielding additional sources for CP-violation.

\subsection{The Unitarity Triangles \hecho{90}}

\begin{figure}[t]
%\vspace{0.10in}
%centerline{
%\epsfysize=5.0truecm
%\epsffile{triangle.eps}}
\makebox[\textwidth][c]{
\framebox[1\textwidth]{
\includegraphics[width=.7\textwidth]{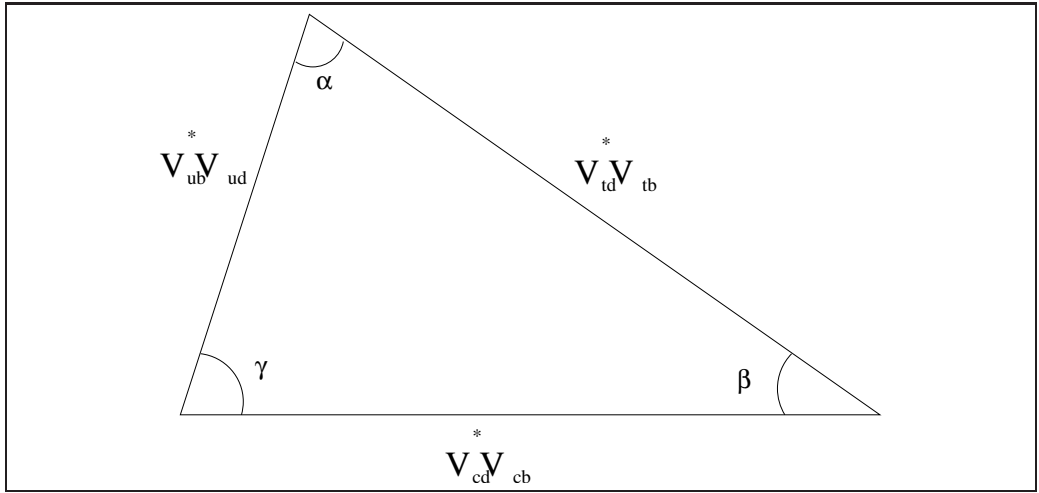}
}}
\caption[]{Unitarity triangle for the $bd$ sector where all sides are the same order of magnitude in the Wolfenstein parameter, namely $\lambda^3$.}
\label{triangle}
\end{figure}

Concerning tests of the Kobayashi--Maskawa picture of CP-violation,
the central targets are the ``unitarity triangles'' of the CKM
matrix. The unitarity of the CKM matrix, which is described by
\begin{equation}
\hat V_{\mbox{{\scriptsize CKM}}}^{\,\,\dagger}\cdot\hat
V_{\mbox{{\scriptsize CKM}}}=
\hat 1=\hat V_{\mbox{{\scriptsize CKM}}}\cdot\hat V_{\mbox{{\scriptsize
CKM}}}^{\,\,\dagger},
\end{equation}
implies a set of 12 equations, consisting of 6 normalization relations
and 6 orthogonality relations.   Of particular interest are the 6 orthogonality relations:
\begin{eqnarray}
V_{ud}V_{us}^\ast+V_{cd}V_{cs}^\ast+V_{td}V_{ts}^\ast & = &
0\quad\mbox{[1st and 2nd column]}\label{UT-ort1}\\
V_{ud}V_{ub}^\ast+V_{cd}V_{cb}^\ast+V_{td}V_{tb}^\ast & = &
0\quad\mbox{[1st and 3nd column]}\label{UT-ort2}\\
V_{us}V_{ub}^\ast+V_{cs}V_{cb}^\ast+V_{ts}V_{tb}^\ast & = &
0\quad\mbox{[2nd and 3rd column]}\label{UT-ort3}\\
&&\nonumber\\
V_{ud}^\ast V_{cd}+V_{us}^\ast V_{cs}+V_{ub}^\ast V_{cb} & = &
0\quad\mbox{[1st and 2nd row]}\label{UT-ort4}\\
V_{ub}^\ast V_{tb}+V_{us}^\ast V_{ts}+V_{ud}^\ast V_{td} & = &
0\quad\mbox{[1st and 3rd row]}\label{UT-ort5}\\
V_{cd}^\ast V_{td}+V_{cs}^\ast V_{ts}+V_{cb}^\ast V_{tb} & = &
0\quad\mbox{[2nd and 3rd row].}\label{UT-ort6}
\end{eqnarray}
These can be represented as six ``unitarity triangles'' in the complex plane \cite{AKL} -- see \fig{triangle}. It should be noted that the set of Eqs.\,(\ref{UT-ort1})--(\ref{UT-ort6}) is invariant under the phase transformations specified in \eq{eq:rephase}. If one performs such transformations, the triangles corresponding to Eqs.\,(\ref{UT-ort1})--(\ref{UT-ort6}) are rotated as a whole in the complex plane. However, the angles and sides of these triangles remain unchanged and are therefore physical observables. It can be shown that all six unitarity triangles have the same area \cite{JS}, which is given by the Jarlskog parameter as follows:
\begin{equation}
A_{\Delta}=\frac{1}{2}J_{\rm CP}.
\end{equation}

The shape of the unitarity triangles can be analyzed with the help of
the Wolfenstein parametrization, implying the following structure for
(\ref{UT-ort1})--(\ref{UT-ort3}) and (\ref{UT-ort4})--(\ref{UT-ort6}) respectively:
\begin{eqnarray}
{\cal O}(\lambda)+{\cal O}(\lambda)+{\cal O}(\lambda^5)&=&0\ \ \ [ds-\mbox{triangle}] \label{ds}\\
{\cal O}(\lambda^3)+{\cal O}(\lambda^3)+{\cal O}(\lambda^3)&=&0\ \ \
[bd-\mbox{triangle}]\label{bd}. \\
{\cal O}(\lambda^4)+{\cal
O}(\lambda^2)+{\cal O}(\lambda^2)&=&0\ \ \
[bs-\mbox{triangle}]\label{bs}
\end{eqnarray}
Where for purposes of this work we have remarked only to which
triangle of the down sector corresponds each relationship.

It is clear that the $ds$ and $bs$-triangles are much flatter than
the $bd$-triangle, and this will have important consequences.  As a
matter of fact we will show in Chapter \ref{cp} how it is possible
to take profit of it working in the Lagrangian only up to a given
order in $\lambda$.  The rest could be worked out in perturbation
theory.

%%%%%%%%%%%%%%%%%%%%%%%%%%%%%%%%%%%%%%%%%%%%%%%%%%%%%%%%%%%%%%%%%%%%%%%%%%%%%%%%
%%%%%%%%%%%%%%%%%%%%%%%%%%%%%%%%%%%%%%%%%%%%%%%%%%%%%%%%%%%%%%%%%%%%%%%%%%%%%%%%%
%%%%%%%%%%%%%%%%%%%%%%%%%%%%%%%%%%%%%%%%%%%%%%%%%%%%%%%%%%%%%%%%%%%%%%%%%%%%%%%%%
\cleardoublepage\chapter{The neutral meson system}
 \label{meson system}

 Throughout history the first neutral meson system to be
known was the Kaon system, first discovered by Rochester and Butler
in 1947 \cite{rb}. These particles, the $K^0$ and ${\bar K}^0$, with
the respective quark composition $d{\bar s}$ and ${\bar d}s$, have a
very special feature: {\it they are strong interactions eigenstates
which decay and may get converted one into the other through weak
interactions}. The consequences of this feature are rich in the
purpose of testing CP-violation in the weak interactions.  Since
these particles decay through weak interactions they have relatively
long lifetimes.  And since they can be mixed also through weak
interactions, any decay of one of the mesons with time evolution will include the
interference diagram of the other one.  In this way, if the charged
current weak Hamiltonian, responsible of the mixing between $K^0$
and $\bar K^0$, is not invariant under $any$ of the CP operators
defined in \eq{cpoperator} then CP-violating decays will occur in
the $K^0-\kk$ system.

With the later discovery of heavier quarks, other similar meson systems were predicted and discovered, the following table summarizes the situation

\begin{center}
\begin{tabular}{ccc}
{\bf meson system} & {\bf ~ particles} & {\bf ~ ~ ~ quark composition}\\
\hline
&&\\
Kaons & $K^0 ,\ \kk$ & $d{\bar s},\ {\bar d}s$\\
charmed mesons & $D^0 ,\ {\bar D}^0$& $c\bar u ,\ \bar c u$ \\
bottom mesons & $\b,\ \bb$ & $d\bar b ,\ \bar d b$\\
bottom strange mesons & $\b_s,\ \bb_s $&$s\bar b ,\ \bar s b$\\
\hline \\
&{\bf Table $3.1$}&
\end{tabular}
\end{center}

\section{The formalism for the decays in the neutral meson system}
\label{formalism} In 1955 Gell-Mann and Pais \cite{gp} developed the
theory for the neutral Kaon system, which can be adapted for all the
neutral meson systems.  The Kaon  decays should exhibit unusual and
peculiar properties arising from the degeneracy of $K^0$  and $\kk$.
Although $K^0$ and $\kk$ are expected to be distinct particles from
the point of view of strong interactions, they could transform into
each other through the action of weak interactions. Consequently,
$K^0$ or $\kk$ particles are not expected to decay in the simple
exponential manner usually associated with unstable particles,
instead they are rather described through the \ww \cite{ww}
formalism.  We describe now the results and tools of such a
formalism.

For the sake of clarity from now on we will not refer to an specific
meson system, but instead we will describe the two particles of any
of the meson system of Table $3.1$  as $\p$ and $\pp$.

\subsection{Time evolution in the \ww approximation \hecho{90}}
\label{evolution}
 The \ww approximation is applicable for two
particles $\p$ and $\pp$ which are unstable and decay through a
different variety of channels. The \ww formalism shows that once the
decay channels that connect the different particles are integrated
out, the picture can be understood with the help of a non-hermitian
effective Hamiltonian, $H$, which is responsible of the time
evolution and decay of these particles.

The different components of the $\H$ $2\times2$-matrix in the $\p,\,
\pp$ basis are
 \bea
 \begin{array}{lc}
 H_{ij} = \langle \p_i | H | \p_j \rangle ,\ \ & \p_1 = \p\, \ \ \p_2=\pp
 \end{array}
 \label{indice1}
 \eea
  Notice at this point that there is a non-global phase
indeterminacy in the $H_{ij}$ elements. This is due to the fact that
the $|\pp\rangle$ state has a free phase which comes from having the
CP-operator defined up to the phase matrices $\Theta$ as far as the
weak Hamiltonian is not included.  In this Chapter this
indeterminacy will remain, meanwhile in Chapter \ref{cp} we will see
how we can get rid of this indeterminacy through a useful choice of
the CP operator.

Since $\H$ contains as well the evolution as the decay information
of the meson system, it must contain an hermitian and anti-hermitian
part, $\H=\M - i/2 \GAMMA$ where \be \M = \frac{\H + \H^\dagger}{2}
\ \ \ \mbox{and} \ \ -i\GAMMA /2 = \frac{\H - \H^\dagger}{2}, \ee
both $\M$ and $\GAMMA$ hermitian matrices.

The time-evolution of any state living in the 2-dimensional
vector-space of $\p,\,\pp$ will be given through the {\it
time-evolving \ww eigen\-states}\footnote{Notice that in the
literature these states are usually identified as
"mass-eigenstates", in this work we will be more circumspect and we
will use the above-mentioned name (see Section \ref{apendice}).}. Id
est, the two eigenvectors\footnote{Notice at this point that the
fact of having the off-diagonal terms in $\H$ different from zero
implies the existence of two linear independent eigenvectors.} of
$\H$, namely $\P_1$ and $\P_2$ with corresponding eigenvalues
$\mu_1$ and $\mu_2$ evolve in time according to the Schro\-din\-ger
equation
 \bea
 H |\P_\alpha \rangle = \mu_\alpha |\P_\alpha \rangle
 \label{schrodinger}
 \eea
  which implies a time evolution through the non-unitary evolution operator
  matrix
 \be
 U(t) \, |\P_\alpha \rangle = e^{-i \mu_\alpha t} \, |\P_\alpha \rangle \ \ \ \ \alpha=1,\, 2.
 \label{mass states}
 \ee
 (The notation is such that Greek indices indicate a \ww eigenstate
meanwhile Latin indices indicate a flavour state.) Any given linear
combination of them evolves as the same linear combination of the
eigenstates evolved.  Of course, since the $\mu_\alpha\equiv
m_\alpha -i \, \Gamma_\alpha /2$ are complex, this time evolution
implies a non-conservation of the probability which is due to the
fact that the states living in this vector-space are unstable.

In this way all the time evolution of the system is reduced to find
$\H$'s eigenvectors and write the state function, $|\psi\rangle$, as
a linear combination of them.  Any confusion that might arise from
the fact of having non-ortho\-nor\-mal eigenvectors will be
clarified in the next paragraphs.

Generally speaking, when working with a non-orthonormal basis it is
necessary to differentiate between $coordinate$ and $component$ of a
vector in a given basis.  Id est, given a state function that may be
written as \be |\psi \rangle = \lambda_1 |\P_1\rangle + \lambda_2
|\P_2\rangle , \label{coordinates} \ee we say that $\lambda_1$ and
$\lambda_2$ are the $coordinates$ of $|\psi\rangle$ on the vector
basis.  Meanwhile, the $components$ of the vector on such a basis
are $\eta_1=\langle \P_1|\psi\rangle$ and $\eta_2= \langle \P_2
|\psi \rangle$.  If the basis $\{ \P_\alpha \}$ is non-orthonormal
then $\lambda_\alpha \neq \eta_\alpha$.

Once this is clear we realize that for the time evolution of any
given state-function it is necessary to extract the $coordinates$ of
it.  For the sake of concreteness in the following we write the
eigenvectors of $\H$ as
\begin{eqnarray}
\left(
\begin{array}{c}
| \P_1 \rangle \\ | \P_2 \rangle
\end{array}
\right)
=
\left(
\begin{array}{cc}
p_1 & q_1\\ p_2 & - q_2
\end{array}
\right)
\
\left(
\begin{array}{c}
| \p \rangle \\ | \pp \rangle
\end{array}
\right)
\equiv
\mbox{\boldmath $X$}^T\
\left(
\begin{array}{c}
| \p \rangle \\ | \pp \rangle
\end{array}
\right).
\label{PaPb}
\end{eqnarray}

As a result, the coordinates of any vector $|\psi\rangle$ in the \ww
basis are
 \be
 \lambda_\alpha = \left( \X^\dagger \X \right)^{-1}_{\alpha\beta}
 \langle \P_\beta | \psi \rangle .
 \label{coordinates-extraction}
 \ee
  The state into which $|\psi\rangle$ is being projected in
\eq{coordinates-extraction},
 \be
 |\tilde \P_\alpha \rangle = \left( \X^T
 \X^* \right)^{-1}_{\alpha\beta} | \P_\beta \rangle = X^{*-1}_{\alpha i} |\p_i \rangle ,
 \label{reciprocal}
 \ee
  may be found in the literature under the name of reciprocal basis
\cite{silva} or out-state \cite{ag}.  We recall that in terms of
$\p$ and $\pp$ the particle composition of $\tilde \P_\alpha$
differs from that of $\P_\alpha$.  In fact, the reciprocal basis, or
out-state basis $\{\tilde \P_\alpha \}$, {\it has no physical
meaning}, it is only a tool to extract the coordinates of any given
vector and that is the only use that shall be given to it.

As it can be easily seen from \eq{reciprocal} the reciprocal basis
will differ from the regular basis only in the case that the set of
eigenvectors of the Hamiltonian is not orthogonal.  Notice that the
non-orthogonality of $\H$'s eigenvectors is not necessarily a
consequence of non-hermiticity; the causes are analyzed in the
following paragraphes.

The operator for the effective Hamiltonian is written as
\bea
 H &=& |\P_1\rangle \mu_1 \langle \tilde \P_1 | + |\P_2 \rangle \mu_2 \langle \tilde \P_2 | \nonumber \\
&~& \nonumber \\
&=& \left( |\P_1\rangle , \, |\P_2 \rangle \right)
\left(
\begin{array}{cc}
\mu_1&0\\
0&\mu_2
\end{array}
\right)
\left(
\begin{array}{c}
\langle\tilde \P_1 | \\
\langle\tilde \P_2 |
\end{array}
\right) .
\label{operador H}
\eea
Id est, the Hamiltonian is being diagonalized through the eigenvectors' matrix as
\bea
\X^{-1} \H \X =
\left(
\begin{array}{cc}
\mu_1&0\\
0&\mu_2
\end{array}
\right)
\eea

Using the reciprocal basis \eq{reciprocal} to extract the
coordinates of the state-function, it is easy now to get the time
evolution due to the Schrodinger Equation (\eq{schrodinger}):
 \be
|\psi (t) \rangle =  \left[ e^{-i\mu_1 t} \langle \tilde \P_1
|\psi(0)\rangle \right] |\P_1 \rangle + \left[ e^{-i\mu_2 t} \langle
\tilde \P_2 |\psi(0)\rangle \right] |\P_2 \rangle . \ee

\subsection{T, CP and CPT analysis in the time evolution (mixing)\hecho{84}}
\label{analysys}
\nota{esta escrito todo.  Falta volver a analizar
bien las logica y las conclusiones (rephasings, Re($\epsilon$), etc).} In the previous section the time evolution of the
neutral meson states has been obtained within the \ww approximation.
All the information for such evolution is contained in the effective
Hamiltonian $\H$ matrix or, alternatively, in the eigenvectors
matrix $\X$.  In this section we study the different properties of
these two matrices and their relationship to the discrete symmetries
in the time evolution, which will prove to be useful for a better
understanding of the system.

To begin the analysis, we first show how the symmetry-operations C,
P, and T on $H$ have immediate consequences on the matrix $\H$
written in the flavour basis.  Acting on each matrix element we
find, \bea
 \CPT\ok \ &\then& \H_{11}=\H_{22} \label{mic1}\\
 \CP\ok \ &\then& \H_{11}=\H_{22} \mbox{ and }|\H_{12} | = |\H_{21}| \label{mic2}\\
 \T \ok &\then& |\H_{12}| = |\H_{21}| \label{mic3}
\eea where the moduli are due to the relative phase indeterminacy
between the $|\p\rangle$ and $|\pp\rangle$ states.

The eigenvectors of $\H$ should be simultaneously eigenstates of CP
if this were a good symmetry. Instead, under CP, T and CPT violation, the
matrix $\X^T$ in \eq{PaPb} can be parameterized as \bea \X^T= \left(
\begin{array}{cc}
p_1 & q_1\\ p_2 & - q_2
\end{array}
\right)
=
\left(
\begin{array}{cc}
1+\epsilon_1 & 1-\epsilon_1\\
1+\epsilon_2 & -(1-\epsilon_2 )
\end{array}
\right) , \label{sandia} \eea where the obvious normalization
constants in each row are not put in order to avoid visual
saturation.  It should be clear that these parameters are
rephasing-variant, id est changes in the relative phases of the
states are reflected in a change in the epsilons.  This
indeterminacy will be removed in Chapter \ref{cp} when the
$\pp(=\bar B ^0)$ state is defined through a given $CP$-operator.

The physical meaning of the parametrization in \eq{sandia} is clear,
assuming $|\pp\rangle = \CP\, | \p \rangle $, $\H$'s eigenvectors
are parameterized as almost $CP$ eigenvectors as far as
$\epsilon_{1,2}$ are small, id est \bea
|\P_1 \rangle &=& |\p_+\rangle + \epsilon_1 \p_- \rangle  \label{e1}\\
|\P_2 \rangle &=& |\p_-\rangle + \epsilon_2 \p_+ \rangle \label{e2}
\eea where
 \be |\p_{^+_-} \rangle = \frac{1}{\sqrt{2}} \left( \p \pm
 \pp ) \right)
 \ee
(It is clear that if $\epsilon_{1,2}$ are not small, the parametrization is still valid and general.)

In order to have a more direct connection between the discrete
symmetries and the parameters, it is convenient to define \be
\epsilon=\frac{\epsilon_1 +\epsilon_2 }{2} \ \ \ \ \ \ \ \ \ \
\delta =\frac{\epsilon_1 -\epsilon_2 }{2}. \ee   In this way the T,
CP and CPT symmetries or violations can be easily parameterized.  It
is convenient to note at this point that if, as supported by
experiments \cite{pdg}, $\delta$ is very small compared to
$\epsilon$, then $Re(\epsilon)\neq0$ is needed to guarantee that
$\epsilon$ cannot be rephased to zero.  In fact, it is easy to see
that if we have $\epsilon = i\, x$ then the rephase

\begin{eqnarray}
\begin{array}{c}
\p\to \p e^{-i \arctan(x)} \\
\pp \to \pp e^{i \arctan(x)}
\end{array}
\label{rephase}
\end{eqnarray} makes $\epsilon=0$ in Eqs.\ (\ref{e1}-\ref{e2}).

To analyze the relationship between the parameters
($\epsilon,\delta$) and the discrete symmetries it is instructive to
compute the evolution operator matrix in the flavour basis
$\left\{\p,\pp\right\}$, which reads
\begin{eqnarray}
U(t) &=& X \left(
\begin{array}{cc}
e^{-i\mu_1 t} & 0 \\
0 & e^{-i\mu_2 t}
\end{array}
\right) X^{-1}
\label{udet}\\
&=& e^{-(i\bar m + \frac{\G}{2})t} \left(
\begin{array}{ll}
\cosh(\alpha t) - K_\delta \sinh (\alpha t) & - K_+ \sinh (\alpha t) \\
 - K_- \sinh (\alpha t) & \cosh(\alpha t) + K_\delta \sinh (\alpha t)\\
\end{array}
\right) \nonumber
\end{eqnarray}
where
 \bea
 \alpha &=& i\frac{\Dm}{2} + \frac{\DG}{4}, \nn\\
 K_{\pm} &=& \frac{(1\pm\epsilon)^2 - \delta^2}{1 - \epsilon^2 + \delta^2} \label{micaela bus 170}\\
 K_\delta &=& \frac{2\delta}{1 - \epsilon^2 + \delta^2}, \label{mica2}
 \eea
 $\DG=\Gamma_1-\Gamma_2,\ \Dm=m_1-m_2,\ \G=(\Gamma_1+\Gamma_2)/2$ and $\bar m = (m_1+m_2)/2$.

 Under a rephasing of the states $\b \mapsto e^{i\gamma},\ \bb \mapsto e^{-i\bar\gamma}$ the parameters rephase as
  \bea
  K_{\pm} &\mapsto& K_{\pm}e^{\pm i(\gamma-\bar \gamma)} \label{rephase1}\\
  K_{\delta} &\mapsto& K_\delta \label{rephase2} .
  \eea
   Hence, there are five rephasing invariant parameters ($|K_+|,\ |K_-|$,
$Re(K_\delta)$, $Im(K_\delta)$ and the phase of $K_+ K_-$) which,
due to the two real relationships contained in
  \be
  K_+\, K_- = 1 - K_\delta ^2 ,
  \ee
  get reduced to {\it three physical parameters}.

  In terms of $\epsilon$ and $\delta$ the number of parameters must
be the same.  In particular this is easily seen for the case of
$B_d$ mesons where a small $\DG$ implies a small $Re(\epsilon)$ (see
the demonstration below).  In this case, we have that to first order
in $Re(\epsilon)$ and $\delta$,
\eqs{micaela bus 170}{mica2} say
 \bea
 |K_{\pm}|^2 &=& 1 \pm \frac{4Re(\epsilon)}{1+\mec}, \nn\\
 K_\delta &=& \frac{2\delta}{1+\mec} \nn,
 \eea
  and hence the three physical parameters which are rephasing
invariant in this approximation (valid for the $B_d$-system) are
 \bea \left\{ \frac{Re(\epsilon)}{1+\mec},\, \frac{Re(\delta)}{1+\mec},\, \frac{Im(\delta)}{1+\mec} \right\} \longrightarrow \mbox{rephasing invariants}
 \label{sakuraII}
 \eea
  It is worth to note at this point that if, as in Chapter \ref{cp},
the relative phase of the states is {\it defined} by a CP-conserving
direction then the {\it four parameters} included in $\epsilon$ and
$\delta$ would have physical meaning; id est, $Im(\epsilon)/1+\mec$
should be added in \eq{sakuraII}.

 With the evolution operator written explicitly in the
flavour basis (see \eq{udet}), the T, CP and CPT transformed expressions
of $U(t)$ are obtained straightforward.  In fact, the T operation
transposes the $U(t)$ matrix; the CP operation transposes it and
exchanges $U_{11}\leftrightarrow U_{22}$; and the CPT operation
exchanges $U_{11}\leftrightarrow U_{22}$.  Therefore, from the
matrix expression of the $U$-operator we obtain the behaviour of the
T, CP and CPT transformed operator as a function of $\epsilon$ and
$\delta$:
\begin{eqnarray}
\T\, U(\epsilon,\delta)\, \T^\dagger &=& U(-\epsilon, \delta) \\
\CP\, U(\epsilon,\delta)\, \CP^\dagger &=& U(-\epsilon, -\delta) \\
\CPT\, U(\epsilon,\delta)\, \CPT^\dagger &=& U(\epsilon, -\delta)
\end{eqnarray}
From here, having into account that $Re(\epsilon)\neq0$ is the
requirement to guarantee that $\epsilon$ cannot be rephased to zero
(see \eq{rephase}), the relationship between the parameters and the
discrete symmetries in the time evolution (mixing) follows
\begin{eqnarray}
\T\ok &\iff& Re(\epsilon)=0, \label{carlita1}\\
\CP\ok &\iff& Re(\epsilon)=0 \and \delta=0, \\
\CPT\ok &\iff& \delta=0 \label{carlitagianninazilli}.
\end{eqnarray}
It is also useful to see the implications of measuring a non-zero value for one
of the parameters,
\begin{eqnarray}
Re(\epsilon)\neq 0 &\then& \tv \and \cpv \\
\delta\neq 0 &\then& \cpv \and \cptv \label{carlita2}.
\end{eqnarray}

Conclusions in Eqs.\ (\ref{carlita1}-\ref{carlita2}) are general and
their understanding implies a good insight into the formalism of the
neutral meson system.

It is also interesting to obtain the Hamiltonian operator
parameterized in the flavour basis.  This is easily done using the
expression for the evolution operator in \eq{udet} and the expansion for small $t$
 \be
 U(t) \approx 1 - i H t;
 \ee
we obtain
 \bea
 H =
 \left(
 \begin{array}{cc}
 \bar m - i\frac{\G}{2} - i K_\delta \,\alpha & - i K_+ \, \alpha \\
 - i K_- \,\alpha & \bar m - i\frac{\G}{2} + i K_\delta \, \alpha
 \end{array}
 \right) .
 \eea

~

Up to here we have worked with $\epsilon$ and $\delta$ at the same
level.  In nature, though, CP and T violations are much greater than
CPT violation (if any).  Hence, in virtue of Eqs.\
(\ref{carlita1}-\ref{carlitagianninazilli}) we can work out
interesting conclusions approximating $\delta\approx0$.  In fact,
when studying T and CP violation this is an excellent approximation.
In the remaining of this Section we assume CPT invariance, and
thus $\epsilon=\epsilon_1=\epsilon_2$.

Notice, in \eq{sandia}, that if CPT is assumed then $p_1=p_2\equiv p$
and $q_1=q_2\equiv q$ and $\X^T$ is simply
 \be
 \X^T = \left(
\begin{array}{cc}
p&q\\
p&-q
\end{array}
\right) . \label{equis}
 \ee
 Moreover $p$ and $q$ values in term of
$\H$'s elements are trivial, \be \frac{q}{p} = \pm \sqrt{
\frac{H_{21}}{H_{12}} } , \label{pirulin} \ee where the $\pm$ sign
comes from the two different eigenvectors.  A later assignment --
through the eigenvalues -- of the eigenvectors to each column of
$\X$ will eliminate this indeterminacy.

Taking the positive sign in \eq{pirulin} and using \eq{sandia} yields
\be
\epsilon = \frac{\sqrt{H_{12}}-\sqrt{H_{21}}}{\sqrt{H_{12}}+\sqrt{H_{21}}} .
\label{epsilon}
\ee
It is interesting to notice that for the case of the neutral mesons
having similar mean-lives -- as with the $B_d$ -- the expression in
\eq{epsilon} can be expanded in terms of the small
rephasing-invariant parameter $\frac{\Gamma_{21}}{2M_{21}}$.  The
result, renormalized to $1+|\epsilon|^2$, is
 \bea
 \frac{Re\, \epsilon}{1+|\epsilon|^2} &=& \frac{1}{2} Im \left( \frac{\Gamma_{21}}{2M_{21}} \right) + O \left(\left[ \frac{\Gamma_{21}}{2M_{21}} \right]^2 \right) \label{real}\\
 \frac{Im\, \epsilon}{1+|\epsilon|^2} &=& - \frac{1}{2} \sin \phi + O
 \left( \frac{\Gamma_{21}}{2M_{21}}  \right)\label{imaginaria}
 \eea where $M_{21}= |M_{21}| e^{i\phi}$.  As we see, we retrieve
again that while \eq{imaginaria} is phase dependent, \eq{real}
is rephasing-invariant.  Also notice that in \eq{real} we have the proof that for
small $\DG$, $Re(\epsilon)$ is also small and proportional to it.

Also a different expression for $\epsilon$ may be obtained through
the elements of $\GAMMA$ and $\M$.  Using that the eigenvalues of
$\H$ accomplish \be 2\sqrt{H_{12} H_{21}} = \mu_1 - \mu_2 = \Delta m
-\frac{i}{2} \Delta\Gamma \label{chucrut} \ee in \eq{epsilon} it is
straightforward to get
 \be
 \epsilon= \frac{Im(\Gamma_{12}) + 2i Im(M_{12})}{2 Re (M_{12}) - i Re (\Gamma_{12}) + \Delta m - \frac{i}{2} \Delta\Gamma} .
 \label{epsilon2}
 \ee
Also, from the real and imaginary part of the square of \eq{chucrut} other two practical relations may be obtained, namely
\bea
|M_{12}|^2 - \frac{1}{4} |\Gamma_{12}|^2 &=& \Delta m \, ^2 - \frac{1}{4} \Delta\Gamma \, ^2 \\
Re \left( \Gamma_{12}M_{12}^* \right) &=& \Delta m \, \Delta \Gamma .
\eea

At last, in order to recompile the relationship between all the
variables and parameters in game, we analyze the non-orthogonality of
the \ww eigenstates \footnote{A complete probabilistic analysis of
this non-orthogonality is given in Section \ref{apendice}.}.
Assuming CPT invariance, the internal product of the two \ww
eigenstates is directly proportional to
 \be \langle \P_1|\P_2\rangle \propto 1 - \left| \frac{H_{12}}{H_{21}} \right| \propto 4 Re \, \epsilon  .
 \label{laurita}
 \ee Hence the real part of $\epsilon$ is also directly related to
the non-orthogonality of the eigenvectors.  Besides this, another relationships can be
found by studying the causes for $|H_{12}| \neq |H_{21}|$ or
$|H_{12}| = |H_{21}|$ with the matrices $\M$ and $\GAMMA$.  It is easy to see that
 \be
 |H_{12}| = |H_{21}| \ \iff \ Im(M_{12}\Gamma_{12}^*) = 0 .
 \label{sakura}
 \ee
Moreover, for CPT$\ok$ it is valid that
 \be
 Im(M_{12}\Gamma_{12}^*) = 0 \iff [M,\Gamma]=0 .
 \ee
On the other hand, for the mass and life-time difference the implication is only one
way,
 \be
 M_{12}\ (or\ \Gamma_{12}) = 0 \then
 \left\{
 \begin{array}{l}
 Im(M_{12}\Gamma_{12}^*) = 0 \\
 \Dm\ (or\ \DG) = 0 .
 \end{array}
 \right.
\label{350}
 \ee

We can summarize all the information by writing
 \bea
 \left.
 \begin{array}{l}
 \cpv \mbox{ in the mixing} \\
 \tv \mbox{ in the mixing} \\
 Re(\epsilon)\neq 0 \\
 \frac{|H_{12}|}{|H_{21}|} \neq 1 \\
 \langle \P_1 | \P_2 \rangle \neq 0 \\
 ~[M,\Gamma] \neq 0 \\
 Im(M_{12}\Gamma_{12}^*) \neq 0
 \end{array}
 \right\} \mbox{ ~ ~ are equivalent (if CPT $\ok$),}
 \label{cinthia}
 \eea
 If, on the contrary,
CPT would be violated then the analysis should be redone on the same steps (easy, but carefully): for instance, we
might have $\cpv$ but $Re(\epsilon)=0$, since it might be T$\ok$.

\section{The correlated neutral meson system \hecho{90}}
\label{correlated}
As a last section in the study of the meson system, we describe and analyze the so called {\it correlated neutral meson system}.  This is a very important topic, since a great part of the Kaon and $B$-meson experiments are performed nowadays in a correlated preparation at the Phi and B factories, respectively.

\nota{el factor 5 es del 2004...}

These factories began operating around 1999, and since then the B
factories (PEP-II and KEK-B) have exceeded their design peak
luminosity and greatly exceeded the expected integrated luminosity,
whereas the DAFNE Phi factory is still about a factor 5 below the
design peak and total integrated luminosity \cite{aceleradores}.  We
study here how the B factories produce correlated pairs of
$B$-mesons, the $\b\bb$ initial state and the notion of tagging; the working of the Phi factories is analogous but,
instead, produces pairs of correlated Kaons.

The essence of a B factory is to create in an asymmetric
electron-positron collider the upsilon $(4S)$ meson,
$$ e \, + \, \bar e \to \Upsilon (4S). $$
This is achieved by running the facility at an energy in the center
of mass system equal to the mass of the $\Upsilon (4S)$.   Once the
$\Upsilon(4S)$ has been created, its decay is more than 96\% of the
times, through the Strong Interactions, to a $B\bar B$ pair, and
from it half of the times corresponds to a correlated neutral $\b\bb$ pair
\cite{pdg}.

The interesting physics for the purposes of this work comes out when
the B mesons are neutral. In this case they mix each other and are
indistinguishable in their weak decays, since there is no
superselection rule to distinguish them, and therefore they must
obey Bose statistics.  The indistinguishability implies that the
physical neutral meson-antimeson state must be $symmetric$ under the
combined operation $C\parity$, with $C$ the charge conjugation and
$\parity$ the operator that permutes the spatial coordinates.
Specifically, assuming conservation of angular momentum, and a
proper existence of the $antiparticle$ $state$, one observes that
for $\b\bb$ states which are $C$ conjugates with $C=(-1)^\ell$ (with
$\ell$ the angular momentum quantum number), the system has to be an
eigenstate of $\parity$ with eigenvalue $(-1)^\ell$.  Hence, for
$\ell=1$ (from the spin of the $\Upsilon$), we have that $C=-$,
implying $\parity=-$. As a result the initial correlated state
$\b\bb$ produced in a B factory can be written as
 \be |i\rangle =
 \frac{1}{\sqrt{2}} \left( |B^0 (-\vec{k}), \bar{B}^0 (\vec{k})
 \rangle - |\bar{B}^0
 (-\vec{k}), B^0 (\vec{k}) \rangle \right),
 \label{initial3}
 \ee
  where $\vec{k}$ is along the direction of the momenta of the
 mesons in the center of mass system.  A system in a state as
 \eq{initial3} is called a correlated neutral $B$-meson system.

To summarize and for future purposes, we remark that \eq{initial3} was obtained using:
\begin{enumerate}
\item conservation of angular momentum in the $\Upsilon(4S)\to \b\, \bb$ decay,
\item  indistinguishability of $\b$ and $\bb$ in their weak decays, and hence the Bose statistics requirement $C\parity = +$.
\end{enumerate}

The correlation in \eq{initial3} makes very rich the experimental
and theoretical research since a first decay acts as a filter in a
quantum mechanical sense and gives a precise preparation for
analyzing the second meson still flying.  In fact, due to the
definite anti-symmetry in \eq{initial3} the time evolution of the
initial state leaves it, up to a global phase which we omit, with its original structure but attenuated
by an exponential factor,
 \bea
 U(t_1) \otimes U(t_1) |i\rangle &=& e^{-\G t_1} |i\rangle \\
 &=& e^{-\G t_1} \left(\frac{1}{\sqrt{2}}  |B^0 (-\vec{k}), \bar{B}^0 (\vec{k}) \rangle - |\bar{B}^0 (-\vec{k}), B^0 (\vec{k})
 \rangle .
 \right)\nn
 \eea
  Now suppose, for instance, a first decay $B \to X$ at time $t_1$,
then the state function of the second B meson at that time is projected to
 \be
 e^{-\G t_1}\, |B_{\bar X} \rangle =\frac{e^{-\G t_1}}{\sqrt{2}}\left[ \langle X|B^0\rangle \, |\bar B^0\rangle -
 \langle X | \bar B^0 \rangle \, |B^0\rangle \right]
 \ee
  which satisfies $\langle X|B_{\bar X} \rangle=0$.  (As required by
Bose statistics, the same decay simultaneously at both sides is
not allowed at any time after the creation of the initial state
\cite{lip68}.)  If the decay product $X$ is flavour
specific, we have a flavour tag and we can establish by the
conservation law which was the complementary first state $|B_X
\rangle$ decaying to $X$.  The goal in Chapter \ref{cp} is to define
an analogous tag, but for some CP-eigenstate decays.  Id est, a
CP-tag.

\section{{\bf Appendix:}\newline On the definition of probability u\-sing \-non-orthogonal basis}
\label{apendice}
\nota{necesita unificar notacion e incluir que ahora CPT $\ok$ se discute, no se asume}
As it was studied in Section \ref{formalism}, several experimental
facts in the neutral meson system may allow to have
non-orthogonality in the \ww eigenstates, $\langle \P_1 | \P_2
\rangle \neq 0$.

This feature needs a deeper discussion, since it means that when the
state is in one of the \ww eigenstates, then it has some projection
on the other such eigenstate.  And thus at first sight, thinking
with the usual Quantum Mechanics laws, there could be a transition.
In this Section we will try to clarify this issue:  we show that if
the \ww states are assumed to be asymptotic states, then the usual
Dirac definition of probability is inconsistent to a
gedanken-experiment we propose. In addition, we also show that some
other possible generalizations of the definition of probability are
also inconsistent,  we finally conclude that the \ww eigenstates
should be taken as {\it intermediate states} and not as {\it
asymptotic observable states}.

\subsection{The Dirac definition, possible generalizations,\\ and their problems \hecho{89}}

\nota{y, esta igual que en la tesina...} Generally speaking, the $\{
\p,\, \pp \}$ space may be thought as an $n$-dimensional Hilbert
space $\h$, and the state function being $u\rangle \in \h$.  Using
the Quantum Mechanic postulates in a gedanken-experiment, we will
suppose that if a measure is taken through the Strong Interactions
Hamiltonian then the result of measuring an observable
distinguishable property of the particle when the state-function is
$u\rangle$ must yield that the particle is either $\p$ $or$ $\pp$.
Analogously, if in a gedanken-experiment we measure a distinguishable
property of the particle then the result must be either $\P_1$ $or$
$\P_2$.  This gedanken-experiment is the basis of this Appendix.

We will call $\{ v_i \rangle \}$ the orthogonal basis which
describes the $\{ \p,\ \pp \}$ basis.  And we will call $\{ v'_i
\rangle \}$ the basis which corresponds to the \ww eigenstate basis
$\{ \P_1,\, \P_2 \}$, which we will assume non-orthogonal.  As it
can be seen, in order to use a more abstract notation, in this
Section we adopt Latin indices for the flavour as well as for the
\ww basis; the difference between them comes then solely from the
not primed and primed vectors, respectively.

In usual Quantum Mechanics, if the system is in the state $u \rangle
$ and $v_i \rangle $ corresponds to an observable, then the
probability of measuring $v_i$ is \be P(v_i|u) = | \langle v_i |
u\rangle|^2 = \langle u|v_i\rangle\langle v_i | u \rangle .
\label{dirac} \ee That is the component of $u\rangle$ on
$v_i\rangle$, moduli square.  This is the case for $\{v_i\rangle\}$
orthogonal basis.  In this case, the probability amplitude is given
by the coordinate, equal to the component, along $v_i \rangle $ of
the state $u\rangle$.

If, instead, we are dealing with a non orthogonal basis, as $\{v'_i \rangle \}$ then there are problems of conservation of probability.  Let $\X$ be the change of basis matrix,
\be
v'_j \rangle = X_{ij} \, v_i \rangle .
\ee
The reciprocal basis, introduced in \eq{reciprocal}, is then
\be
\tilde v '_j = X^{*\, -1}_{ji} \, v_i \rangle ,
\ee
and furnishes
\be
\langle \tilde v '_i, v'_j \rangle = \delta _{ij} .
\ee

Once introduced the notation, we propose the following two gedanken-experiments:

{\it \noindent $a)$ The state of the system is $u\rangle=v'_j\rangle$ and we want to have that if we measure one of the $\{v_i\rangle\}$ observables with the basis of the Strong Hamiltonian, then we will have {\it either} $v_1\rangle $ or$ \ v_2\rangle$ or $ \ldots$ or $v_n\rangle$.}  Therefore if all the probabilities are to sum one,
\be
\sum_i |\langle v_i|v'_j\rangle|^2 \equiv 1 \then \sum_i |X_{ij}|^2=1 \ \ \forall j
\label{4}
\ee

{\it \noindent $b)$ The other way around, $u\rangle = v_j\rangle$
and we want to have that if we measure observables that correspond
to the basis of the \ww effective Hamiltonian, then we will have
{\it either} $v'_1\rangle $ or$ \ v'_2\rangle$ or $ \ldots$ or
$v'_n\rangle$.}  Therefore \be \sum_i |\langle v'_i|v_j\rangle|^2
\equiv 1 \then \sum_i |X_{ji}|^2=1 \ \ \forall j. \label{5} \ee

If $\X$ is unitary then satisfies \eq{4} and \eq{5}, but if $\{ v'_i \rangle \}$ is non-orthogonal then $\X$ is non-unitary and --for the case of the neutral mesons-- it does not satisfy such conditions.  To see this we prove the anti-reciprocal:
\begin{theorem} If \eq{4} and \eq{5} are accomplished with the $2\times 2$ neutral mesons $\X$, \eq{equis}, then $Re\{\epsilon\}=0$ and thus $\X$ is unitary \end{theorem}

\noindent {\bf Dem:} quite simple, \eq{4} imposes the known condition for normalization of the eigenstates which, assuming CPT invariance, read
\be
|p|^2+|q|^2 =1 \then p,q = \frac{1\pm\epsilon}{\sqrt{2(1+|\epsilon |^2)}} .
\ee
Meanwhile \eq{5} imposes $2|p|^2=2|q|^2=1$ therefore $Re\{\epsilon\}=0$ as it was to prove. (q.e.d.)

Notice that this proposition is not true for any non-unitary $X$, there exist even $2\times 2$ non-unitary matrices which accomplish Eqs$.\,$(\ref{4},\ref{5}).

\vskip.6cm
Therefore we see that, for the neutral meson system, the usual Dirac definition of probability \eq{dirac} does not conserve the probability in the gedanken-experiment proposed.  The sum of the probabilities is not unity.

In order to generalize the definition of probability to non-orthogonal basis the following four reasonable conditions shall be required:
\renewcommand{\labelenumi}{\roman{enumi})}
\begin{enumerate}
\item let this definition be reduced to the usual Quantum Mechanics' Dirac probability when the basis happens to be orthogonal;
\item let the probability of any measurement be real;
\item let the probability of any measurement be non-negative;
\item let the experiments $a)$ and $b)$ behave well in the sense that the sum of the probabilities is one.
\end{enumerate}

Since it has been shown that Dirac's probability does not work for
non-orthogonal basis, a second choice is to replace \eq{dirac} and
define the probability as the moduli square of the $coordinates$ as
in \cite{silva}, \bea
\mbox{P}&=&coordinate \cdot coordinate^* \nonumber \\
P(v'_i|u) &\equiv& |\langle \tilde{v}'_i|u\rangle|^2 \label{dasilva}
\eea In this case \eq{4}, concerning experiment $a)$, remains the
same, but experiment $b)$ instead now implies \be \sum_i |\langle
\tilde{v}'_i|v_j\rangle|^2=1 \rightarrow \sum_i |X^{-1}_{ij}|^2 =1.
\label{8} \ee But if this is true, the following proposition shows
that $\X$ must be unitary, therefore by reduction ad absurd is
proved that this second choice for the definition of probability
\eq{dasilva} does not conserve probability for non-unitary $\X$.

\begin{theorem} If $\sum_i |X_{ij}|^2 = \sum_i |X^{-1}_{ij}|^2 =1$ for all $j$, then $\X$ is unitary. \end{theorem}

\noindent{\bf Dem:} first note that
\begin{equation}
\left.
\begin{array}{rcl}
(v,v)&=&1 \nonumber \\
(v,w)&=&1 \nonumber \\
v&\neq&w \nonumber
\end{array}
\right\}
\then \ (w,w) > 1.
\label{teo1}
\end{equation}
In fact, since $w \neq v$ is valid the strict inequality $|w|^2 + |v|^2 > 2 Re\{(w,v)\}$.  Therefore using the other two hypothesis we get $|w|^2=(w,w)>1$, as it was to show in \eq{teo1}.

In order to prove the proposition it is enough to prove the following:
\be
\left.
\begin{array}{l}
i) \X \mbox{ is not unitary} \nonumber \\
ii) \sum_i |X_{ij}|^2 = 1 \ \forall j \nonumber
\end{array}
\right\}
\then \exists j / \sum_i |X^{-1}_{ij}| \neq 1.
\label{teo2}
\ee
Using $i)$ we know that exist $i_0$ and $j_0$ such that
\be
X^{-1}_{i_0j_0} \neq X^*_{j_0i_0} .
\label{locas}
\ee
Therefore if $v_j = X^*_{ji_0}$ and $w_j = X^{-1}_{i_0j}$ then $ii)$ implies $(v,v)=1$, and $X^{-1} \cdot X =1$ implies $(v,w)=1$.  Since $v\neq w$ because of \eq{locas}, using \eq{teo1} we get that $(w,w)>1$, that is
\be
\sum_j |X^{-1}_{i_0j}|^2 > 1 .
\label{mayor}
\ee
If instead is $i'_0$ such that $X^{-1}_{i'_0j} = X^*_{ji'_0}\ (j=1\ldots n)$ then
\be
\sum_j |X^{-1}_{i'_0j}|^2 =1.
\label{igual}
\ee

Summing over all the cases of \eq{mayor} and \eq{igual} we get
\be
\sum_{i,j} |X^{-1}_{ij}|^2 > n.
\ee
Therefore exists at least one $j$ such that
\be
\sum_i |X^{-1}_{ij}|^2 \neq 1.
\ee
And this proves the proposition (q.e.d.).

\vskip.6cm
Motivated by this failure, we attempt a third try to define the probability in such a way that condition $iv)$ is furnished.  The next possibility which, in a symmetrical way automatically furnishes condition $ii)$, is
\bea
\mbox{P} &=&\frac{1}{2}(coordinate \cdot component^* + coordinate^* \cdot component )\nonumber \\
 P(v_i|u) &\equiv & {1\over 2} \left( \langle u | v_i\rangle\langle\tilde{v}'_i|u\rangle + \langle u | \tilde{v}'_i \rangle \langle v_i | u \rangle \right) .
\label{3rd try}
\eea
With this definition we see that experiment $a)$ gives again \eq{4} and thus imposes the normalization on the non-orthogonal states.  On the other hand experiment $b)$ imposes
\be
Re \left\{ \sum_i \langle v_j|v'_i\rangle\langle \tilde{v}'_i|v_j\rangle \right\} = 1 \then Re\left\{ \sum_i X_{ji}X^{-1}_{ij} \right\} = 1,
\label{11}
\ee
which is automatically accomplished, i.e. $a)$ and $b)$ are well behaved.  Thus the only condition that remains to be checked is $iii)$, i.e. the non-negativity of the probability.

We will show, for the neutral meson system, that given any state-function
\be
u\rangle = \lambda_i \, v_i\rangle ,
\label{u}
\ee
if $\X$ is non-unitary then there might be cases where the probability to measure the $\{v'_i\rangle \}$ observables will be negative.

\begin{theorem}With the definition of probability in \eq{3rd try} and given a state-vector as in \eq{u}, if for all $\lambda$ and $j$ is $Re\{ P(v'_j|u )\} \geq 0$ then the neutral meson matrix $\X$ is unitary \end{theorem}

\noindent{\bf Dem:} We have that the components and coordinates of $u\rangle$ on $v'_j\rangle$ are
\begin{eqnarray}
components: \lambda_i\langle v'_j,v_i\rangle = \lambda_i X^* _{ij}, \ \ \  (j)
\label{11bis} \\
coordinates:  \lambda_i\langle \tilde{v}'_j,v_i\rangle = \lambda_i X^{-1} _{ij}, \ \ \  (j)
\label{12}
\end{eqnarray}
where $(j)$ means that the index $j$ shall not be summed.  From it we have that
\be
P(v'_j | u ) = Re \left\{ \lambda^*_i X_{ij} X^{-1}_{jl} \lambda_l \right\} \equiv Re \left\{ \lambda^*_i P^{(j)}_{il} \lambda_l \right\} .
\label{tt}
\ee
Where we have defined
\be
P^{(j)}_{il} = X_{ij}X^{-1}_{jl}
\ee
which furnish
\be
\left.
\begin{array}{l}
 P^{(j)} \cdot P^{(k)} = \delta^{jk} P^{(j)} \nonumber \\
\sum_j P^{(j)}=1
\end{array}
\right\}
\label{14}
\ee
In order to extract the real part of $\lambda^*_i P^{(j)}_{il} \lambda_l$ we write, since
we are supported on the complex, $P$ as a sum of an hermitian and an anti-hermitian matrix, $H$ and $A$ (for the time being we are not writing the index $j$ on the $P$s)
\be
P=H+A, \ \  H=H^\dag, \ \ A=-A^\dag.
\label{19}
\ee
Therefore
\be
P(v'_j | u ) = Re\left\{ \lambda^*_i P_{il} \lambda_l \right\} = \lambda^*_i H_{il} \lambda_l ,
\label{20}
\ee
since $H$ is hermitian it can be diagonalized by a unitary matrix $C$,
\be
H=C^\dag D C,
\ee
$D$ diagonal with the eigenvalues of $H$ as the diagonal elements.  Therefore  \eq{20} becomes
\be
 P(v'_j | u ) = Re\left\{ \lambda^*_i P_{il} \lambda_l \right\} = \tilde{\lambda}^*_r D_{rs} \tilde{\lambda}_s = \sum_l D_{ll} |\tilde{\lambda}|^2
\label{21}
\ee
where $\tilde{\lambda}_s = C_{sl} \lambda_l$.  Observe that the $D_{ll}$ are real, thus we have to prove that if $D_{ll} \geq 0$ then $\X$ is unitary.

At this point the demonstration will get reduced to the case of the Kaons or $B$-mesons $2\times 2$ $\X$-matrix assuming CPT invariance.  In this case is valid
\be
P^{(1)} = \left( \matrix{1 \over 2&p\over 2q \cr q\over 2p&1\over 2} \right)
\then H = \left( \matrix{1\over 2 & {1\over4}(\frac{p}{q}+\frac{q^*}{p^*}) \cr
{1\over4}(\frac{p^*}{q^*}+\frac{q}{p})&1\over 2} \right) .
\ee
It is straightforward to see that both eigenvalues are greater or equal to zero iff $\left| \frac{p^*}{q^*}+\frac{q}{p} \right| < 2$, which is possible only if
\be
\left| \frac{p}{q} \right| = \left| \frac{1+\epsilon}{1-\epsilon} \right| = 1,
\ee
id est iff $Re\{\epsilon\} = 0$.  But this implies $\X$ being unitary, reduction ad absurd.  (q.e.d.)

 \vskip.6cm
\subsection{Conclusions \hecho{89}}
Through a gedanken-experiment, and four reasonable requirements for
a probability, in the last three propositions we have shown, using
different combinations of $components$ and $coordinates$, that they
cannot generalize the usual Dirac definition of probability for the
case of non-orthogonal -- claimed to be -- observables.  As it is
the case of the \ww eigenstates of the Kaons and $B$-meson system.
Of course, this should not be a surprise, since in Quantum Mechanics
the observables correspond to hermitian operators whose eigenvectors
are orthogonal.

Therefore, besides some other sophisticated combination of
components and coordinates, we conclude that the eigenstates of the
\ww approximation should be taken as $intermediate$ states which make
evolve the system, and not as asymptotic states -- at least for
$Re\{\epsilon\}\neq 0$.

\nota{esto es aventurado, pero me gusta}As an alternative point of
view, observe that in the \ww picture the time-evolving eigenstates
$|\P_j\rangle$ are the eigenstates of $H=M-\frac{i}{2}\Gamma$ with
complex eigenvalues, $\mu_j$.  On the other hand, we have the
eigenstates of the operator $M$, id est \be M |P^M_j\rangle =
\mu^M_j |P^M_j\rangle ; \ee and analogously the eigenstates of the
hermitian operator $\Gamma$, \be \Gamma |P^\Gamma_j\rangle =
\mu^\Gamma_j |P^\Gamma_j\rangle . \ee Using \eq{cinthia} and some
algebra, it is easy to show that \bea Re\{\epsilon\} \neq 0 \then
\left\{
\begin{array}{l}
i) \left\{ |P^\Gamma_j \rangle \right\} \neq \left\{ |P^M_j \rangle \right\} \neq \left\{ |\P_j \rangle \right\} \neq \left\{ |P^\Gamma_j \rangle \right\} \\
ii) Re\{\mu_j\} \neq \mu^M_j \nonumber \\
iii) Im\{\mu_j\} \neq -\frac{1}{2}\mu^\Gamma_j .
\end{array}
\right.
\eea
 Therefore the real and imaginary part of the \ww eigenvalues,
$\mu_j$, do not correspond to the eigenvalues of the mass ($M$) and
width ($\Gamma$) operators.  Since $Re\{\epsilon\}\neq 0$, we cannot
have a state with definite mass $and$ width simultaneously.   When
usually one refers to a particle with definite mass $and$ width, one
actually is regarding a definite complex eigenvalue $\mu_j$
associated with the time evolution:  if $[M,\Gamma]=0$, i.e. in
absence of either CP violation or non-vanishing $\DG$, then $Re\{\mu_j\}$ coincides with $\mu^M_j$
whereas $Im\{\mu_j\}$ equals $-\frac{1}{2}\mu^\Gamma_j$.  But this
is not the case for the neutral meson system, in general.   This
agrees and strengths our conclusion that the \ww time-evolving
eigenstates are to be taken as intermediate states responsible of
the time evolution of the system.
%%%%%%%%%%%%%%%%%%%%%%%%%%%%%%%%%%%%%%%%%%%%%%%%%%%%%%%%%%%%%%%%%%%%%%%%%%%%%%%%%
%%%%%%%%%%%%%%%%%%%%%%%%%%%%%%%%%%%%%%%%%%%%%%%%%%%%%%%%%%%%%%%%%%%%%%%%%%%%%%%%%
%%%%%%%%%%%%%%%%%%%%%%%%%%%%%%%%%%%%%%%%%%%%%%%%%%%%%%%%%%%%%%%%%%%%%%%%%%%%%%%%%
\cleardoublepage\chapter{T, CP and CPT violation, state-of-the-art
review}
 \label{stateoftheart}
 In the previous chapters we have studied the violation of the
discrete symmetries T, CP and CPT focused to  their effects in the
neutral meson system, in particular to the B-meson system. These
studies are the essential tool to develop and understand the theory
of the CP-tag and the $\w$-effect in the forth-coming chapters, where
a variety of observables which explore the violation of the
above-mentioned discrete symmetries are presented.  In this Chapter,
as an intermediate point, we present a concise review of the T, CP
and CPT violation experimental state-of-the-art,
focused on the purposes aimed in this work.

 In this Chapter we describe the present picture of the experimental
constraints on the violation of the discrete symmetries.  This is
accompanied with a corresponding discussion of the theoretical
framework. The information here contained should serve as a
reference point for the prospects of future constraints or
measurements of the respective observables.

In the following sections we study separately each one of the discrete
symmetries discussed above.

\section{Time reversal violation \hecho{90}}

From a formal point of view, {\it time reversal} is an operation
which changes the direction in which time flows.  Since this is of
course not possible from the practical point of view, we analyze its
mathematical equivalence, to change the sign of $t$. From here we
see that if the operator T exists, then it must furnish
 \bea
 T U(t) T^{-1} &=& U(-t).
 \eea
 From the existence of the inverse operator,
and a one-to-one mapping, the T operator shall be either linear and
unitary, or anti-linear and unitary, called anti-unitary.  Considering the free
particle case, it is immediate to obtain that T is anti-unitary
\cite{martin and spearman}.

Experimental direct measurements of T violation were not actually done
until 1998 by the CPLEAR collaboration \cite{Angelopoulos:1998dv}.
(Up to the date it remains as the only widely accepted direct measure
of T violation with positive results.)  In this experiment Kaons are
produced through
 \bea
 p \bar p \to
 \begin{array}{l}
 K^-\pi^+K^0 \\
 K^+\pi^-\bar K^0 ,
 \end{array}
 \eea
where the strangeness of the neutral Kaon is tagged by the sign of
the accompanying charged particles.  If the neutral Kaon
subsequently decays to $e\pi\nu$, its strangeness could also be
tagged at the decay time by the charge of the decay electron:
neutral Kaons decay to $e^+$ if the strangeness is positive at the
decay time and to $e^-$ if is negative.

With this experimental arrangement the T-asymmetry
 \bea
 A_{T [K^0]} (t) &=& \frac{|\langle K^0 |U(t)|\bar K^0\rangle|^2 - |\langle \bar K^0 |U(t)|K^0\rangle|^2}{|\langle K^0 |U(t)|\bar K^0\rangle|^2 + |\langle \bar K^0
 |U(t)|K^0\rangle|^2}
 \label{pipi}
 \eea
  was measured.  The \ww analysis for this asymmetry is done in
  Chapter \ref{cp}, \eq{kkkk}, and its prediction is
 \bea
 A_{T [K^0]} (t) &=& \frac{4Re(\epsilon_K)}{1+|\epsilon_K|^2},
 \eea
id est independent of time.  The measured result, fitted to a
constant, is
 \bea
 \langle A_{T [K^0]}\rangle = (6.6 \pm 1.3_{stat} \pm 1.0_{syst} ) \times
 10^{-3} .
 \eea

In the neutral B-meson system, the similar asymmetry has been
measured by the asymmetric B-factories Babar and Belle, but in this
case with results which are still compatible with zero.  In this
experiment the flavour tagging is easier, since a first
flavour-specific channel, as $B^0 \to \ell^+ X$ or $\bar B^0 \to
\ell^- \bar X$ tags the second meson as $\bar B^0$ or $B^0$,
respectively.   After a time $\Dt$, a flavour-specific channel (not
necessarily the same) determines the flavour of the second meson at the
time of the decay.

The asymmetry observed in this case is
 \bea
 A_{T [B^0]} (\Dt) &=&  \frac{|\langle B^0 |U(\Dt)|\bar B^0\rangle|^2 - |\langle \bar B^0 |U(\Dt)|B^0\rangle|^2}{|\langle B^0 |U(\Dt)|\bar B^0\rangle|^2 + |\langle \bar B^0
 |U(\Dt)|B^0\rangle|^2} .
 \label{hola don jose}
 \eea
The predicted result is the same as for the Kaons, but with the $\epsilon$ that corresponds to the B-system, \eq{kkkk}.
The measured results are, fitted to a constant,
 \bea
 \langle A_{T [B^0]} \rangle =
 \begin{array}{ll}
 (0.50 \pm 1.20_{stat} \pm 1.40_{syst}) \times 10^{-2} &\mbox{BABAR
  \cite{Aubert:2002mn}} \\
 (-0.11 \pm 0.79_{stat} \pm 0.70_{syst}) \times 10^{-2} &\mbox{BELLE
  \cite{Nakano:2005jb} .}
  \end{array}
  \label{vecina}
  \eea

Observe that the asymmetry in \eq{hola don jose} is also a
CP-asymmetry.  As a matter of fact, it is an indicator of both CP
and T violation in the B-mixing, and is usually referred as $A_{sl}$.

\section{CP violation \hecho{90}}
\label{que ganas de una chupada}
Although the phenomenology of CP violation is superficially
different in $K$, $D$, $B$ and $B_s$ decays due to the different
balance between decay rates, oscillations and lifetime splitting
that govern each system, the underlying mechanism of CP violation is
the same for all pseudoscalar mesons; as studied in the previous
chapters.

In order to classify the three different kinds of CP
violation existing, we first define the decay amplitudes of
the meson $P$ (neutral or charged) and its CP conjugate $\bar P$ to a
multi-particle final state $f$ and its CP conjugate $\bar f$ as
 \bea
 A_f = \langle f | H | P \rangle &,\qquad& \bar A _f = \langle f | H | \bar P
 \rangle \\
 A_{\bar f} = \langle \bar f | H | P \rangle &,\qquad& \bar A _{\bar f} = \langle \bar f | H | \bar P
 \rangle . \\
 \eea
With this in mind the CP-violating effects are classified in three
categories:
\begin{itemize}
\item[I.] CP violation in decay, defined by
 \bea
 |\bar A _{\bar f} / A_f | \neq 1.
 \eea
 In charged meson decays, where there is no mixing, this is the only
 possible source of CP-asymmetries.
 \item[II.] CP (and T) violation in mixing, defined by
 \bea
 Re(\epsilon) \neq 0 &\quad or \quad & |q/p| \neq 1.
 \eea
 In charged-current semi-leptonic neutral meson decays $P,\ \bar P
 \to \ell^\pm X$, this is the only source of CP violation, and can
 be measured via the semi-leptonic asymmetry $A_{sl}$ as in
 Eqs.\ (\ref{pipi},\ref{hola don jose}).
 \item[III.] CP violation in interference between a decay without
 mixing, $P^0 \to f$, and a decay with mixing, $P^0\to\bar P^0 \to
 f$ (such an effect occurs only in decays to final states that are
 common to $P^0$ and $\bar P^0$, including all CP-eigenstates).
 This CP violation is measured through
 \bea
 Im(\lambda_f) \neq 0,
 \eea
 with
 \bea
 \lambda_f \equiv \frac{q}{p}\, \frac{\bar A _f}{A_f} .
 \label{carlita estas caliente}
 \eea
In Chapter \ref{cp}, however, we demonstrate that this kind of CP
violation is equivalent to another therein defined CP$_A$-violation that
occurs solely in the single meson transition in the mixing.  As
 explained below, this comes out from a right choice of the B-basis
at the beginning and the end of the $\Dt$-oscillation, which
determines the phase of the states.
\end{itemize}

From its discovery in 1964, CP violation has been established
experimentally in the neutral K and B meson decays.  The current
state-of-the-art is the following,
\begin{enumerate}
\item All three types of CP violation have been observed in
$K\to\pi\pi$ decays \cite{pdg}:
 \bea
 \begin{array}{rcll}
 Re(\epsilon'_K)&=&\frac{1}{6} \left( \left| \frac{\bar A _{\pi^0 \pi^0}}{A _{\pi^0 \pi^0}}
 \right| - \left| \frac{\bar A _{\pi^+ \pi^-}}{A _{\pi^+ \pi^-}}
 \right|\right) = (2.5 \pm 0.4) \times 10^{-6} & \mbox{ (type I)}  \\
 Re(\epsilon_K) &=& \frac{1}{2} \left( 1 - \left| \frac{q}{p} \right|
 \right) = (1.657 \pm 0.021) \times 10^{-3} & \mbox{ (type II)} \\
 Im(\epsilon_K) &=& -\frac{1}{2} Im (\lambda_{(\pi\pi)_{I=0}}) =
 (1.572 \pm 0.022) \times 10 ^{-3} & \mbox{ (type III)}
 \end{array}
 \nn
 \eea
 Notice that the parameters $\epsilon_K$ and $\epsilon'_K$ --here defined-- correspond to the
 historical notation used to describe CP violation in the Kaon
 system.

\item In the B-system it has been observed CP violation in interference of decays with and without
mixing (type III), and also recently direct CP violation in the
decay (type I).

CP violation of type III has been observed in $B\to J/\psi K_S$ and
related $b\to c \bar c s$ modes \footnote{see the B-meson listing in
Ref.\ \cite{pdg}}. Within the \sm the predicted value for the
coefficient accompanying $\sin(\Dm t)$ in the corresponding CP
asymmetry is, to an approximation that is better than one percent,
 \bea
 S_{\psi K} &=& \sin (2\beta).
 \eea
  (This is computed in Chapter \ref{cp}, \eq{cp5}).  The measured
  result for this coefficient is \cite{pdg}
 \bea
 S_{\psi K} &=& Im(\lambda_{\psi K}) = 0.731 \pm 0.056 \qquad~\qquad~\qquad\mbox{ (type
 III) .}\nn
 \eea

Direct CP violation has been observed in the CP asymmetry of the
decay $B^0 \to K^+\pi^-$.  In this decay the CP violation is due to
the interference between the tree diagram and the penguin diagrams,
therefore there are large uncertainties related to the hadronic
effects in the theoretical predictions \cite{Keum:2002vi}. The
experimental results from Babar \cite{direct cpv1} and Belle
\cite{direct cpv2}, respectively, are
 \bea
 \nn
 A_{CP}(K^+ \pi^- ) =
 \begin{array}{l}
 -0.133 \pm 0.030_{stat} \pm 0.009_{syst} \\
 -0.101 \pm 0.025_{stat} \pm 0.005_{syst}
 \end{array}
 \qquad \mbox{(type I)}.
 \eea
\end{enumerate}

\vskip .5cm
 Searches for additional CP violation are ongoing in B, D
and K decay, and current limits are consistent with \sm
expectations. However, a dynamically generated matter-antimatter
asymmetry of the Universe requires additional sources of CP
violation,  and such sources should be generated by extensions to
the Standard Model.

\section{CPT violation \hecho{89}}
CPT violation, at difference of the other discrete symmetries, is
protected by the CPT-theorem \cite{pauliCPT}, which states that a
Lorentz invariant local quantum field theory must be CPT invariant.
This implies that the observation of CPT violation would
be a sensitive signal for unconventional physics from the
theoretical and experimental point of view.  For these reason,
several works in this direction explore the possible violation of
this symmetry.

In particle physics, one of the most common experimental test of CPT
invariance is the mass difference between a particle and its
antiparticle.  The most accurate measure corresponds to the Kaon
system \cite{pdg},
\bea
\frac{|m_{K^0}-m_{\bar K^0}|}{m_{K^0}} < 10^{-18} .
\eea
This limit, of course, corresponds to the K-system and does not
constrain a possible CPT violation in other systems, as for instance the B-meson
or high energy neutrinos where there is considerable work going on in this
direction.

In B physics there are experimental and theoretical efforts to test
CPT invariance.  The latest experiments seeking for CPT violation in
the $\b\bb$ mixing have used the flavour-flavour and flavour-CP
intensities results to put limits on $K_\delta$.  The predicted
results for each one of these intensities are found in Section
\ref{intensity for correlated}, Table \ref{f-cp}.  The experimental
results, fitted to these expressions for the intensities, have put
the following limits on the CPT violating parameter, \bea
\begin{array}{lr}
\begin{array}{rcl}
Re(K_\delta) &=& 0.020 \pm 0.051_{stat} \pm 0.050_{syst}, \\
Im(K_\delta) &=& 0.038 \pm 0.029_{stat} \pm 0.025_{syst},
\end{array}
&\quad \mbox{BABAR \cite{Aubert:2003hd}}\\
~&~\\
\begin{array}{rcl}
Re(K_\delta) &=& 0.00 \pm 0.12_{stat} \pm 0.01_{syst}, \\
Im(K_\delta) &=& - 0.03 \pm 0.01_{stat} \pm 0.03_{syst} .
\end{array}
&\quad \mbox{BELLE \cite{Hastings:2002ff}}
\end{array}
\eea
Where the relationship between our $K_\delta$ and Babar and Belle's CPT violating
parameter $z$ and $\cos(\theta)$, respectively, is the following
\bea
K_\delta = z = -\cos(\theta).
\eea
  \parami{

  In any case we have to use that
  $Re(\lambda)/|\lambda|=\cos(2\beta)=0.68$ .
  }%fin de nota parami

It is mandatory to notice at this point that CPT violation through
the loss of indistinguishability of particle-antiparticle, as the
one studied in Chapter \ref{w} and Ref.\ \cite{plb,prl}, has not
been analyzed experimentally yet.  In any case, we show in this work
(Section \ref{por fin}) that existing experimental data on the
equal-sign dilepton asymmetry constrains indirectly the value of
$\w$.   To do this we compute the time-dependent $A_{sl}(\w,\Dt)$
asymmetry and compare it to the fitted-to-a-constant experimental
result $A_{sl}^{exp}$.  By setting their weighted difference equal
to two standard deviations we get the allowed values for $\w$,
namely
 \bea
 -0.0084 \leq Re(\w) \leq 0.0100 \qquad ~ \quad 95\%\mbox{C.L.}
 \eea

%%%%%%%%%%%%%%%%%%%%%%%%%%%%%%%%%%%%%%%%%%%%%%%%%%%%%%%%%%%%%%%%%%%%%%%%%%%%%%%%%
%%%%%%%%%%%%%%%%%%%%%%%%%%%%%%%%%%%%%%%%%%%%%%%%%%%%%%%%%%%%%%%%%%%%%%%%%%%%%%%%%
%%%%%%%%%%%%%%%%%%%%%%%%%%%%%%%%%%%%%%%%%%%%%%%%%%%%%%%%%%%%%%%%%%%%%%%%%%%%%%%%%
\cleardoublepage\chapter{T, CP and CPT violation in correlated B
meson mixing and decays through the CP-tag} \label{cp} In this
Chapter we analyze in depth the $B$-meson Golden Plate decay and the
relationship of its structure to the definition of the CP$_A$
operator (see below). Through the use of the CP$_A$ operator we show
how in the two correlated $B$-meson experiments the study of the C, P
and CPT symmetries may get reduced to the $B$-mixing for the cases
of CP Golden Plate and flavour-specific decays.  We compute all the
possible intensities relating these possible decays. Most of the
results herein are found also in reference \cite{sequi}.

\section{The $B$ meson system: the Golden Plate decay \hecho{90}}
In this section we will devote to study one of the most important
decays in the subject of CP-violation in $B$ physics.  The Golden
Plate decay, \be B^0 \mbox{ or } \bar B ^0 \to J/\psi K_{S,L}
\nonumber \ee has numerous virtues, one of the most important is
that the QCD penguin correction has the same CKM phase as the
electro-weak tree contribution \cite{gpdecay}.  Allowing this
feature to work it out as if the decay would be through the
electro-weak tree diagram, forgetting about any interference
pollution with QCD penguin diagram with different weak phases.
Having a branching ratio of $8.7\times 10^{-4}$ \cite{pdg} it has
been the star of the first generation of $B$-meson experiments in
the search for CP-violation. It was through the Golden Plate decay
that the first measurements of CP-violation in the $B$ sector
 were made \cite{Bsector}.

\begin{figure}[t]
%\vspace{0.10in}
%\centerline{
%\epsfysize=5.0truecm
%\epsffile{gpdecay.eps}}
\makebox[\textwidth][c]{
\framebox[1\textwidth]{
\includegraphics[width=0.9\textwidth]{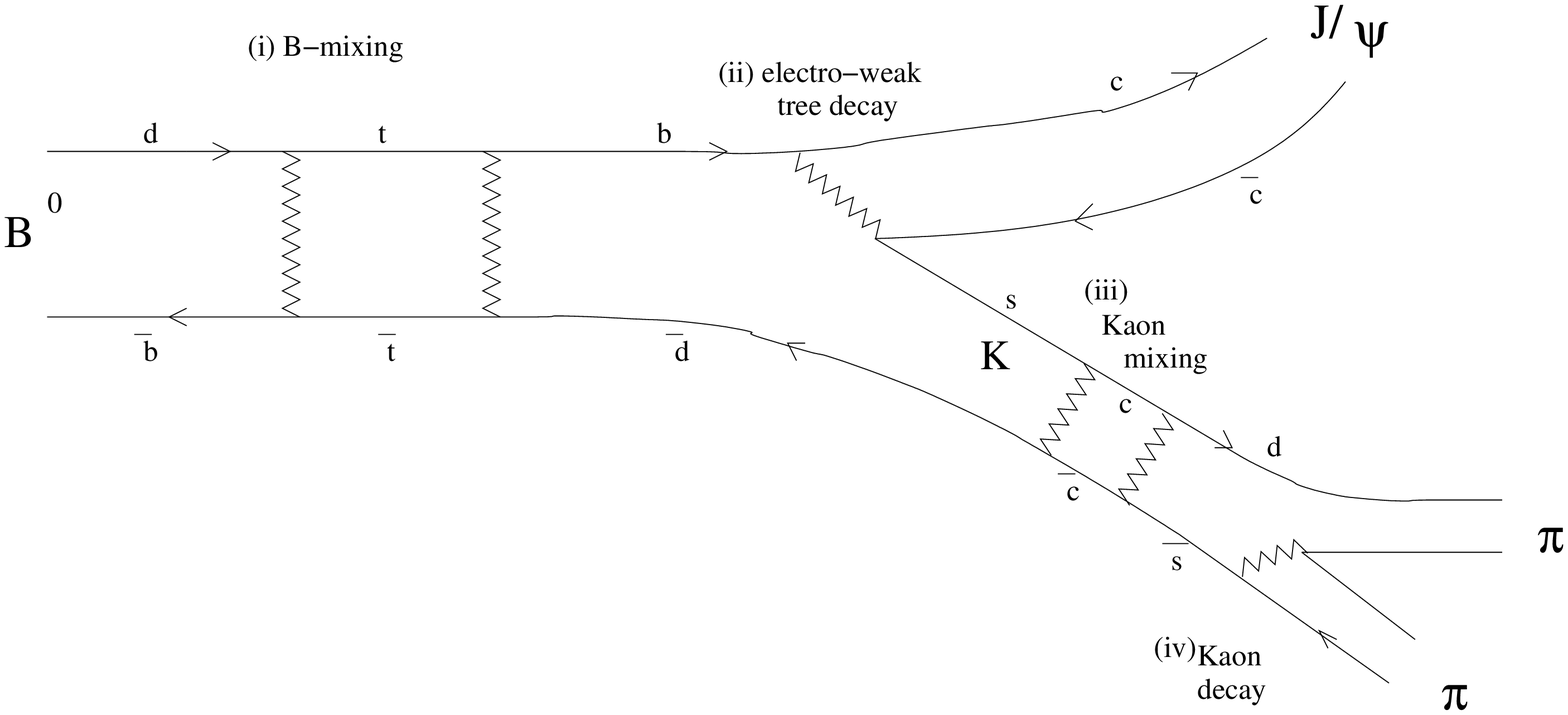}
}}
\caption[]{The Golden Plate Decay.  The piece of the Hamiltonian that acts in the decay may be sub-divided and consequently find a $\cp$ operator that leaves invariant, up to $\lambda^4$, every part but the $B$-mixing.}
\label{gpd}
\end{figure}

The Golden Plate decay, from the creation of the $B$ until the decay
to $J/\psi K_S (K_S\to 2\pi)$ may be divided into four different
parts, $(i)$ the mixing of the $B$-meson, $(ii)$ the electroweak
tree decay into $J/\psi$ and a neutral Kaon, $(iii)$ the mixing of
the Kaon before it decays, and $(iv)$ the decay of the Kaon in
$2\pi$ or $3\pi$. These divisions are sketched in \fig{gpd}.

\section{The determination of the $\cp$ operator and the CP-tag \hecho{90}}
We will show in this section that, when working up to $\lambda^4$ in
the Wolfenstein parametrization \cite{wolf}, exists a choice of
$\ti$ that leaves invariant the Hamiltonian responsible of the
Golden Plate decay without the $B$-mixing piece. This will prove to
be useful to place CP-tags in the two correlated $B$-meson
experiment, as it will be discussed in next section.  To this order
in $\lambda$, the reasoning of the CP-tag is parallel to that made for the
flavour-tag.

As it was mentioned in last section, the Hamiltonian responsible of
the decay may be sub-divided into four pieces.  The requirement of
invariance of the pieces responsible of the electro-weak tree decay
and the Kaon mixing, when the $\cp$ operator acts, will determine
the values of $\Theta '$.  (The decay of the Kaon will be negligible
in the approximation of $\lambda^4$.) Notice also that the value of
$\Theta$, the phases corresponding to the up sector, will not be
determined; anyway, in next sections it will prove to be unimportant
for the purposes we follow.

\begin{figure}[t]
%\vspace{0.10in}
%\centerline{
%\epsfysize=5.0truecm
%\epsffile{tree.eps}}
\makebox[\textwidth][c]{
\framebox[1\textwidth]{
\includegraphics[width=.8\textwidth]{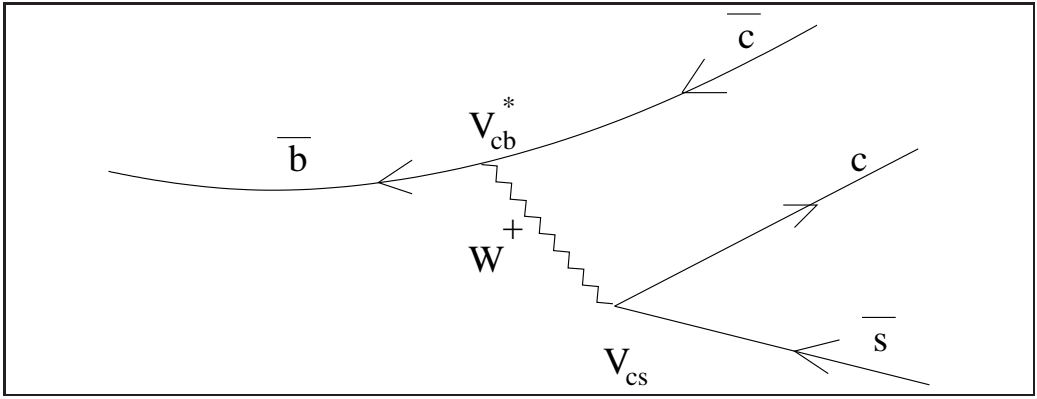}
}}
\caption[]{The electro-weak tree decay for the case $B^0 \2 J/\psi K^0$.}
\label{tree}
\end{figure}

The piece of Hamiltonian responsible of the electro-weak tree decay
shown in \fig{tree} is \bea
H_{tree} &=& \left( W_\mu^- \bar c \gamma^\mu (1-\gamma_5) V_{cs} s \right) \left( W_\nu^+ \bar b \gamma^\nu (1-\gamma_5) V^*_{cb} c \right) + \nonumber \\
&& \left( W_\mu^+ \bar s \gamma^\mu (1-\gamma_5) V^*_{cs} c \right)
\left( W_\nu^- \bar c \gamma^\nu (1-\gamma_5) V_{cb} b \right)
\label{one} \eea where $s$ and $c$ are the quark field operators,
the first line corresponds to the decay $B^0 \2 J/\psi K^0$ and the
second to $\bar B^0 \2 J/\psi \bar K^0$.

In order to find the values of $\ti$ that satisfy $[\cp, H_{tree}] =
0$, we compute \bea
\begin{array}{lrr}
\cp^\dagger H_{tree} \cp =&\ \ \ \ \ \ \ \ \ \ &\ \ \ \ \ \ \ \ \ \ \ \ \ \ \ \ \ \ \ \ \ \ \ \ \ \ \ \ \ \ \ \ \ \ \ \ \ \ \ \ \ \ \ \ \ \ \ \
\end{array}
\nonumber \\
\begin{array}{rll}
&&\left( W_\mu^+ \bar s \gamma^\mu (1-\gamma_5) V_{cs} c \right) \left( W_\nu^- \bar c \gamma^\nu (1-\gamma_5) V^*_{cb} b \right) e^{-2i(\theta'_b - \theta '_s)} + \\
&&\left( W_\mu^- \bar c \gamma^\mu (1-\gamma_5) V^*_{cs} s \right) \left( W_\nu^+ \bar b \gamma^\nu (1-\gamma_5) V_{cb} c \right) e^{2i(\theta '_b - \theta '_s)} ,
\end{array}
\label{two}
\eea
where \eq{cpoperator} and the results of Chapter 2 have been used.

It is straightforward to see which is the condition to be satisfied
in order to have \eq{one} equal to \eq{two}: \be [\cp, H_{tree}] = 0
\iff e^{i(\theta '_b - \theta '_s )} \equiv  \frac{V_{cs}\,
V_{cb}^*}{|V_{cs}\, V_{cb}^*|} . \label{condicion bs} \ee

On the other hand the piece of Hamiltonian involved in the Kaons
mixing is a product of four charged current terms.   A standard
calculation \cite{kaonsmixing} yields the effective current-current
Hamiltonian for the mixing, \bea
H_{K-mix} &=& -\frac{G_F}{\sqrt 2} \frac{\alpha}{16\pi} \mbox{csc}^2 \theta_W \int_0^\infty dx \left[ \sum_i \frac{V^*_{is} V_{id} \epsilon_i }{x+\epsilon_i} \right] ^2 \cdot \nonumber \\
&\cdot & \left( \bar s \gamma_\mu (1-\gamma_5) d \right) \left( \bar
s \gamma^\mu (1-\gamma_5) d \right) + h.c. \label{k-mix} \eea Where
$G_F$ is the Fermi coupling constant, $\theta_W$ the Weinberg angle
and $\epsilon_i = (m_i /m_W)^2$.  The sum index $i$ goes through the
$up$ quarks.

When trying to find $\theta '_d - \theta '_s$ requiring $[ \cp ,
H_{K-mix}] =0$ it is clear that the three sides of the
$ds$-unitarity triangle, $V^*_{is} V_{id}$ will not permit it.
Anyway, a simple observation on \eq{ds} shows up that if we work an
approximation up to $\lambda^4$ in the sides of the triangle, then
the $ds$ triangle collapses to a line.  Getting in this way that all
the sides have the same complex phase, the one determined, for
instance, by the $c$ side, $\left. \frac{V^*_{cs} V_{cd}}{|V^*_{cs}
V_{cd}|} \right|_{O(\lambda^4)}$.

With this approximation it is straightforward to get, analogous to
the electro-weak tree calculation that lead to \eq{condicion bs} ,
the condition for invariance of $H_{K-mix}$: \be [\cp, H_{K-mix}]=0
\ \iff \ e^{i(\theta '_s - \theta '_d )} \equiv \left.
\frac{V_{cd}\, V_{cs}^*}{|V_{cd}\, V_{cs}^*|} \right|_{O(\lambda^4)}
\label{condicion ds} \ee

As we see, approximating $\left. VV^*\right|_{O(\lambda^4)}$ implies
taking no CP-violation in the $K$-mixing.  As a matter of fact this
accounts for the fact of having the CP-violation in the $ds$ sector
much smaller that in the $bd$ sector.  Notice also that the direct
CP-violation in the Kaons, which comes from the final decay of the
Kaon into the Pions, may also be neglected since it is already
smaller than the CP-violation in the $K$-mixing.  The order of
magnitude in the rate of the direct to indirect CP-violation in the
Kaons comes from the result $\epsilon '/\epsilon \approx 10^{-3}$
(see Ref.~\cite{bertola} and Section \ref{que ganas de una
chupada}).

Therefore we have already set the shape of the $\cp$ operator that
leaves invariant, up to $\lambda^4$, the Hamiltonian in charge of
the Golden Plate decay {\it after} the $B$-mixing.  For the sake of
brevity we will call $CP_A$ the operator with the definitions of
$\ti$ given above.  For consistency the approximation $O(\lambda^4)$
should be also taken in \eq{condicion bs}.  Thus the $CP_A$ operator
is defined through \be e^{i(\theta '_b - \theta '_s )} \equiv \left.
\frac{V_{cs}\, V_{cb}^*}{|V_{cs}\, V_{cb}^*|}\right|_{O(\lambda^4)}\
\ \ \ \  e^{i(\theta '_s - \theta '_d )} \equiv \left.
\frac{V_{cd}\, V_{cs}^*}{|V_{cd}\, V_{cs}^*|} \right|_{O(\lambda^4)}
\ee which simultaneously define the value of the phases for the $bd$
sector, \be e^{i(\theta '_b - \theta '_d )} \equiv  \left.
\frac{V_{cd}\, V_{cb}^*}{|V_{cd}\, V_{cb}^*|}\right|_{O(\lambda^4)}
. \label{condicion bd} \ee With this definition of the $CP_A$
operator one can define the $CP_A$ eigenstates as \bea | B_{\pm _A}
\rangle \equiv \frac{1}{\sqrt{2}} \left( |B^0\rangle \pm CP_A | B^0
\rangle \right) ;\ \ \ CP_A | B_{\pm _A}  \rangle = \pm | B_{\pm _A}
\rangle. \label{cadorna} \eea The decay after the $B$-mixing,
conserves $CP_A$.

\subsubsection{The CP-tag}
The results in \eq{cadorna} make now direct and natural the
implementation of the CP-tag.  To see this, we first write the
initial state of a B-factory, \eq{initial3}, in the basis defined in
\eq{cadorna}:
 \be
 |i\rangle = \frac{1}{\sqrt{2}} \left( |B_{-_A} (-\vec{k}), B_{+_A} (\vec{k}) \rangle - |B_{+_A} (-\vec{k}), B_{-_A} (\vec{k}) \rangle \right).
 \label{dactari}
 \ee
  With the initial state written in this way, we repeat the
reasoning of the flavour tag (see Section \ref{correlated}).  We let
the state evolve in time until the time $t_1$ of the first decay.  As before, its
definite anti-symmetry makes it keep its original structure and the only
modification comes from an attenuation factor and an unimportant global
phase which we omit,
 \bea
 U(t_1) \otimes U(t_1) |i\rangle &=& e^{-\G t_1} |i\rangle .
 \eea
Suppose now a first decay $B\to J/\psi K_S$ at time $t_1$, then the
state function is projected at that time to
 \bea
 |B_{\overline{\j/\psi K_S}} \rangle &=& \frac{e^{-\G t_1}}{\sqrt{2}}\left[ \langle J/\psi K_S | B_{-A}\rangle \, |B_{+A}\rangle - \langle J/\psi K_S | B_{+A} \rangle \, |B_{-A}\rangle
 \right] \nn \\
 &=& \frac{e^{-\G t_1}}{\sqrt{2}} A_{J/\psi K_S} \, |B_{+A}\rangle .
 \eea
  Where in the last step we have used $\langle J/\psi K_S | B_{+A}
\rangle = 0$, since the $CP_A$ operator commutes with the piece of
Hamiltonian in charge of the decay and $J/\psi K_S$ is a $CP_A=-$
eigenstate. In this way we placed a CP$_A$-tag (CP-tag for short) in
the second meson flying. (If the first decay is to $J/\psi K_L$ the
reasoning is analogous but the second meson flying will be
$B_{-A}$.)

\section{Time evolution with the rephasing-in\-va\-riant $\epsilon$ and $\delta$\hecho{90}}
We have studied in Chapter \ref{meson system} the time evolution of
the neutral meson system using the \ww approximation. At that level,
 though, $\epsilon$ and $\delta$ were rephasing
variant quantities due to the ambiguity in the relative phase
between $\p$ and $\pp$. In this Chapter, instead, we will take
profit of the existence of the $CP_A$ operator to define $\epsilon$
and $\delta$ through the \ww eigenstates for the $B$ system as
follows, \bea \left (
\begin{array}{c}
B_L \\
B_H \\
\end{array}
\right ) &=&  \left (
\begin{array}{c c}
\frac{1}{\sqrt{1+|\epsilon+\delta|^2}} & \frac{\epsilon+\delta}{\sqrt{1+|\epsilon+\delta|^2}} \\
\frac{\epsilon-\delta}{\sqrt{1+|\epsilon-\delta|^2}} & \frac{1}{\sqrt{1+|\epsilon-\delta|^2}} \\
\end{array}
\right ) \left (
\begin{array}{c}
B_{+_A} \\
B_{-_A} \\
\end{array} \right ) \nonumber \\
&=& \frac{1}{\sqrt 2}\left (
\begin{array}{c c}
\frac{1+\epsilon+\delta}{\sqrt{1+|\epsilon+\delta|^2}} & \frac{1-\epsilon-\delta}{\sqrt{1+|\epsilon+\delta|^2}}\\
\frac{1+\epsilon-\delta}{\sqrt{1+|\epsilon-\delta|^2}} & \frac{-1+\epsilon-\delta}{\sqrt{1+|\epsilon-\delta|^2}} \\
\end{array}
\right ) \left (
\begin{array}{c}
B^0\\
CP_A \, B^0 \\
\end{array}
\right ) . \label{estados de masas} \eea  As it can be seen, with
this definition of $|\bar B^0\rangle=$CP$_A\, |B^0\rangle$ there is no
more a relative phase ambiguity between $B^0$ and $\bar B^0$, and
therefore the parameters $\epsilon$ and $\delta$ in \eq{estados de
masas} are rephasing-$invariant$. The time evolution of the \ww
eigenstates is
 \be
 U(\Delta t) | B_{L,H} \rangle = e^{-i \mu_{L,H}
 \Delta t} | B_{L,H} \rangle ,
 \ee
where $\mu_k = m_k -\frac{i}{2} \Gamma_k$, but now the CP$_A$
information is contained in the eigenvalues and eigenvectors through
the parameters.

Observe that the evolution operator constructed with these
eigenstates has the exact same form as the one obtained in
\eq{udet}, but now $\epsilon$ and $\delta$ refer to
rephasing-invariant quantities which include the information of the
CP$_A$-operator's definition. This means that if the same analysis
which leads to Eqs.\ (\ref{carlita1}-\ref{carlita2}) is performed
here, then we must make two important modifications in the
conclusions: $(i)$ we must refer to $\epsilon$ (and not only to
$Re(\epsilon)$) in the logical implications in Eqs.\
(\ref{carlita1}-\ref{carlita2}); and $(ii)$ the conclusions now
refer to CP$_A$-violation in the time evolution, or equivalent, to
CP violation in the interference of the time evolution and the
decay.  This is because the choice of the CP$_A$ operator through
the Golden Plate decay incorporates the information of the
CP-direction of the decay into the definition of the CP$_A$
eigenstates which evolve in the mixing.

Although the $\delta$ parameter plays a major role in discussing the
CPT violating observables, it is useless to keep it while analyzing
exclusively CP and T violation, since its effect in those cases is
negligible. Hence, it is useful for future purposes (Section
\ref{intensity for correlated}) to obtain important results
concerning CP and T violation using the approximation
$\delta\approx0$. This is what is done in the remain of this
Section.

First we would like to point out the relationship between this
$\epsilon$-rephasing invariant picture and the also common used
notation in CP-violation of the $\lambda$-parameter \cite{lambda},
which resumes the information of mixing times decay.  If CPT is
assumed then $q/p$ is defined to be the mixing parameter between the
rephasing-variant $B^0 -\bar B^0$, the decay amplitude is ${\cal A}
= \langle f_{CP} | \delta H_A | B^0 \rangle$, and the value of
$\lambda$ is defined to be\be \lambda = \frac{q}{p} \frac{\bar {\cal
A}}{{\cal A}} = \mbox{ rephasing invariant quantity}. \ee Taking
into account the invariance of $\delta H_A$ under CP$_A$, one gets
the connection \be \pm \lambda = \frac{1-\epsilon}{1+\epsilon}= -
\sqrt{\frac{H_{21}}{H_{12}}}.
%un saludito a la muchachada...
\label{por la proxima chica} \ee where the sign depends on the final
CP-eigenstate, and $1=\b$ and $2=CP_A\b$.  Although the equivalence is clear in \eq{por la proxima
chica}, it is shown in this Chapter that using the
$\epsilon$-picture makes possible to reduce the CP-violation to
transitions between well defined neutral $B$-meson states.

Assuming CPT and $\DG=0$ it is easy to get two profitable relationships
which are obtained by using \eq{por la proxima chica}, the choice of
$CP_A$ in \eq{condicion bd} and the top-quark dominance in the
mixing:
 \bea
 \sin (2\beta) &=& -\frac{2 Im\ \epsilon}{1+|\epsilon|^2} \label{sonia2} \\
 \cos (2\beta) &=& -\frac{1-|\epsilon|^2}{1+|\epsilon|^2} ,
 \label{rosa2}
 \eea
  where $\beta$ is the well known CP-phase between the top and charm
sides of the $b-d$ unitary triangle, $\beta=\mbox{arg} \left(
-\frac{ V_{cd}V^*_{cb} }{V_{td}V^*_{tb}} \right) $.  It is obvious
that these important relationships can only be obtained if $\epsilon$
is rephasing-invariant.  The sign connection of Eq.$\,$(\ref{rosa2})
depends on the $\epsilon$ definition in Eq.$\,$(\ref{estados de
masas}), id est the sign of $\cos(2\beta)$ is sensible to the
preferred CP-parity of each \ww eigenstate.  (Recall that it is
possible to distinguish and label the light eigenstate $(B_L)$ by
having the smaller real part of the eigenvalue.)  The Standard Model
expectation $\cos(2\beta) > 0$, measured in Ref.~\cite{cos}, leads through
Eq.$\,$(\ref{rosa2}) to a $B_L$ state with preferred CP-parity
equal to CP-.

\section{The Intensity and its reduction to \\ $B$-transitions \hecho{90}}
We now proceed to describe the experimental variable to be measured
in a correlated $B$ meson decay, the intensity \cite{intensity}.
Under the assumption of an initial state as in
Eq.$\,$(\ref{initial3}) the probability density of having a decay
into $X$ with momentum $-\vec{k}$ at $t=t_1$ and a decay into $Y$
with momentum $\vec{k}$ at $t=t_2 >t_1$ is \be \rho \left( X(t_1),
Y(t_2) \right) = | \langle X, Y | U(t_1) \otimes U(t_2) | i \rangle
|^2 , \ee $U(t)$ being the (non-unitary) evolution operator that
corresponds to a single $B$ meson, defined in Section
\ref{evolution}. Since the important experimental variable is
$\Delta t= t_2 - t_1$ we define the so-called intensity,
$I(X,Y,\Delta t) =$ $\Delta t$-variable probability density of
having an $X$ decay and, after a time $\Delta t$, a $Y$ decay on the
other side.  After performing a change of variables to $\Delta t$
and $t_1$ and integrating in $t_1$ it results
 \be I(X,Y,\Delta t) =
 \int_{0}^{\infty} d t_1 | \langle X, Y|U(t_1) \otimes U(\Delta t
 +t_1 ) | i \rangle |^2 .
 \label{intensidad}
 \ee
  In order to solve this integral we reason as follows.  At $t=t_1$,
when the first decay to $X$, if $X$ is a flavour or Golden Plate
decay then we know that the decay comes from a $|B_X\rangle$.  The
projected remaining state is $|B_{\bar X}\rangle$.  In this way the
only appearance of the variable $t_1$ will be in the attenuation of
the norm of the state due to the elapsed time since the creation of
the $B$-mesons. Therefore we may re-write the integral in
Eq.$\,$(\ref{intensidad}) and solve it as follows
 \bea
 I(X,Y,\Delta t) &=&  \int_{0}^{\infty} d t_1 e^{-2\G t_1} |\langle X,Y| 1 \otimes U(\Delta t)| B_X,B_{\bar X} \rangle |^2\nonumber \\
 &=&  \frac{1}{2 \bar \Gamma} |A_X|^2 |A_Y|^2 |T_{B_Y B_{\bar X}}
 (\Delta t) |^2.
 \label{naif}
 \eea
  Here $\G$ is the average
width of B-mesons, and $T_{B_Y B_{\bar X}} (\Delta t) = \langle B_Y
| U(\Delta t) | B_{\bar X} \rangle$ is the transition amplitude in
the mixing of the corresponding states.  Notice that in writing
Eq.$\,$(\ref{naif}) we have assumed a conservation law in the $Y$
decay, id est, it may be flavour or Golden Plate decay.   In this
way we have transformed the integral expression of
Eq.$\,$(\ref{intensidad}) in the explicit result of
Eq.$\,$(\ref{naif}) in terms of a $B$-meson transition, developing a
powerful tool for the calculation of intensities.

\section{The Intensity for correlated neutral\- $B$-meson decays \hecho{89}}
 \label{intensity for correlated}
 \nota{estan las cuentas revisadas: tablas revisadas, intensidad
 revisada, pero algunas asimetrias que dan resultados raros no
 estaria de mas revisarlas.  Ademas de pegarle una leida en fresco
 para ver como quedo. Ademas: hay que chequear con Pepe lo de $Im(\delta)$!!}
 In this Section we compute the
intensities of all possible correlated decays concerning CP and
flavour channels.  In order to explore the T, CP and CPT symmetries
we express them as a function of the above studied rephasing
invariant parameters $\epsilon$ and $\delta$.

The general expression for the intensity reads ($\Delta m=m_H - m_L$
and $\Delta \Gamma=\Gamma_H - \Gamma_L$)
 \bea
 I(X,Y,\Delta t) = \frac{1}{8} \frac {e^{-\bar{\Gamma}\Delta t}}{\bar \Gamma} &|A_X|^2 |A_Y|^2 \left\{ a \cosh \left( \frac{\Delta\Gamma \Delta t}{2}\right) \right. + \left. b \cos (\Delta m \Delta t) \right. \nonumber \\
 &+\left. c \sinh \left( \frac{\Delta\Gamma \Delta t}{2}\right) + d \sin (\Delta m \Delta t) \right\} .
 \label{laura altea}
 \eea
The coefficients are easily calculated using Eq.$\,$(\ref{naif})
once the $T_{ij}(\Delta t)$ are computed using the time evolution of
the \ww eigenstates.  For the sake of clarity we write the curly
bracket expression valid up to first order in $\Delta \Gamma$ and
$\delta$ and zeroth order in their product, $\DG\,\delta\approx 0$. In the $\Delta\Gamma$ expansion one has that if CPT is
assumed then --see \eq{real} --
 \be \frac{Re\, \epsilon}{1+|\epsilon
 |^2} \propto Im (\Gamma_{21} / M_{21} ),
 \label{jana mrazkova}
 \ee
which is proportional to $\Delta\Gamma$, thus we can parameterize in
this case $$\frac{Re(\epsilon)}{1+|\epsilon |^2} \equiv x
\Delta\Gamma$$. \comentariodelta{On the contrary, $Im\, \epsilon$ contains
$\Delta\Gamma$-independent contributions and
$\Delta\Gamma$-corrections, but an explicit expansion is not needed.
On the other hand, if CPT is not assumed it can be shown \cite{mcb3}
that the imaginary part of the CPT parameter is proportional to
$\DG$, $Im(\delta)\sim\DG$.}

In Eq.$\,$(\ref{laura altea}) we will ap\-prox\-i\-mate  that $\cosh
(\frac{\Delta\Gamma}{2} \Delta t) \approx 1$, $\sinh(
\frac{\Delta\Gamma}{2} \Delta t) \approx \frac{\Delta\Gamma}{2}
\Delta t$, and keep only first order terms in $Re(\epsilon)$ and
$\delta$ in the coefficients $a$, $b$ and $d$, and zeroth order
in the coefficient $c$.  The product $\delta\,Re(\epsilon)$ will be neglected according to \eq{jana mrazkova}.  The results for the different possible
decays (flavour-flavour, flavour-CP, CP-flavour and CP-CP) are found
in Tables \ref{f-f}-\ref{cp-cp}.  We analyze separately each case.

\subsubsection{Flavour-flavour decays}
\begin{table}[ht]
\begin{tabular*}{\columnwidth}{@{\extracolsep{\fill}}|c|c|c|c|c|}
\hline
decays & ~$(l^+,l^+)$ & ~$(l^-,l^-)$ & ~$(l^+,l^-)$ & ~$(l^-,l^+)$ \\
\hline
transition & $B_2 \to B_1$ & $B_1 \to B_2$ & $B_2 \to B_2$ & $B_1 \to B_1$ \\
\hline
\hline
$a$ & $ 1 + 4 \frac{Re\,  \epsilon}{1+|\epsilon |^2}$ & $ 1 - 4 \frac{Re\,  \epsilon}{1+|\epsilon |^2}$ & 1 &
1 \\
\hline
$b$ & $ -\left( 1 + 4 \frac{Re\,  \epsilon}{1+|\epsilon |^2}\right)$ & $ -\left( 1 - 4 \frac{Re\,  \epsilon}{1+|\epsilon |^2} \right)$ & 1 & 1 \\
\hline
$c$ & 0&0&0&0 \\
\hline
$d$ & 0&0&$\frac{2Im(\delta)}{1+\mec}$&$-\frac{2Im(\delta)}{1+\mec}$ \\
\hline
\end{tabular*}
\caption{Correlated dilepton decays, described by flavour-flavour
transitions. The notation is self evident, the first row indicates
the two final decay products, the second row shows the one meson
transition that happens in the mixing ($1=\b,\ 2=CP_A \b$); the
remaining four rows are the coefficients for the respective
intensity, see Eq.$\,$(\ref{laura altea}).} \label{f-f}
\end{table}
The coefficients in \eq{laura altea} for correlated flavour-flavour
decays are found in Table \ref{f-f}. Here the first two columns are
conjugated under CP and also under T. The corresponding equal-sign
dilepton charge
asymmetry, also known as Kabir asymmetry \cite{kabir}, becomes $exactly$ $\Delta t$-independent and
determined by
\begin{eqnarray}
A_{sl} &=&
\frac{I(\ell^+,\ell^+,\Dt)-I(\ell^-,\ell^-,\Dt)}{I(\ell^+,\ell^+,\Dt)+I(\ell^-,\ell^-,\Dt)}\nonumber\\
&=&\frac{4Re(\epsilon)}{1+\mec}=4x\,\DG .
\label{kkkk}
\end{eqnarray}
This is a genuine CP and T violating observable, which needs both
the violation of the symmetry and the absorptive part $\Delta\Gamma
\neq 0$.

The asymmetry built from the third and fourth columns of Table $5.1$
is a CPT (and CP) observable:
 \bea
 \frac{I(\ell^+,\ell^-,\Dt)-I(\ell^-,\ell^+,\Dt)}{I(\ell^+,\ell^-,\Dt)+I(\ell^-,\ell^+,\Dt)}&=&
 \frac{\frac{4Im(\delta)}{1+\mec}\sin(\Dm\Dt)}{1+\cos(\Dm\Dt)}.\
 \
 \eea
Any deviation from zero in this observable would be a signal of CPT
violation.  \comentariodelta{Notice that although this is an appealing CPT
observable, considering the high branching ratio of the leptonic
decays, it is double suppressed by $\delta$ and by $\DG$, since
$Im(\delta)\sim\DG$.}

\subsubsection{Flavour-CP and CP-flavour decays}
\begin{table}[ht]
\begin{flushleft}
{\tiny
\begin{tabular}{|c|c|c|c}
\hline
decays &$(l^-,K_L)/(K_S,l^-)$ & $(l^-,K_S)/(K_L,l^-)$&~\\
\hline
transition& $B_1 \to B_{+ A}\ /\ B_{+A} \to B_2$ & $B_1 \to B_{-A}\ /\ B_{-A} \to B_2$ & ~\\
\hline
\hline
$a$ & $1- \frac{2 Re\ \epsilon}{1+|\epsilon|^2} \pm 2\frac{1-\mec}{1+\mec}\frac{Re(\delta)}{1+\mec} \mp \frac{4Im(\epsilon)}{1+\mec}\frac{Im(\delta)}{1+\mec}$
  & $1- \frac{2 Re\ \epsilon}{1+|\epsilon|^2} \mp 2\frac{1-\mec}{1+\mec}\frac{Re(\delta)}{1+\mec} \pm \frac{4Im(\epsilon)}{1+\mec}\frac{Im(\delta)}{1+\mec}$ & ~  \\
\hline
$b$ &$  \frac{2 Re\, \epsilon}{1+|\epsilon|^2} \mp 2\frac{1-\mec}{1+\mec}\frac{Re(\delta)}{1+\mec} \pm \frac{4Im(\epsilon)}{1+\mec}\frac{Im(\delta)}{1+\mec}$ &
  $ \frac{2 Re\, \epsilon}{1+|\epsilon|^2} \pm 2\frac{1-\mec}{1+\mec}\frac{Re(\delta)}{1+\mec} \mp \frac{4Im(\epsilon)}{1+\mec}\frac{Im(\delta)}{1+\mec}$& ~  \\
\hline
$c$& $  \frac{1-| \epsilon|^2 }{1+| \epsilon |^2}  $& $ - \frac{1-| \epsilon|^2 }{1+| \epsilon |^2}  $& ~\\
\hline
$d$& $  \frac{2 Im \, \epsilon}{1+|\epsilon |^2} \left( 1- \frac{2Re\ \epsilon}{1+|\epsilon|^2} \right) \pm \frac{Im(\delta)}{1+\mec}$ &
  $ -\frac{2 Im \, \epsilon}{1+|\epsilon |^2} \left( 1- \frac{2Re\ \epsilon}{1+|\epsilon|^2} \right) \pm \frac{Im(\delta)}{1+\mec} $& ~\\
\hline
\end{tabular}
}
\end{flushleft} \vskip.3cm
\begin{flushright}
{\tiny
\begin{tabular}{c|c|}
\hline
$(l^+,K_L)/(K_S,l^+)$&$(l^+,K_S)/(K_L,l^+)$\\
\hline
$B_2 \to B_{+ A}\ /\ B_{+A} \to B_1$ & $B_2 \to B_{-A}\ /\ B_{-A} \to B_1$\\
\hline
\hline
$1+ \frac{2 Re\ \epsilon}{1+|\epsilon|^2} \mp 2\frac{1-\mec}{1+\mec}\frac{Re(\delta)}{1+\mec} \mp \frac{4Im(\epsilon)}{1+\mec}\frac{Im(\delta)}{1+\mec}$ &
  $1+ \frac{2 Re\ \epsilon}{1+|\epsilon|^2} \pm 2\frac{1-\mec}{1+\mec}\frac{Re(\delta)}{1+\mec} \pm \frac{4Im(\epsilon)}{1+\mec}\frac{Im(\delta)}{1+\mec}$ \\
\hline $ -\frac{2 Re\,\epsilon}{1+|\epsilon|^2} \pm
2\frac{1-\mec}{1+\mec}\frac{Re(\delta)}{1+\mec} \pm
\frac{4Im(\epsilon)}{1+\mec}\frac{Im(\delta)}{1+\mec}$ &
  $ -\frac{2 Re\, \epsilon}{1+|\epsilon|^2} \mp 2\frac{1-\mec}{1+\mec}\frac{Re(\delta)}{1+\mec} \mp \frac{4Im(\epsilon)}{1+\mec}\frac{Im(\delta)}{1+\mec}$  \\
\hline
$  \frac{1-| \epsilon|^2 }{1+| \epsilon |^2}  $& $ - \frac{1-| \epsilon|^2 }{1+| \epsilon |^2}  $\\
\hline
$- \frac{2 Im\, \epsilon}{1+|\epsilon |^2} \left( 1+ \frac{2Re\ \epsilon}{1+|\epsilon|^2} \right) \mp \frac{Im(\delta)}{1+\mec}$&
 $ \frac{2 Im \, \epsilon}{1+|\epsilon |^2} \left( 1+ \frac{2Re\ \epsilon}{1+|\epsilon|^2} \right) \mp \frac{Im(\delta)}{1+\mec}$\\
\hline
\end{tabular}
}
\end{flushright}
\caption{Correlated flavour-CP and CP-flavour decays.  Each column
contains the information of two decays which are connected by CPT.
The first decay $(flavour,CP)$ is described by the upper sign, and the
second $(CP,flavour)$ by the lower sign.  As expected, both decays in each column
are related by the $\delta\to-\delta$ replacement. Notation is as in
Table \ref{f-f}.} \label{f-cp}
\end{table}

The coefficients for the flavour-CP and CP-flavour correlated
channels are found in Table \ref{f-cp}. If CPT is not violated then
for each flavour-CP decay, its CPT transformed CP-flavour decay
would have the same intensity. This is seen in each column by
setting $\delta=0$.

If we assume CPT invariance ($\delta=0$), the comparison between the
second and fourth columns allows to build a CP-odd asymmetry. (An
alternative CP-odd asymmetry can be obtained from the comparison
between the first and third columns.) In the limit of $\Delta \Gamma
=0$, they are equivalent and correspond to the well known
measurement of $\sin (2\beta)$ in the Standard Model:
 \be
 A_{CP}=
 \left.
 \frac{I(l^- ,K_S , \Delta t) - I(l^+ , K_S , \Delta t)}{I(l^- ,K_S ,
 \Delta t) + I(l^+ , K_S , \Delta t)} \right|_{\Delta\Gamma=\delta=0} = \sin
 (2\beta ) \sin (\Delta m \Delta t) ,
 \label{cp5}
 \ee
where Eq.$\,$(\ref{sonia2}) has been used.  The measurement of this
asymmetry by the $B$ factories has been the major success of CP
violation studies in $B$ physics by testing the CKM sector of the
\sm. From its first measurements in 2000 \cite{Bsector} up to
present day, the value of $\beta$ has acquired a striking precision
(which is still improving):
 \bea
 \begin{array}{ccc}
 \begin{array}{l}
 \sin(2\beta)=
  \begin{array}{lr}
  0.34\pm0.25 & \mbox{ \bf (Babar 2000)}\\
  0.99\pm0.20 & \mbox{ \bf (Belle 2000)}
  \end{array}
 \end{array}
 &
 \longrightarrow
 &
 \begin{array}{c}
 \sin(2\beta) = 0.731\pm0.056 \\
 \mbox{{\bf (Babar+Belle 2005)}}
 \end{array}
 \end{array}
 \nonumber
 \eea

A direct test of T-violation can be performed also using the
advantages of the CP-tag.  Regarding the meson transition in the
mixing it is easy to see that $I(\ell^-,K_S,\Dt)
\stackrel{T}{\longrightarrow} I(K_L,\ell^+,\Dt)$, and hence a T
asymmetry is easily built and computed as
 \nota{revisar una vez mas}
 \bea
 A_T =
 \left.
 \frac{I(\ell^-,K_S,\Dt)-I(K_L,\ell^+,\Dt)}{I(\ell^-,K_S,\Dt)+I(K_L,\ell^+,\Dt)}\right|_{\DG=\delta=0} = \sin(2\beta)\sin(\Dm\Dt)
 .
 \label{AT}
 \eea
   A measurement of $A_T\neq0$ is a {\it direct
measurement} of $T$ violation.

A non-genuine asymmetry can be built using the $\Dt$-operation,
consisting in the exchange in the order of appearance of the decay
products $X$ and $Y$.  The result, in the limit of CPT invariance
and vanishing $\DG$, is
 \bea
 A_{\Dt} = \left.
 \frac{I(\ell^-,K_S,\Dt)-I(K_S,\ell^-,\Dt)}{I(\ell^-,K_S,\Dt)+I(K_S,\ell^-,\Dt)}\right|_{\DG=\delta=0}
 = \sin(2\beta)\sin(\Dm\Dt) .
 \label{ADT}
 \eea
The equality $A_{\Dt}=A_T$ should not be a surprise within the
approximations here performed: for CPT invariance, in the exact
limit $\DG=0$, $\Dt$ and T operations are equivalent.  This is valid
for Hamiltonians with the property of hermiticity up to a global
(proportional to the identity) absorptive part.

If we re-analyze the three asymmetries $A_{CP}$, $A_T$ and $A_{\Dt}$
relaxing the approximations of CPT invariance and vanishing $\DG$,
we find that their equality does not hold any more. Using the
results in Table \ref{f-cp} it is easy to show that
 \bea
 \begin{array}{rcl}
 \left.
   \begin{array}{c}
   \DG \neq 0 \\
   \delta = 0
   \end{array}
 \right\}
 & \then
 & A_{CP}=A_T \neq A_{\Dt} \\
 ~&~&~\\
 \left.
   \begin{array}{c}
   \DG \neq 0 \\
   \delta \neq 0
   \end{array}
 \right\}
 & \then & A_{CP} \neq A_T \neq A_{\Dt} \neq A_{CP}
 \end{array}
 \label{paja}
 \eea
From here, we see that $\delta = 0$ implies $A_T=A_{CP}$; this is an
{\it explicit} demonstration that the violation of CP balances the
violation of T if there is CPT invariance.  On the contrary, if
$\delta \neq 0$, we have
 \bea
 A_{CP}-A_T = \left( 1 - \frac{2Im(\epsilon)}{1+\mec}
 \sin(\Dm\Dt)\right)\times
 \left[ \frac{Im(\delta)}{1+\mec} \sin(\Dm\Dt) \right. \nn \\
     \left.
          \left(
          \frac{8 Im(\epsilon) Im(\delta)}{\left(1+\mec\right)^2} +
           \frac{1-\mec}{1+\mec} 4 \frac{Re(\delta)}{1+\mec}
          \right)
          \sin^2
            \left(
            \frac{\Dm\Dt}{2}
            \right)
 \right] + {\cal O}(\DG\,\delta,\delta^2),
 \nonumber
 \eea
which is another observable for CPT-violation.

At last we will extract from table \ref{f-cp} an observable to
measure $\DG$, assuming CPT invariance.  This is easily done once we
notice that, if $\delta=0$, the first and fourth columns (or the
second and third) are connected by the CP$\Delta t$-operation. They
should also be equal by CPT-invariance if $\Delta \Gamma =0$. The
presence of $\Delta \Gamma \neq 0$ will induce a CP$\Delta t$ -odd
asymmetry which is linear in $\Delta \Gamma$, i.e. a fake CPT-odd
asymmetry induced by absorptive parts. Explicitly \cite{berni} one
has
 \bea
 \left.\frac{I(K_S , l^+ , \Delta t) - I(l^- , K_S ,\Delta t)}{I(K_S , l^+ , \Delta t) + I(l^-, K_S, \Delta t)}\right|_{\delta=0} =\frac{\Delta\Gamma}{1 + \sin(2\beta ) \sin (\Delta m \Delta t) } \cdot \ \ \ \ \ \ \ \ \ \ \ \ \ \ \ \   \label{maite}
 \\
 \left\{ - \frac{\Delta t}{2} \cos (2\beta) +   4 x \sin^2 \left( \frac{\Delta m \Delta t}{2} \right) + 2 x \sin (2\beta)  \sin (\Delta m \Delta t)
 \right\}\ \ \nn
 \eea
where Eqs.$\,$(\ref{sonia2}) and (\ref{rosa2}) have been used.

The three terms of Eq.$\,$(\ref{maite}) contain different $\Delta
t$-dependence, therefore a good time resolution would allow the
determination of the parameters.  Notice that --as expected-- the
asymmetry is linear in $\Delta \Gamma$.  In addition, $\cos
(2\beta)$ --a quantity of high interest to remove the two-fold
ambiguity in the measurement of $\beta$-- is accompanied by
$\Delta\Gamma$.  We conclude that the comparison between the $(K_S,
l^+)$ and $(l^-, K_S)$ channels is a good method to obtain
information on $\Delta\Gamma$, due to the absence of any
non-vanishing difference when $\Delta\Gamma=0$.

\subsubsection{CP-CP decays}
\nota{no estaria mal darle una revisada a esta tabla de mierda}
\begin{table}[ht]
\begin{flushleft}
\begin{tabular}{|c|c|c|l}
\hline
decays & $(K_L,K_L)$ & $(K_S,K_S)$ & ~ \\
\hline
transition & $B_{-A} \to B_{+A}$ & $B_{+A} \to B_{-A}$ & ~ \\
\hline
\hline
$a$&$ \left( \frac{2 Im(\epsilon)}{1+|\epsilon|^2}\right)^2 - \frac{8Im(\epsilon)}{1+\mec} \frac{Im(\delta)}{1+\mec}$ &
  $\left( \frac{2 Im(\epsilon)}{1+|\epsilon|^2}\right)^2 + \frac{8Im(\epsilon)}{1+\mec} \frac{Im(\delta)}{1+\mec}$ &
  ~\\
\hline
$b$&$-\left( \frac{2 Im(\epsilon)}{1+|\epsilon|^2}\right)^2
+\frac{8Im(\epsilon)}{1+\mec} \frac{Im(\delta)}{1+\mec}$ &
  $-\left( \frac{2 Im(\epsilon)}{1+|\epsilon|^2}\right)^2 - \frac{8Im(\epsilon)}{1+\mec} \frac{Im(\delta)}{1+\mec}$ &
  ~\\
\hline $c$&0&
  0&
  ~\\
\hline $d$&$0$&
  $0$&
  ~\\
\hline
\end{tabular}
\end{flushleft}
\vskip .3cm
\begin{flushright}
\begin{tabular}{r|c|c|}
\hline
~ & $(K_L,K_S)$ & $(K_S,K_L)$ \\
\hline
~ & $B_{-A} \to B_{-A}$ & $B_{+A} \to B_{+A}$ \\
\hline \hline
 ~&
  $1+\left( \frac{1-\mec}{1+\mec} \right)^2$ &
  $1+\left( \frac{1-\mec}{1+\mec} \right)^2$ \\
\hline
 ~&
  $\left( \frac{2 Im(\epsilon)}{1+|\epsilon|^2}\right)^2$ &
  $\left( \frac{2 Im(\epsilon)}{1+|\epsilon|^2}\right)^2$ \\
\hline
 ~&
  $2\frac{1-|\epsilon|^2}{1+|\epsilon|^2}$&
  $-2\frac{1-|\epsilon|^2}{1+|\epsilon|^2}$\\
\hline
 ~&
  $+8\frac{Re\, \epsilon \ Im\, \epsilon}{(1+|\epsilon|^2)^2}$ &
  $-8\frac{Re\, \epsilon \ Im\, \epsilon}{(1+|\epsilon|^2)^2}$ \\
\hline
\end{tabular}
\end{flushright}
\caption{Correlated CP-CP decays.  The notation is as in Table
\ref{f-f}.} \label{cp-cp}
\end{table}

Although the case of correlated decays to CP eigenstates on both
sides is suppressed by a branching ratio factor of $10^{-8}$, the
high accumulated statistic at present day ($\sim 2\times10^8\ B\bar
B$ pairs) and the possible future upgrades to Super B-factories
\cite{10} makes worth their study.  The coefficients in
\eq{laura altea} for these decays, which need the CP-tag tool, are
found in Table \ref{cp-cp}.

The transitions of the first two columns are connected by T and, as
these channels consist of CP-eigenstates, by CPT-operation.  Their
asymmetry, at leading order in $\delta$, reads
 \nota{revisar una vez mas esta cuenta}
 \be
 \frac{I(K_S,K_S,\Dt)-I(K_L,K_L,\Dt)}{I(K_S,K_S,\Dt)+I(K_L,K_L,\Dt)}
 = 2 Im(\delta)/Im(\epsilon) .
\label{pucholarotonda}
 \ee
A deviation from zero of this observable would be a signal of CPT
violation. \comentariodelta{In any case, notice that $Im(\delta)$ provides a double
suppression, by CPT violation and by its proportionality to $\DG$.}
Note that the $Im(\epsilon)$ factor in the denominator of
\eq{pucholarotonda} should not be a surprise, since each intensity
by itself represents a transition which is forbidden unless there is
CP$_A$ violation in the mixing.

The decays of the last two columns are connected by $\Delta t$ and
$CP\Delta t$-operations.  Hence, assuming CPT invariance
($\delta=0$), we can reason as in the previous Section (see
\eq{maite}) to obtain an observable for $\DG$.  The asymmetry
between the third and fourth columns reads
 \bea
 \begin{array}{rl}
 \left. \frac{I(K_S, K_L, \Delta t) - I(K_L, K_S,\Delta t )}{I( K_S,  K_L, \Delta t) + I( K_L, K_S,\Delta t)} \right|_{\delta=0}
 =& \frac{\Delta \Gamma}{\left( 1 + \cos ^2 (2\beta ) \right)+  \sin ^2 (2\beta ) \cos (\Delta m \, \Delta t)} \cdot \\
 &\cdot  \left\{ 2\cos (2\beta) \frac{\Delta t}{2} + 4 x \sin (2\beta ) \sin (\Delta m \, \Delta t )
 \right\},
 \end{array}
 \label{maitules}
 \eea
which is linear in $\Delta \Gamma$, as expected.  As before,
$\cos(2\beta)$ shows up together with $\Delta\Gamma$.  This makes
impossible to measure the sign of one factor without having an
independent knowledge of the sign of the other one.  The individual
intensities shown in Table \ref{cp-cp} have additional dependence in
$\cos(2\beta)$ as shown in the denominator of
Eq.$\,$(\ref{maitules}): the $a$ coefficient of $I(K_S,K_L, \Delta
t)$ goes as $ 1 + \cos ^2 (2\beta) $, which once more is
unable to see the sign of $\cos(2\beta)$.

The correlated channel associated to the first, or second, column
corresponds to a CP-forbidden transition for all $\Delta t$.  The
decays to $(K_L ,\, K_L )$ or $(K_S , \, K_S )$ are associated with
transitions of $B$-states with opposite CP$_A$ eigenvalues. Assuming
CPT invariance and no absorptive parts, the ratio to the dilepton
intensity is
 \bea
 \left. \frac{I(K_S,K_S,\Delta t)}{I(l^+,l^+,\Delta t)} \right|_{\DG=\delta=0}
 %&=& \frac{|A_{J/\psi K_S}|^4}{|A_{l^+}|^4} \frac{\left(\frac{2Im(\epsilon)}{1+\mec}\right)^2 - \frac{2 Im(\epsilon)}{1+\mec} \frac{2Im(\delta)}{1+\mec}}{\left( 1 + \frac{4 Re\ \epsilon}{1+|\epsilon|^2} \right)} \nonumber \\
 &=& \frac{|A_{J/\psi K_S}|^4}{|A_{l^+}|^4}  \sin^2 (2\beta).
\label{la milanesa}
 \eea
It is natural to find that $I(K_S,K_S,\Dt)$ comes as the square of
the CP violation measure in \eq{cp5}, since it is the modulus square
of a CP-forbidden transition.  It is also expected to find no time
dependence in this ratio, since the transition amplitude between any
two orthogonal states, like $B_{+A} \rightarrow B_{-A}$ or $B_2
\rightarrow B_1$, has always the same time dependence, $e^{-i\mu_L
\Delta t} - e^{-i\mu_H \Delta t}$.

\section{Final remarks \hecho{90}}
 In this Chapter we have given a precise definition of the CP-tag
and have seen all the advantages of it.  In Section \ref{intensity
for correlated} we have computed with this tool all the possible CP
and flavour correlated intensities, in particular the CP-flavour and
CP-CP intensities.  We have shown that appropriate comparisons
between the intensities can give rise to precise T, CP and CPT
violation observables that can be used as consistency tests for the
Standard Model, as well as to explore beyond It.

Several of these observables here proposed can be measured in the
B-factories Babar and Belle with the present statistics.

As a close to this Chapter  --and a prelude to the next one-- it is
interesting to see in Tables \ref{f-f} and \ref{cp-cp} how,
independent of the details of the correlated $B$ meson state
evolution, the EPR correlation imposed by Bose statistics prevails
from the $\Upsilon$ decay up to the two final decays.  This can be
seen by setting $\Delta t=0$ and verifying the impossibility of
having the same decay at both sides, independent of the
CP-properties.  In next Chapter, where we add an extra CPT violation
ingredient relaxing the definite antisymmetry of the initial state,
this feature will prove to be possibly violated.

%%%%%%%%%%%%%%%%%%%%%%%%%%%%%%%%%%%%%%%%%%%%%%%%%%%%%%%%%%%%%%%%%%%%%%%%%%%%%%%%%
%%%%%%%%%%%%%%%%%%%%%%%%%%%%%%%%%%%%%%%%%%%%%%%%%%%%%%%%%%%%%%%%%%%%%%%%%%%%%%%%%
%%%%%%%%%%%%%%%%%%%%%%%%%%%%%%%%%%%%%%%%%%%%%%%%%%%%%%%%%%%%%%%%%%%%%%%%%%%%%%%%%
\cleardoublepage\chapter{CPT violation through distinguishability of particle and antiparticle}
\label{w}
\section{Introduction \hecho{90}}
In this Chapter we study a novel kind of CPT violation which occurs
in the initial state of the B-factories through the loss of
indistinguishability of particle-antiparticle.  It is worth noticing
that this CPT violation has no direct connection with the one that
could occur in the time evolution from B-mixing, studied in the
previous Chapter.

This novel idea was introduced by Bernab\'eu, Mavromatos and
Papavassiliou in 2003 \cite{prl}, and as theoretical work
 on it is developing, the first experimental tests should be done in the
fore-coming years at the B-factories.  In this Chapter we do a
theoretical study of the consequences of this CPT violation on the
equal-sign dilepton events.  We find several modifications to the usual
observables, which we propose to re-measure in a different region.  We also do the
experimental and statistical analysis for the measurements here proposed.
As a completion of the work, we set the first indirect limits on this new
effect by re-analyzing existing data on the equal-sign
dilepton charge asymmetry.

\subsection{Modification of the initial state due to distinguishability of particle-antiparticle \hecho{90}}
We now analyze how the implementation of CPT violation, as a
breakdown of the indistinguishability of particle-antiparticle,  can
modify the initial state in the B factories.

In the usual formulations of $entangled$ meson states
\cite{lip68,usual} in the $B$ factories, one imposes the requirement
of {\it Bose statistics} for the state $\b \bb$, which implies that
the physical system must be $symmetric$ under the combined operation
of charge conjugation ($C$) and permutation of the spatial
coordinates ($\parity$), $C\parity$.   As careful discussed in
Section \ref{correlated}, this argument plus conservation of angular
momentum ($l=1$) and the proper existence of the antiparticle state,
yields to an initial state after the $\Upsilon(4S)\to\b\bb$ decay
which can be written as
 \be |\psi(0)\rangle_{\w=0} = \sq \left( | \b (+\vec k),\bb(-\vec k)\rangle - |\bb(+\vec k), \b(-\vec k)\rangle \right) .
 \label{initial}
 \ee
Where the $\vec k$ vector is along the direction of the momenta of the
mesons in the center of mass system.

Although not directly evident, a detailed analysis of the previous
paragraph shows that among the assumptions leading to the initial
state, \eq{initial}, it is $CPT$-invariance.  As was first pointed
out in \cite{prl}, and later developed in \cite{plb,nik}.  In fact,
as mentioned above, if $CPT$ symmetry is violated then $\b$ cannot
be considered as indistinguishable to $\bb$, and hence the
requirement of $C\parity=+$ imposed by Bose statistics must be
relaxed. As a result, we may rewrite the initial state,
\eq{initial}, as
 \bea
 |\psi(0)\rangle = \frac{1}{\sqrt{2(1+|\w|^2)}} \times \left\{ | \b (+\vec k),\bb(-\vec k)\rangle - |\bb(+\vec k), \b(-\vec k)\rangle + \right. \nonumber \\
 \left.\w \left( | \b (+\vec k),\bb(-\vec k)\rangle + |\bb(+\vec k), \b(-\vec k)\rangle \right) \right\}.
 \label{i}
 \eea
Where $\w=|\w| e^{i\Omega}$ is a complex CPT violating parameter, associated
with the non-indistinguishable particle nature of the neutral meson
and antimeson states, which here parameterizes the loss of Bose
symmetry.  Observe the symmetry in \eq{i} in which a change in the
sign of $\w$ is equivalent to the exchange of the particles $\b
\leftrightarrow \bb$.  Moreover, once defined through \eq{i} the
modulus and phase of $\w$ have physical meaning which could be in
principle measured.

We interpret \eq{i} as a description of a state of two {\it
distinguishable} particles at first-order in perturbation theory,
written in terms of correlations of the zeroth-order one-particle
states.  As it is easily seen, the probabilities for the two states
connected by a permutation are different due to the presence of
$\w$.  Therefore, in order to give physical meaning to $\w$ we {\it
define}, in \eq{i}, $+\vec k$ as the direction of the {\it first}
decay; and viceversa, $-\vec k$ as the direction of the second
decay.  Of course, when both decays are simultaneous there is no way
to distinguish the particles; although the effect of $\w$ is still
present (see Section \ref{Demise}).

Notice that an equation such as the one given in \eq{i} could also
be produced as a result of deviations from the law of quantum
mechanics, during the initial decay of the $\Upsilon$ state.  Thus
\eq{i} could receive contributions from two different effects, and
can be thought of as simultaneously parameterizing both of them.  In
the present work we will assume that \eq{i} arises $solely$ due to
deviations from the indistinguishable-particle nature of the $\b -
\bb$ states, while the Hamiltonian for the initial decay of the
$\Upsilon$ and the subsequent evolution of the entangled state is
the usual in quantum mechanics.

It is clear that the modification of the initial state, \eq{i}, will
introduce modifications in all the observables of the $B$ factories,
since it is the departure point of every analysis.  In some
observables this change may be minor, whereas in others may produce
considerable modifications in its behaviour.  Throughout this
Chapter we study these two cases: in Section \ref{Demise} we analyze
observables which need of $\w^2$, but conceptually important since
constitute the demise of flavour-tagging; whereas in Section
\ref{time dependent} we advocate to the study of time dependent
observables, we find in them a linear dependence in $\w$ and an
enhanced peak in the $A_{sl}$ asymmetry if $\w\neq0$.

\section{The demise of flavour tagging and the equal-time decay
observables \hecho{90}}
\label{Demise}

In this Section we study a direct consequence of the
modification of the initial state (\eq{i}), the demise of flavour
tagging \cite{plb}.  This is a conceptual modification for the B factories,
which use the flavour-tag in a great part of their analysis.

\subsection{Time evolution and conceptual changes due to the
appearance of 'forbidden' states \hecho{90}}
As it is easily seen, in passing from \eq{initial} to \eq{i} there
is a loss in the definite antisymmetry of the state.  This
antisymmetry, when $\w=0$, forbids the system to have the same decay
at the same time on both sides,
 \bea
 \langle X,X |U(t)\otimes U(t)|\psi(0)\rangle_{\w=0} = 0,
 \eea
since a permutation of the particles introduces a minus sign in
$|\psi(0)\rangle_{\w=0}$, whereas the rest remains the same. If, on
the contrary, we have $\w\neq0$ as in \eq{i}, then there is nothing
to be said and the same decay on both sides at the same time it is,
in principle, allowed.  This fact constitutes a breakdown of any tag
as known to the date, since having a decay $X$ on one side does not
assure that the meson on the other side is the complementary, $\bar X$.
In particular, we study here the demise of flavour-tagging.

To analyze in depth the argument above exposed it is mandatory to
study the time evolution of the initial state in \eq{i}.  Using the
results of the \ww formalism in Section \ref{evolution} it is
straightforward to obtain, up to a global phase, the time-evolved
state function in the flavour basis:
 \bea
 |\psi(t)\rangle=\frac{e^{-\G}}{\sqrt{2\left(1+\abs{\omega}^2\right)}}\left\{
 C_{0\bar 0}(t)\BBb + C_{\bar 0 0}(t)\BbB + \right. \nn \\
 \left. C_{00}(t)\BB +C_{\bar 0\bar 0}(t)\BbBb\right\}
; \label{t}
 \eea
where
 \bea
 C_{0\bar 0} &=& 1 + \w \left[ \cosh (\alpha t) + 2 K_\delta ^2  \cosh^2\left(\frac{\alpha t}{2}\right) \right] \nn \\
 C_{\bar{0} 0} &=& - 1 + \w \left[ \cosh (\alpha t) + 2 K_\delta ^2  \cosh^2\left(\frac{\alpha t}{2}\right) \right] \nn \\
 C_{00} &=& \w K_+ \left[ -\sinh (\alpha t) + 2 K_\delta \sinh^2
 \left(\frac{\alpha t}{2}
 \right) \right] \nn \\
 C_{\bar 0 \bar 0} &=& \w K_- \left[ -\sinh (\alpha t) - 2 K_\delta
 \sinh^2 \left(\frac{\alpha t}{2} \right) \right] , \label{cs}
 \eea
and $K_{\delta,\pm}$ and $\alpha$ are defined in Section
\ref{analysys}.  We emphasize that the above expressions are exact;
no expansion with respect to any of the parameters has taken place.
We also notice again at this point that the CPT violating parameter $\delta$
measures CPT violation in the $B$-mixing, and is totally independent
of the parameter $\w$, whose origin lies in the
particle-antiparticle distinguishability nature.

It is worth to notice that phase redefinitions of the single
$B$-meson states such as $\b \mapsto e^{i\gamma} \b ,\, \bb \mapsto
e^{-i\bar \gamma} \bb$ are easily handled through the
transformations of the $K_i$-expressions in Eqs.\ (\ref{rephase1}-\ref{rephase2}).  They lead
to explicitly rephasing invariant $C_{0\bar 0}(t)$ and $C_{\bar 0
0}(t)$ coefficients, whereas $C_{00}(t)\mapsto
e^{i(\gamma-\bar\gamma)}C_{00}(t)$ and $C_{\bar 0\bar 0}(t)\mapsto
e^{i(\bar\gamma-\gamma)}C_{\bar 0\bar 0}(t)$ are individually
rephasing-variant, but their dependence on the phase is such that
the considered physical observables are rephasing invariant, as they
should.

Observe how the loss of the definite anti-symmetry in the initial
state due to $\w$ gives rise to the appearance of the states $\BB$
and $\BbBb$ in the time evolved $|\psi(t)\rangle$, \eq{t}.  These
states are {\it forbidden} at any time if $\w=0$.  They are the
responsible of the demise of flavour tagging: a flavour specific
decay on one side at time $t_1$ does not determine uniquely the
flavour of the other meson at the same time $t_1$.  In fact, due to
the same flavour states in \eq{t}, a first flavour specific $\b$
decay filters at that time the wave function of the meson on the
other side to be $\sim {\cal O}(1) |\bb\rangle + {\cal O}(\w)
|\b\rangle$, and vice-verse.  Therefore, the probability of having
the same flavour specific decay at the same time on both sides goes
as
 \be
 I(\ell^\pm, \ell^\pm, \Dt=0) \sim |\w|^2 .
 \label{demise}
 \ee

Having concluded the demise of the concept of tagging in the
presence of $\w$, we follow to propose observables
which would actually measure the deviation, if any, from the basic
tagging assumption.

\subsection{The experimental observables \hecho{90}}
We will focus on observables involving {\it simultaneous} $\b$ or
$\bb$ flavour specific decays on both sides.  In what follow we will
restrict our attention to the most characteristic case of flavour
specific channels, namely semileptonic decays.  The main reason for
this choice is the fact that the flavour specificity of such decays
relies on a minimum number of assumptions, in particular solely on
the equality $\Delta B=\Delta Q$, and is completely independent on
whether the CP and CPT symmetries are exact \cite{15}.  We emphasize
that other flavour specific channels may not share this property
when there is CP or CPT violation in the decay.  Notice also that
any effects stemming from the possibly decoherent (i.e. non
quantum-mechanical) evolution of the initial state can be
unambiguously separated from the $\omega$ effect through the
difference in the symmetry properties of their contributions to the
density matrix \cite{prl}.

We define the equal time intensity as $I_{ab}(t) = \abs{\langle X_{ab}|\psi(t)\rangle}^2$.  We find that
 \be
 I_{ab}(t)=\abs{\langle Y_{a}| B^{a}\rangle }^2
 \abs{\langle Z_{b}| B^{b}\rangle }^2
 \frac{e^{-2\Gamma t}}{2(1+\abs{\omega}^2)}\abs{C_{ab}(t)}^2 ,
 \label{eq:FLAVOURintensities:01}
 \ee
where the state $|X_{ab}\rangle$ has been decomposed into the
two single-meson flavour-specific decay states, $Y_{a}$ and $Z_{b}$, i.e.
$|X_{ab}\rangle = |Y_{a},Z_{b}\rangle $.  (For instance,
$|X_{\b\b}\rangle = |X\ell^+,X'\ell^+\rangle$.)
These equal-time intensities can be easily time-integrated:
\be
{\cal{I}}_{ab}=\int_0^\infty \!\!\!\!\! dt~ I_{ab}(t).
\label{eq:FLAVOURintensities:02}
\ee

In term of these equal time intensities, $\omega\neq 0$ allows
\[
I_{00}(t)\neq 0\quad;\quad  I_{\bar 0\bar 0}(t)\neq 0~.
\]
It is through these otherwise-forbidden intensities that we can
explore the presence of $\omega\neq 0$. As we can see in \eq{cs} and
\eq{eq:FLAVOURintensities:01},
 what one hopes to observe  is an $\abs{\omega}^2$ vs. $0$ effect.
This would be an {\it unambiguous} manifestation of our effect,
independently of any other source of symmetry violation.

In the hypothetical situation of  non-vanishing values for
$I_{00}(t)$ and  $I_{\bar 0\bar 0}(t)$ one could
consider a CP-type asymmetry of the form
\begin{eqnarray}
  A_{CP}(t)&=&\frac{I_{00}(t)-I_{\bar 0\bar 0}(t)}{I_{00}(t)+I_{\bar 0\bar
  0}(t)},
  \\ \addtocounter{equation}{-1}
{\cal A}_{CP}&=&\frac{{\cal{I}}_{00}-{\cal{I}}_{\bar 0\bar 0}}
{{\cal{I}}_{00}+{\cal{I}}_{\bar 0\bar 0}} ~~ . \label{eq:Asymmetries:01}
\end{eqnarray}
The asymmetries $A_{CP}(t)$ and ${\cal A}_{CP}$ express the
difference between the decay rates of  $B^0 \rightarrow X_0$ and
$\bar B^0 \rightarrow \bar X_{0}$, where, $X_0$ is a specific
flavour channel  and  $\bar X_{0}$  its  C-conjugate  state.  In
order to isolate the physics associated with $C_{00}$ and $C_{\bar
0\bar 0}$ through an observable such as $A_{CP}(t)$, one must
eliminate its dependence on  the decay amplitudes $\abs{\langle
Y_{a}| B^{a}\rangle }^2 \abs{\langle Z_{b}| B^{b}\rangle }^2$
entering through  \eq{eq:FLAVOURintensities:01}. If the physics
governing the decay is CPT-invariant  (as in  the  Standard Model),
the  use of inclusive channels   guarantees  the   cancelation  of
the decay amplitudes  in \eq{eq:Asymmetries:01}. If  we consider
exclusive channels instead,  CP violation  in   the decays prevents
in general the aforementioned cancelation from taking place, thus
restricting the usefulness of $A_{CP}(t)$. In addition to these
standard considerations, quantum gravity itself may affect  the CPT
invariance in the decays; nevertheless, such contributions will be
subleading, and we will neglect them in what follows.

Interestingly enough,  $A_{CP}(t)$ and ${\cal A}_{CP}$ are
independent of the value of $\omega$, since the latter clearly
cancels out when forming the corresponding ratios, to leading order,
when quantum gravity induced CPT violating effects in the decays are
ignored. For $\delta=0$ and $\DG$ small, terms of order
$\omega \DG$ can be safely neglected, and hence \eq{eq:Asymmetries:01}
simplifies to
 \begin{equation}
 A_{CP}(t)={\cal A}_{CP}=\frac{\abs{1+\epsilon}^4-\abs{1-\epsilon}^4}{\abs{1+\epsilon}^4+\abs{1-\epsilon}^4}=
 \frac{4~(1+\abs{\epsilon}^2)~Re(\epsilon)}{(1+\abs{\epsilon}^2)^2+(2~Re(\epsilon))^2} ~.% \sim {\mathcal O}(\Delta \Gamma)~.
 \label{eq:asymm2}
 \end{equation}
In terms of the standard mixing parameters $p$ and $q$,
\[
\frac{\abs{1+\epsilon}^4-\abs{1-\epsilon}^4}{\abs{1+\epsilon}^4+\abs{1-\epsilon}^4}=
\frac{\abs{p}^4-\abs{q}^4}{\abs{p}^4+\abs{q}^4}=\frac{2\Delta_B}{1+\Delta_B^2}~,
\]
where
\[
\Delta_B=\frac{2~Im(M_{12}^* \Gamma_{12})}{(\Dm)^2+\abs{\Gamma_{12}}^2}.
\]
According to present measurements the only limit to $A_{CP}$ and
${\cal A}_{CP}$ comes from the $t$-integrated and $\Dt$-averaged
equal-sign dilepton charge asymmetry $A_{sl}$, see \eq{vecina}.

The algebraic cancelation of all the $\omega$ dependence in
\eq{eq:Asymmetries:01} can be physically understood by realizing
that $\omega\neq 0$ allows the equal time presence of $\BB$ and
$\BbBb$ terms, and it has nothing to do with $B^0$--$\bar B^0$
 mixing  or  $B^0,{\bar  B}^0$
decays.
%As mentioned previously, possible quantum gravity effects in the decays contribute to higher order (at least linear in $\omega$-like parameters) terms in \eq{eq:asymm2}.
The CP asymmetries in \eq{eq:Asymmetries:01} are thus
\emph{conventional} CP asymmetries between states which are both
CPT-forbidden; this cancelation is an explicit proof of both
effects. This provides an additional way of testing the
self-consistency of the entire procedure: once non-vanishing
$I_{00}(t)$ and  $I_{\bar 0\bar 0}(t)$ have been established one
should extract the experimental value of ${\cal A}_{CP}$, which
should coincide with the theoretical expression  of
\eq{eq:Asymmetries:01}; for the calculation of the latter one needs
as input only the standard value for the parameter $\epsilon$, with
no reference to the actual value of $\omega$.

It is worth noticing at this point that a CP asymmetry built from
the equal-time intensities $I_{0 \bar 0}$ and $I_{\bar 0 0}$ is not
possible in this framework, since they are experimentally indistinguishable.  However,
if the distinction of the particles would have not been defined
through their order of decay (c.f.\ \eq{i}), then this asymmetry
could have been possible.  Of course, in this case the value of $\w$
would have not been the same \cite{plb}.

\section{Linear in $\w$ time-dependent observables \hecho{90}}
\label{time dependent}

In this Section we study the time dependent observables, where the
interference terms will give rise to linear terms in $\w$. We
analyze correlated flavour specific equal-sign decays on both sides.
As a first step we explore the corrections to the equal-sign
dilepton intensities, and from them we study the modification in the
behaviour of their asymmetry, $A_{sl}$.

\subsection{Corrections to the equal-sign dilepton intensity \hecho{90}}
Restricting the analysis to equal-sign semi-leptonic final states, the intensity is
written as
 \be I(X \ell^\pm,X' \ell^\pm , \Dt) = \int_0^\infty \left| \langle X \ell^\pm ,X'\ell^\pm |U(t_1) \otimes U(t_1+\Dt) |\psi(0)\rangle \right|^2 dt_1 .
 \label{int}
 \ee
The computation of \eq{int} is straightforward using
Eq.~(\ref{i}) and the results of Section \ref{evolution}; its
explicit calculation yields
\newpage
 {\small
 \bea
 I(X\ell^\pm, X'\ell^\pm, \Delta t) = \frac{1}{8} e^{-\G \Dt} \, |A_X|^2 |A_{X'}|^2 \, \left|\frac{(1+s_\epsilon\, \epsilon)^2-\delta^2/4}{1-\epsilon^2+\delta^2/4}\right|^2 \qquad ~ \ \qquad\ \quad\nonumber \\
 %&&\left\{ \Bigg[ \frac{1}{\G} + a_\w \frac{8\G}{4\G^2 + \Dm^2} Re(\w) + \frac{4\Gamma}{4\Gamma^2 - \DG^2} |\w|^2 \Bigg] \cosh \left(\frac{\DG \Dt}{2}\right) + \right. \nonumber \\(esta era la linea siguiente antes de aproximar $\w\DG \sim 0$ -------
 \qquad ~\qquad
 \begin{array}{l}
 \left\{ \Bigg[ \frac{1}{\G} + a_\w \frac{8\G}{4\G^2 + \Dm^2} Re(\w) + \frac{1}{\Gamma} |\w|^2 \Bigg] \cosh \left(\frac{\DG \Dt}{2}\right) + \right. \\
 \Bigg[ - \frac{1}{\G} + b_\w \frac{8\G}{4\G^2 + \Dm^2} Re(\w) -\frac{\Gamma}{\Gamma^2 + \Dm^2} |\w|^2 \Bigg] \cos (\Dm \Dt) +  \\
 %&&\Bigg[ \frac{2 \DG}{4\Gamma^2 - \DG^2} |\w|^2 \Bigg] \sinh \left( \frac{\DG \Dt}{2} \right)  \nonumber \\
  \left. \Bigg[ d_\w \frac{4 \Dm}{4\G^2 + \Dm^2}  Re(\w)  + \frac{\Dm}{\Gamma^2 + \Dm^2} |\w|^2 \Bigg] \sin (\Dm \Dt) \right\},
 \end{array}
 \label{piropo2}
 \eea
 }
\noindent where we have approximated $\w K_\delta
\sim 0$, and terms containing $\w \DG$ have been neglected within
the square brackets -- therefore the terms coming from the expansion
of the cosine hyperbolic that go as $\w\DG$ must be neglected to
maintain the consistency.  The coefficients $s_\epsilon,\ a_\w,\ b_\w$ and $d_\w$ in \eq{piropo2} for the
different decays are found in Table \ref{pavese}.

\begin{center}
\begin{table}[h]
\begin{center}
\begin{tabular*}{\columnwidth}{@{\extracolsep{\fill}}|c|c|c|}
\hline
Decay & ~$\left( \ell^+,\ell^+, \Dt \right)$ &  ~$\left( \ell^-,\ell^- ,\Dt \right)$  \\
\hline
{\tiny aprox.~transition} & {\tiny $\sim (1+\w) \langle \b |T|\bb \rangle + \w \langle \b |T| \b \rangle$} & {\tiny $\sim (1+\w) \langle \bb |T| \b \rangle + \w \langle \bb |T| \bb \rangle$} \\
\hline
\hline
$s_\epsilon$   &  1 & -1 \\
\hline
$a_\w$         &  1 & -1 \\
\hline
$b_\w$         & -1 &  1 \\
\hline
$d_\w$         &  1 & -1 \\
\hline
\end{tabular*}
\end{center}
\caption{Sign coefficients for the different intensities in \eq{piropo2}. The second
row indicates the transition amplitude for the corresponding single
B-processes written in an 'order of magnitude'--estimate, id est, '1'
stands for ${\cal O}(1)$, and '$\w$' for ${\cal O}(\w)$.}
\label{pavese}
\end{table}
\end{center}

%{\it ...Posible parrafo que hable de las vicisitudes de la
%\eq{piropo2}... de por que solo aparece $Re(\w)$ (esto no pasa si no
%se aproxima $\w\delta\sim\w\DG\sim 0$!), de por que aparece $\G$ en
%los terminos pares en $\Dt$, mientras que $\Dm$ con los impares,
%etc.. } \vskip .5cm

\begin{figure}
 \framebox[\textwidth]{
 \subfloat[\label{fig1a}]{\includegraphics[width=.45\textwidth]{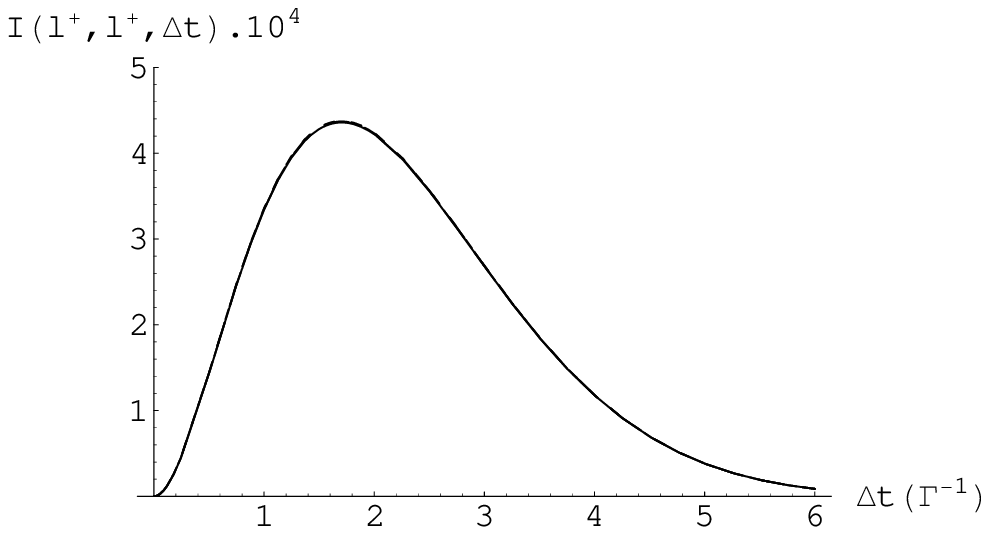}}\quad\subfloat[zoom\label{fig1b}]{\includegraphics[width=.45\textwidth]{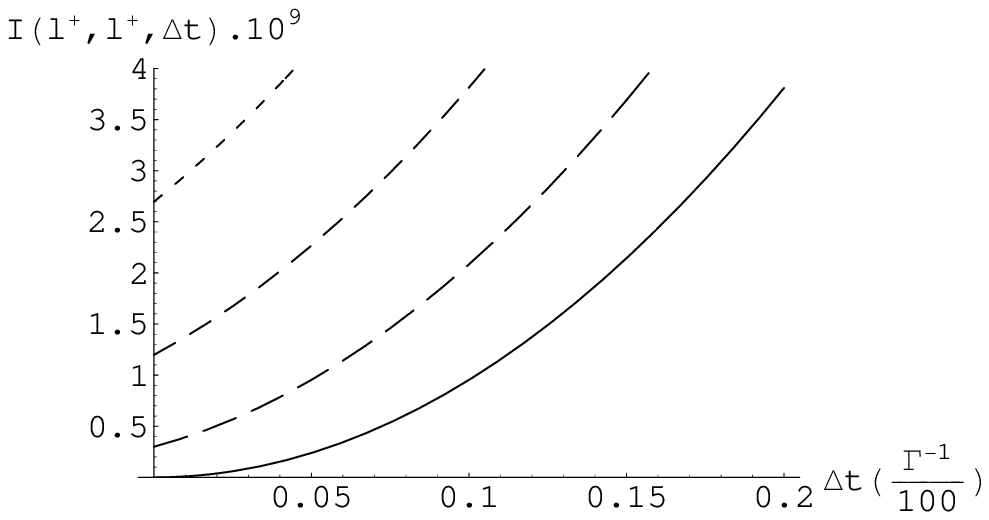}}
 }
 \framebox[\textwidth]{
 \subfloat[\label{fig1c}]{\includegraphics[width=.45\textwidth]{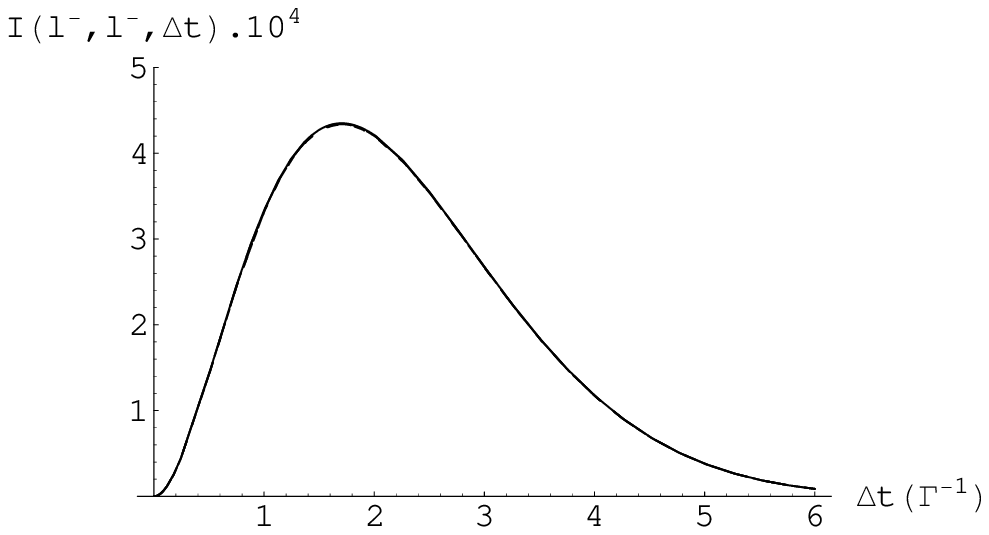}}\quad\subfloat[zoom\label{fig1d}]{\includegraphics[width=.45\textwidth]{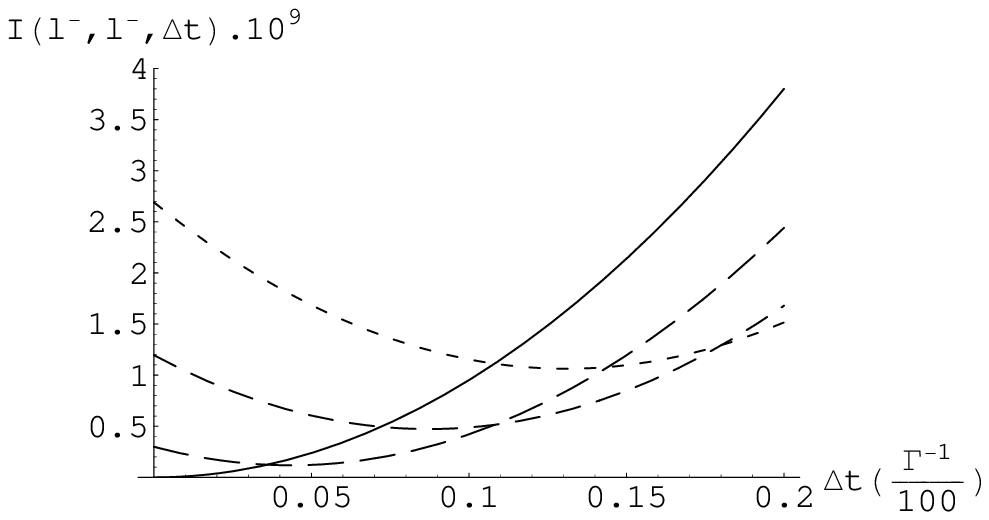}}
 }
 \caption{Modification of the equal-sign dilepton intensities
due to the presence of $\w$; $\w=0$ (solid-line), $\w=0.0005$
(long-dashed), $\w=0.001$ (medium-dashed) and $\w=0.0015$
(short-dashed).  The zoom figures, $(b)$ and $(d)$, plot the demise
of flavour tagging: for $\Dt\to 0$ the intensity goes as $|\w|^2$.
Notice also the different behaviour of the intensity as a function
of $\w$ in each of the zoom-figures.  As it is shown below, this
produces a maximum in their asymmetry for small $\Dt$.} \label{fig1}
\end{figure}

In the next paragraphs we study how the new $\w$-terms in
\eq{piropo2} change the usual $\Dt$-dependent observables.  We first
analyze the variations that occur in the intensities, and then focus
in the time dependent and enhanced modifications which the
$\w$-effect introduces in their asymmetry.  For the sake of
simplicity, without compromising the physical results, from this
point forward we take $\delta \approx \DG \approx 0$, and
$4Re(\epsilon)/(1+|\epsilon|^2) \approx 1 \times 10^{-3}$ in accordance
with its \sm expected value \cite{pdg}.

The change in the intensity due to the presence of $\w$ is easily
analyzed in \eq{piropo2}, as well as visualized in \fig{fig1}.  In the
region $\Dm\Dt \gtrsim 1$, a linear correction in $Re(\w)$ takes
place in both intensities (see Figs.\ \ref{fig1a} and \ref{fig1c}).  This change in the
intensity would be hardly detectable, since there is no enhancement
nor a profitable comparison to measure it.  On the other hand, observe that in the
limit $\Dt \to 0$ the aforementioned linear terms cancel each other
and the leading terms in \eq{piropo2} go as $\sim |\w|^2$ (see
Figs.\ \ref{fig1b} and \ref{fig1d}).  Although this is a much smaller correction to the
intensity, it is worth to study it due to the following reasons:
$(i)$ the modification in the intensity in this case is to be
compared with zero, which is the usual result for $\w=0$; and $(ii)$
it introduces {\it conceptual changes} in the analysis of the
problem since it constitutes, as studied in the previous Section,
the demise of flavour tagging.  Finally, observe the
interesting phenomena that is produced in the $\Dm\Dt<<1$ region due
to $\w$.  In this region, for sufficiently small $\Dt$, the
behaviour of the intensity is dominated by the term in the last line
of \eq{piropo2} which is proportional to $\sim d_\w\,
Re(\w)\sin(\Dm\Dt)$.  Therefore we have that, due to $d_w$, a
different behaviour will be seen in each of the intensities (see
Figs.~\ref{fig1b} and \ref{fig1d}): for these values of $\w$ we have that
$I(\ell^+,\ell^+,\Dt)$ begins growing, whereas
$I(\ell^-,\ell^-,\Dt)$ begins decreasing and then grows.  As
analyzed in next Section, this difference between both intensities will produce
a minimum in the denominator of their asymmetry, and hence a
peak for short $\Dt$'s could be expected.  Notice also that this characteristic
behaviour in each of the intensities depends on the sign of $Re(\w)$,
which is measurable.   This was expected, since the initial state in
\eq{i} possesses the symmetry through which the exchange
$\b\leftrightarrow\bb$ is equivalent to the change
$+\w\leftrightarrow -\w$.

\subsection{Behaviour modification in the equal-sign dilepton charge asymmetry \hecho{90}}
We now study how the presence of $\w$ in the intensity modifies the
behaviour of the equal-sign dilepton charge asymmetry, also known as
Kabir asymmetry.  As studied in Chapter \ref{cp}, this asymmetry for
$\w=0$ is of order $Re(\epsilon)$ and {\it exactly}
time-independent,
 \be
 A_{sl} = \left. \frac{I(\ell^+,\ell^+,\Dt)-I(\ell^-,\ell^-,\Dt)}{I(\ell^+,\ell^+,\Dt)+I(\ell^-,\ell^-,\Dt)}\right|_{\w=0} = 4 \re + {\cal O}((Re\ \epsilon)^2),
 \label{ak}
 \ee
  as it can be also easily seen by setting $\w=0$ in
\eq{piropo2}. The time independence in \eq{ak} comes from the fact
that when $\w=0$ both intensities have the same time dependence,
which cancels out exactly in the asymmetry. However, if $\w \neq 0$
then the time dependence of both intensities are not equal any more,
and hence the $A_{sl}$ asymmetry would acquire a time dependence.
Moreover, as explained below, if $Re(\w) \neq  0$ then the asymmetry
will have an enhanced peak for small $\Dt$ and hence there will be
an optimal region where to search for experimental evidence of $\w$.

\pepe{esto es lo que charlamos la ultima vez.  Ahora esta mas claro}
In order to see qualitatively how the equal-sign dilepton charge
asymmetry acquires a peak for small $\Dm\Dt$, one can compute the
asymmetry using \eq{piropo2} keeping in numerator and denominator
terms quadratic in $(\Dm\Dt)$.  In this expression it can be seen
that, in the limit $\Dm\Dt \ll |\w|$, the value of the asymmetry is
$4Re(\epsilon)/1+|\epsilon|^2$ at $\Dt=0$ and then increases or
decreases its values with $\Dt$ depending whether $Re(\w)$ is
positive or negative, respectively.  On the other hand, in the limit
$\Dm\Dt \gg |\w|$, the asymmetry decreases or increases depending
whether $Re(\w)$ is positive or negative, respectively
\comentariodelta{; converging again to the asymptotic value
$4Re(\epsilon)/1+|\epsilon|^2$ (within the validity of this
expansion)}. Therefore a maximum (minimum) is expected for $\Dm\Dt
\sim |\w|$ when $Re(\w)$ is positive (negative).  In fact, an
expansion of numerator and denominator in the asymmetry at the
region $\Dm\Dt\sim\w\ll 1$ yields
\newcommand{\er}{\frac{4Re(\epsilon)}{1+|\epsilon|^2}}
 \be
 A_{sl} \approx \frac{\er\left( 1-\frac{1}{1+x_d^2} \right) |\w|^2 + \frac{x_d}{1+x_d^2/4}\, Re(\w) \,\Dm\Dt + \er \, \frac{1}{2} (\Dm\Dt)^2}{\left( 1-\frac{1}{1+x_d^2} \right) |\w|^2 + \er \frac{x_d}{1+x_d^2/4}\, Re(\w) \, \Dm\Dt + \frac{1}{2}
 (\Dm\Dt)^2} ,
 \label{anatandarica}
 \ee where $x_d=\Dm/\G$.  As it can be easily analyzed, this
expression possess a pronounced maximum
(minimum) for $Re(\w)>0$ ($<0$) for small positive $\Dm\Dt$.  This
analysis shows that the only existence of $Re(\w)\neq 0$ makes a
drastic change in the behaviour of $A_{sl}$ for small $\Dm\Dt$.
Although the measurement of this peak might require a fantastic time
precision (and the exploration of a {\it dirty} region from the
experimental point of view), it is shown below that the tail of
the peaks gives an still optimal region where to look for experimental
effects related to $\w$.

\begin{figure}
\centering \framebox[1\textwidth]{
\subfloat[$\Omega=0$\label{kabir1}]{\includegraphics[width=.45\textwidth]{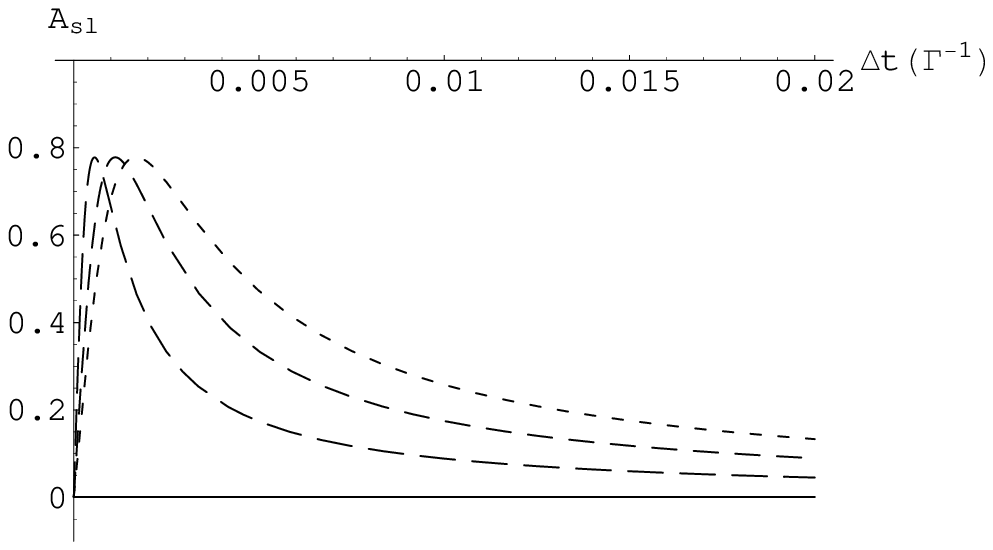}}\quad\subfloat[$\Omega=\frac{3}{2}\pi$
\label{kabir2}]{\includegraphics[width=.45\textwidth]{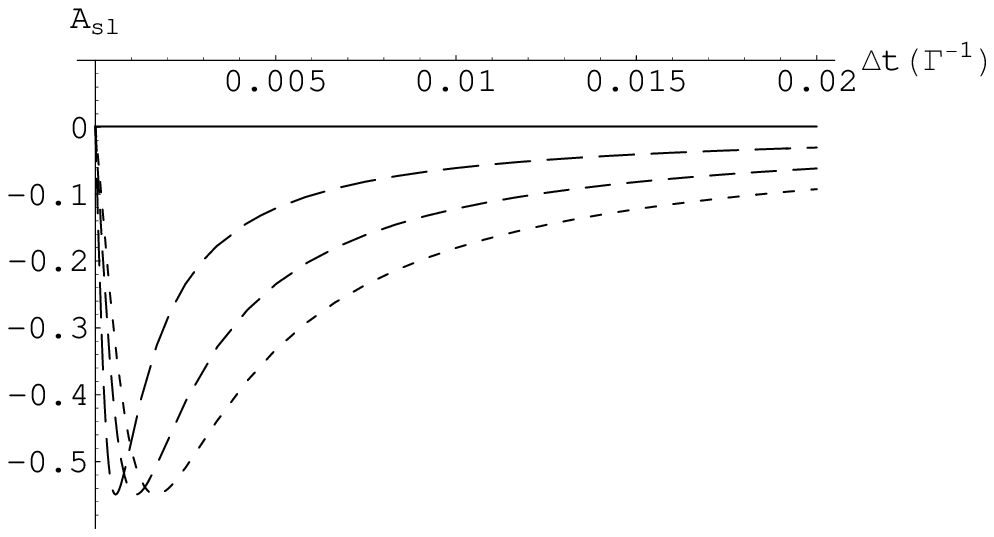}} }
\caption{Equal-sign dilepton charge asymmetry for different values
of $\w$; $|\w|=0$ (solid line), $|\w|=0.0005$ (long-dashed),
$|\w|=0.001$ (medium-dashed), $|\w|=0.0015$ (short-dashed).  When
$\w\neq0$ a peak of height $A_{sl}(peak) = 0.77 \cos(\Omega)$ appears
at $\Dt(peak)=1.12\, |\w| \, \frac{1}{\Gamma}$, producing a drastic
difference, in particular in its time dependence.  Observe that the
peak, independently of the value of $|\w|$, can reach enhancements
of order $10^3$ the value of the asymmetry when $\w=0$.}
\label{fig2}
\end{figure}

For a better study of the equal-sign dilepton charge asymmetry its plot is shown for
different values of $\w$ in \fig{fig2}.  As it can be seen, the
asymmetry in the $\w\neq0$ case is not only time dependent but also
has a pronounced peak, as it was expected from the discussion in the
previous paragraph. Using \eq{piropo2} one obtains that the peak is
at the value
 \be
 \Dt_{peak} = \frac{1}{\Gamma} \, \sqrt{\frac{2}{1+x_d^2}} \, |\w| + {\cal O}(\w^2) \approx \frac{1}{\Gamma}\, 1.12 \, |\w| ,
 \label{tmax}
 \ee
where in the last step we have set $x_d=0.77$ \cite{pdg}. The
height of the peak is
 \bea
 A_{sl}(\Dt_{peak}) &=& \frac{\cos(\Omega) + \sqrt{2} \, \frac{4+x_d^2}{\sqrt{1+x_d^2}} \, \re}{4\, \cos(\Omega) \re + \frac{1}{2\sqrt{2}} \, \frac{4+x_d^2}{\sqrt{1+x_d^2}}} + {\cal O}(w, \, (Re\ \epsilon)^2)  \nonumber \\
 &=& 0.77 \cos (\Omega) + {\cal O}(\w, \, Re(\epsilon))
 \label{hmax}
 \eea
 We see that although the position of the peak is linearly dependent
on the absolute value of $\w$ (\eq{tmax}), {\it its height is independent of the
moduli of $\w$ and only depends on its phase} (\eq{hmax}).  Moreover, the
enhancing numerical factor $0.77$ in the leading term of the height
of the peak, \eq{hmax}, is about three orders of magnitude
bigger than the expected value of the asymmetry if $\w=0$.

In the region $\Dm\Dt \sim 1$ we expect a quasi time-independent
behaviour of $A_{sl}(\Dt)$ for small $\w$.  In fact, analyzing the
time-derivative at the typical time $\Dt=1/\G$ we find
\pepe{el analisis que comienza ahora antes no estaba.  Me parecio
adecuado, por completitud y porque da una idea mejor para comparar
con los datos experimentales existentes}
\bea
\frac{1}{\G}\,\frac{d\, A_{sl}}{d\, \Dt} (\Dt=1/\G) = -0.78 Re(\w) +{\cal O}(\w^2),
\label{pavesetrola}
\eea
a result which is independent of $\epsilon$ and $\DG$ (in its
allowed range $\DG/\G\sim10^{-3}$).  However, a small shift with respect to
the $\w=0$ case there will be always present.  The computing of the
asymmetry at this time gives an estimate of the expected shift:
\bea
A_{sl}(\Dt=1/\G) &=& \frac{4Re(\epsilon)}{1+\mec} + 3.39 Re(\w) + {\cal O}(\w^2).
\label{shift}
\eea

\begin{figure}[h]
\begin{center}
\framebox[1\textwidth]{
\subfloat[\label{moco}]{\includegraphics[width=.45\textwidth]{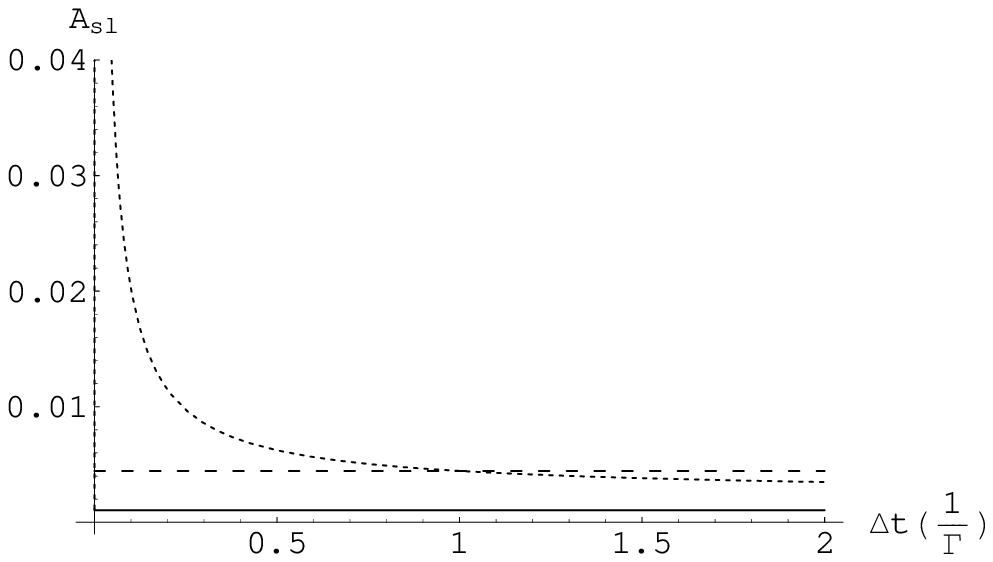}} \quad
\subfloat[\label{moco2}]{\includegraphics[width=.45\textwidth]{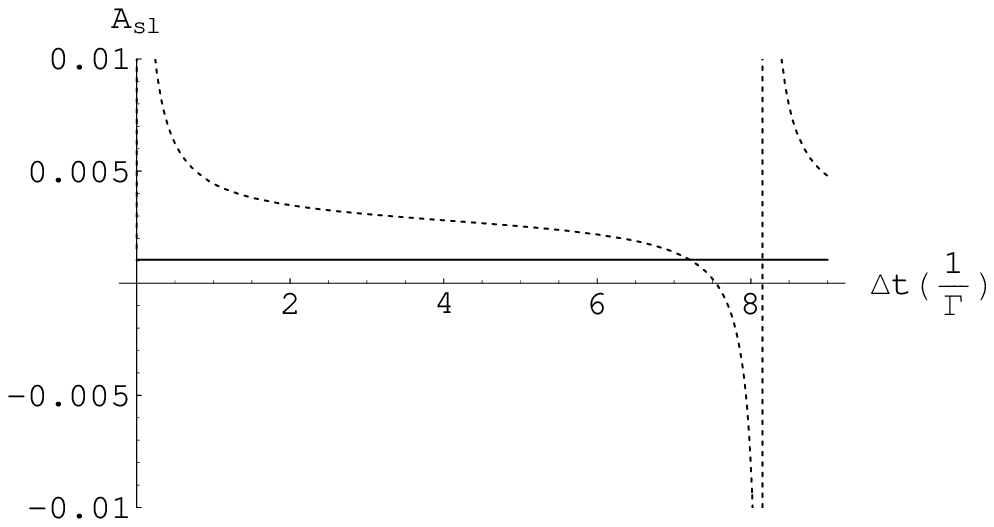}}
}
\end{center}
\caption{The $A_{sl}(\Dt)$ asymmetry at times $\Dt \gtrsim 1/\G$.
The solid line represents the $\w=0$ case, the short-dashed line is
for $\w=0.001$, and the long-dashed line (in Fig.~(a)) is the
shift-approximation for this range of times according to \eq{shift}.
In Fig.~(a) we represent the region of times where the asymmetry is
quasi time-independent but shifted due to $\w$.  In Fig.~(b) we plot
including $\Dm\Dt=2\pi$ to show the second peak, due to the quasi periodicity
of the asymmetry. }
\end{figure}

From this discussion we see that for small enough $\w$, in a range
of times around $\Dt \gtrsim 1/\G$, a good approximation for the
shift in $A_{sl}$ goes linear in $Re(\w)$, as it is seen in
\eq{shift}.  This shift-effect is displayed in \fig{moco}.

\pepe{el objetivo de la \fig{moco2} es para que se pueda comparar
con la \fig{asl}, ya que ambas cubren la misma region de tiempos}

As a last remark on the modification of the behaviour of the
$A_{sl}$ asymmetry it is worth to point out not only its time
dependence, but also its quasi-periodicity.  In fact, the asymmetry
is constructed from the intensities in \eq{piropo2}, where the only
non-periodical terms are those that go with $\cosh(\DG\Dt/2)$, but
these terms are quasi-constant (and hence periodic) for small enough
$\DG$.  In virtue of this, we expect that the asymmetry \pepe{esto
puede ser una opcion si no se pueden explorar $\Dt$ pequenos} will
repeat its behaviour at $\Dm\Dt=2\pi$. This means that the
above-studied peak for small $\Dt$ will be found again at times
around $\Dt \approx 2\pi/\Dm = 8.2\G^{-1}$; this time accompanied by
a counter peak with opposite-sign right before $\Dm\Dt=2\pi$. We
notice, however, that at these times the statistic available is
suppressed by a factor $e^{-8.2}\sim 10^{-4}$. In any case we point
out that, although with still low statistic and low time resolution,
the available data is beginning to explore also this region
\cite{Nakano:2005jb}. This second peak is plotted in \fig{moco2}.
(We note that the smoothen of this second peak due to $\DG$ is
hardly noticeable for $\DG/\G \leq 0.005$.)

\subsection{Observing the $\w$-effect in the $A_{sl}$ asymmetry \hecho{90}}
\label{por fin}

\begin{figure}[t]
\begin{center}
\framebox[1\textwidth]{
%\resizebox{10cm}{16.3cm}{\includegraphics{levelcurve-tesis.eps}}
\subfloat[$|\w|\ vs.\ \Dt(\G^{-1})$; for $\Omega=0$\label{level1}]{\includegraphics[width=.40\textwidth]{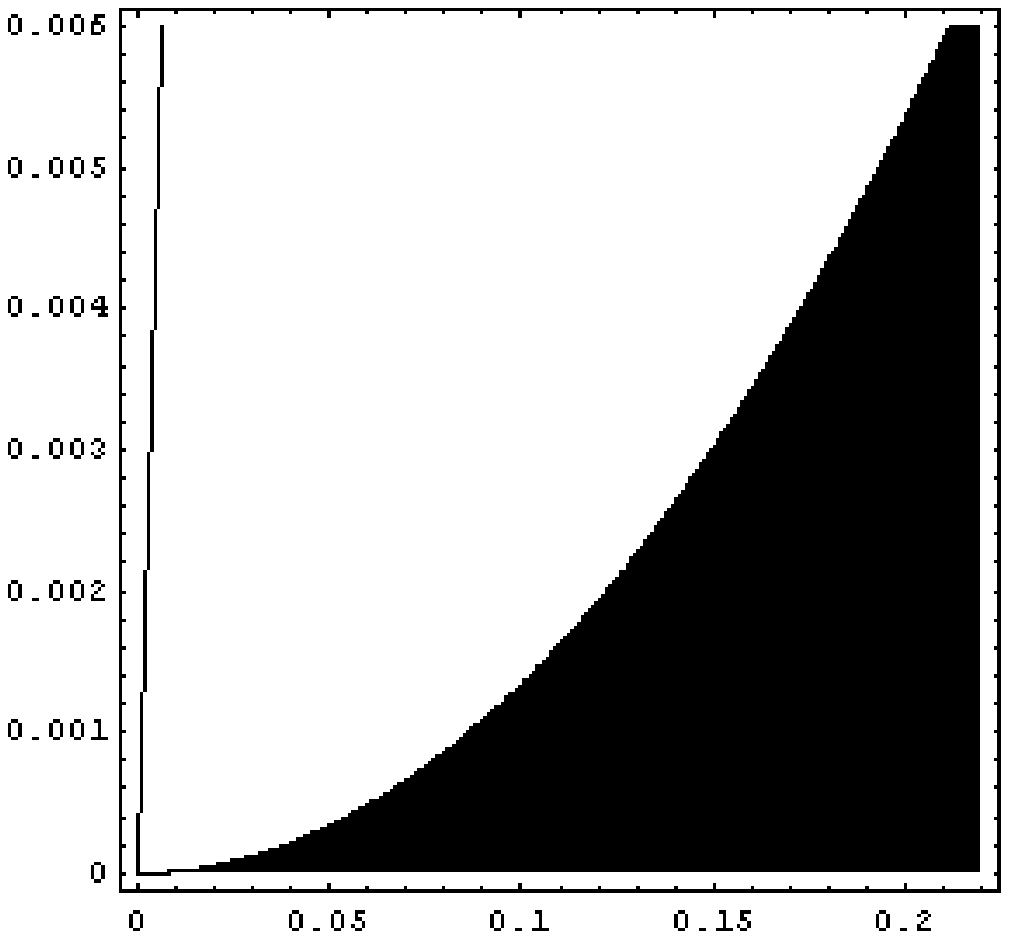}}\quad\subfloat[$\Omega\ vs.\ \Dt(\G^{-1})$; for $|\w|=0.001$ \label{level2}]{\includegraphics[width=.40\textwidth]{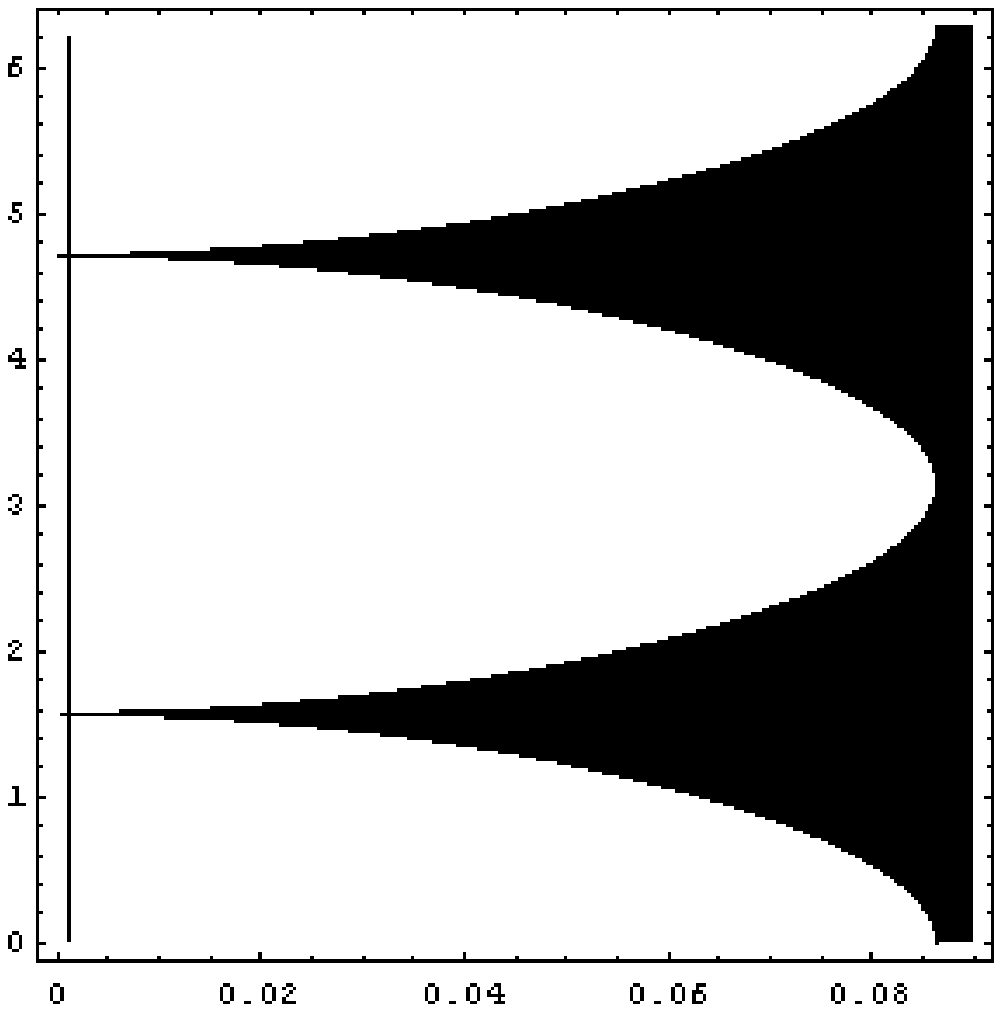}}
}
\end{center}
\caption{Level curves for $\frac{1}{\G}|dA_{sl} / d\Dt |= 0.1$, the white area
represents the points where $\frac{1}{\G}|dA_{sl} / d\Dt | > 0.1$, and hence the
time variation is (expected to be) experimentally detectable. Notice
the tiny dark line on the left of each graph which represents the
peak of the asymmetry, where of course the derivative also
goes to zero.  Fig.\ (a) plots $|\w|\ vs. \ \Dt$ for a fixed
$\Omega=0$, observe that although to see the peak in $A_{sl}$ it is
required a very high $\Dt$-resolution, the region where the time
variation is detectable might be more accessible experimentally.
Fig.\ (b) plots the phase $\Omega \ vs.\ \Dt$ for a fixed value of
$|\w|=0.001$, note that disregarding the values of the phase around
$\pi/2$ and $3\pi/2$, the measurable region (white) is quite
favoured in $\Dt$.} \label{fig3}
\end{figure}

We study now the detectability of the time-dependence in
$A_{sl}$ provided by the tail of the peak.  Since the main difference
between the $\w=0$ and $\w\neq0$ cases is that in the latter the
 asymmetry is time-dependent we may define, with operative purposes, a criterion of
experimental observability regarding the value of the time
derivative $d A_{sl} /d \Dt$.  We may {\it expect} that the limit detectable
time furnishes
 \pepe{al final puse igual a 0.1 , los resultados quedan mas lindos.
  Te parece que es mejor volver a poner 0.5?}
 \be
 \frac{1}{\G}\left| \frac{d A_{sl} }{d \Dt}(\Dt_{limit} ) \right|= 0.1,
 \label{carlinha}
 \ee and hence for $\Dt < \Dt_{limit}$ it is (expected to be)
possible to observe the effect of $\w$ as a time-dependence in the $A_{sl}$ asymmetry. In
\fig{level1} it is plotted the level curve of $\frac{1}{\G}|d A_{sl} /d
\Dt| = 0.1$ as a function of $|\w|$ and $\Dt$ for $\Omega=0$.  As it
can be seen, a value for instance of $\w \sim5\times 10^{-3}$ gives a
$\Dt_{limit} \sim 0.2 \Gamma^{-1}$ as compared to
$\Dt_{peak}=0.005\G^{-1}$.  We conclude that, in much later $\Dt$'s
than $\Dt_{peak}$ the time dependence is still detectable.  In
\fig{level2} it is shown the same level curve, but now plotted as a
function of $\Omega$ and $\Dt$, whereas the modulus is set to $|\w|=
0.001$.  In the figure it can be seen that, disregarding the
values of $\Omega$ close to $\pi/2$ or $3\pi/2$, the measurement of
the equal-sign charge asymmetry $A_{sl}$ at small $\Dt$ represents a
quite good observable to test the $\w$-effect.

\subsubsection{Present limits on $\w$ using existing data}

Although the best region where to see the effects of $\w$ is at small
$\Dt$, the existing data on the $A_{sl}$ asymmetry has not been
taken in that region, but instead at $\Dt \gtrsim 0.8\frac{1}{\G}$
\cite{Aubert:2002mn, Nakano:2005jb} (see \eq{vecina}).  In fact,
since in these experiments at last the $A_{sl}$ would be fitted to a
constant, the small $\Dt$ region was discarded to avoid the
complications which come from the experimental background.  On the other hand, we
have shown in this work that to see the pronounced possible effects of $\w$ it is
mandatory to explore the small $\Dt$ region. In any case, this existing
data could be used to put limits on $\w$, since its existence would
produce a shift in the $\Dt\sim 1/\G$ region, as shown in
\eq{shift}.

We point out here that putting limits to $\w$ using the available fit
of $A_{sl}$ to a constant is not a {\it direct} constraint on $\w$, but
rather {\it indirect}.  It is clear that a direct constrain on $\w$
should be done using the original complete experimental data and fitting it to
the expression $A_{sl}(\w,\Dt)$ that comes from constructing the
asymmetry using \eq{piropo2} with its time dependence.

In order to find indirectly the allowed values for $\w$ using the existing
measurements of $A_{sl}$ we proceed as follows. Using \eq{piropo2}
we compute the asymmetry as a function of $\w$, and we allow
$\epsilon$ to vary in its \sm expectation range \cite{pdg},
 \bea
 \left| \frac{4Re(\epsilon)}{1+\mec}\right| = {\cal O}\left( \frac{m_c^2}{m_t^2} \sin(2\beta) \right) \lesssim 0.001 .
 \label{el che}
 \eea
 The values of $\w$ which are within the $95\%$ C.L.\ are those such
 that $A_{sl}(\Dt,\w)$ is within the two standard deviation of the measured
value of the asymmetry,
 \bea
 A_{sl}^{exp} = 0.0019 \pm 0.0105 .
 \eea
(This value comes from the equally weighted average of Refs.\
\cite{Aubert:2002mn, Nakano:2005jb}, see \eq{vecina}.)  Of course that we
need to define what is {\it 'within'}, since $A_{sl}(\Dt,\w)$ is
time-dependent, whereas $A_{sl}^{exp}$ is constant and, besides, the
density of experimental events, $\rho(\Dt)$, is highly
$\Dt$-dependent.  With this purpose, we define a density-weighted average
difference between the experimental and $\w$-theoretical asymmetries
as
\pepe{te parece bien? Dado que solo tengo el valor del fitting constante, a mi me parecio adecuado asi}
 \bea
 \DA (\w) &=& \frac{\int_{\Dt_1}^{\Dt_2} \rho(\Dt)(A_{sl}(\Dt,\w)-A_{sl}^{exp})\, d\Dt}{\int_{\Dt_1}^{\Dt_2}\rho(\Dt)\, d\Dt} ,
 \label{pendorcha}
 \eea
 where $(\Dt_1,\Dt_2)=(0.8\frac{1}{\G},8\frac{1}{\G})$ is the range
of times where the measurements have been performed, and the density
of events is
 \bea
 \rho(\Dt) &=& e^{-\G\Dt} (1-\cos(\Dm\Dt)).
 \label{que conchita me chupe}
 \eea
  Where in the density function $\rho(\Dt)$ we
have approximated $\DG\approx\w\approx0$ since at this point they
constitute second order contributions. From the definition of
weighted-average-difference in \eq{pendorcha} we obtain the $95\%$
C.L.\ allowed values for $\w$ as those that furnish
 \bea
 | \DA (\w) | \leq 2\sigma = 0.0210 .
 \eea
 The fulfilment of this equation, neglecting quadratic contributions in $\w$, gives the final result:
 \bea
 -0.0084 \leq Re(\w) \leq 0.0100 \qquad ~ \quad 95\%\mbox{C.L.}
\label{pitufina}
 \eea
  Where the values of $\epsilon$ used to define the upper and
 lower limits correspond in each case to those in \eq{el che} which
 extend the most the allowed range for $\w$.  These are the first
 known limits on $\w$.

It is interesting to notice that if this same calculation is performed
approximating $A_{sl}(\Dt,\w)$ by its value at the typical time $\Dt=1/\G$, see
\eq{shift}, we obtain the approximated limits $-0.0059\lesssim Re(\w)
\lesssim 0.0070$ which is a good --and simpler-- {\it estimation} compared to the
exact value in \eq{pitufina}.

\begin{figure}
\framebox[1\textwidth]{ \subfloat[Babar, $\Dt=\frac{"\Dt (ps)"}{1.53
ps}\G^{-1}$]{\includegraphics[width=.50\textwidth]{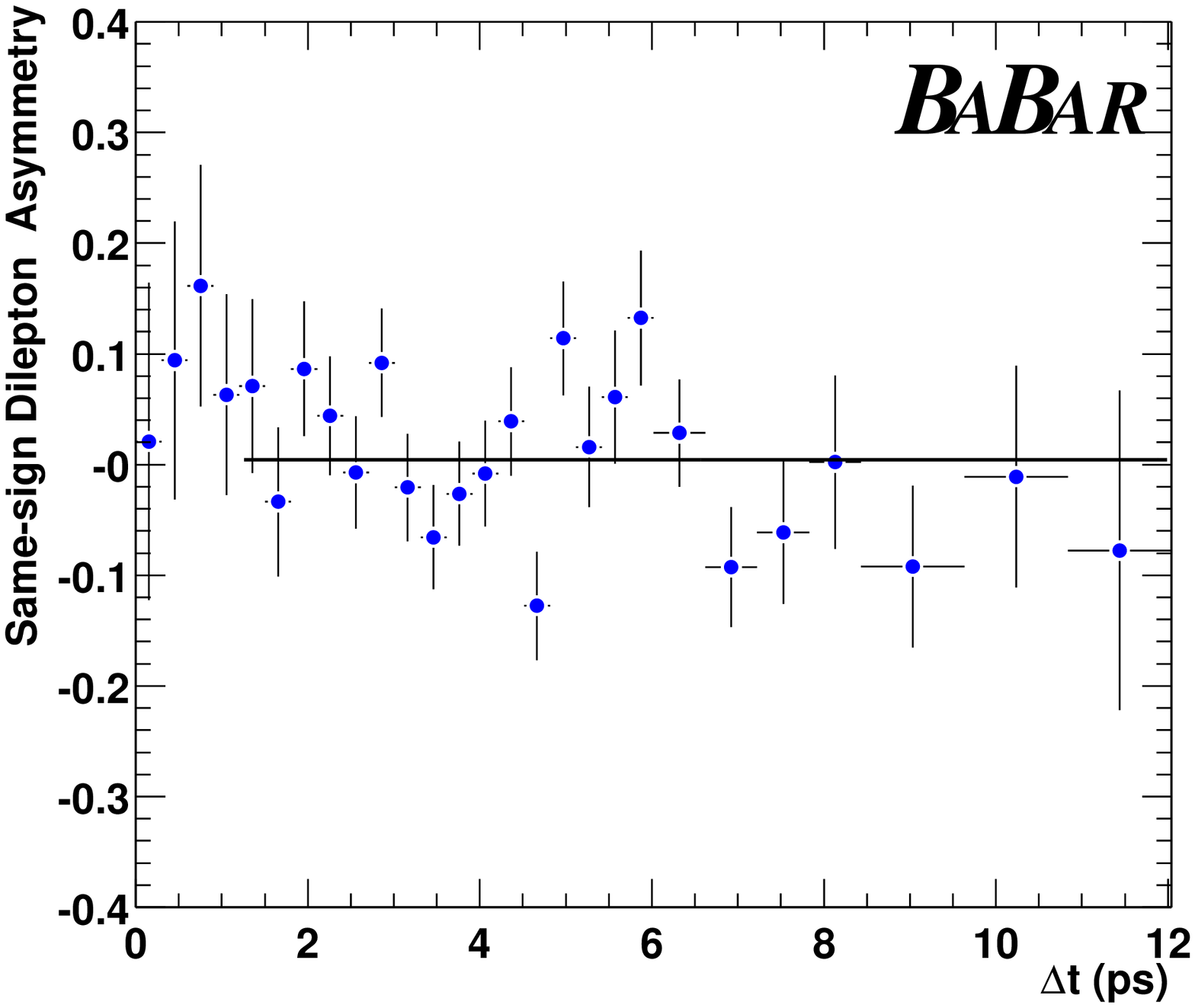}}
\subfloat[Belle,  $\Dt=\frac{|\Delta
z|}{0.0186cm}\G^{-1}$]{\includegraphics[width=.50\textwidth]{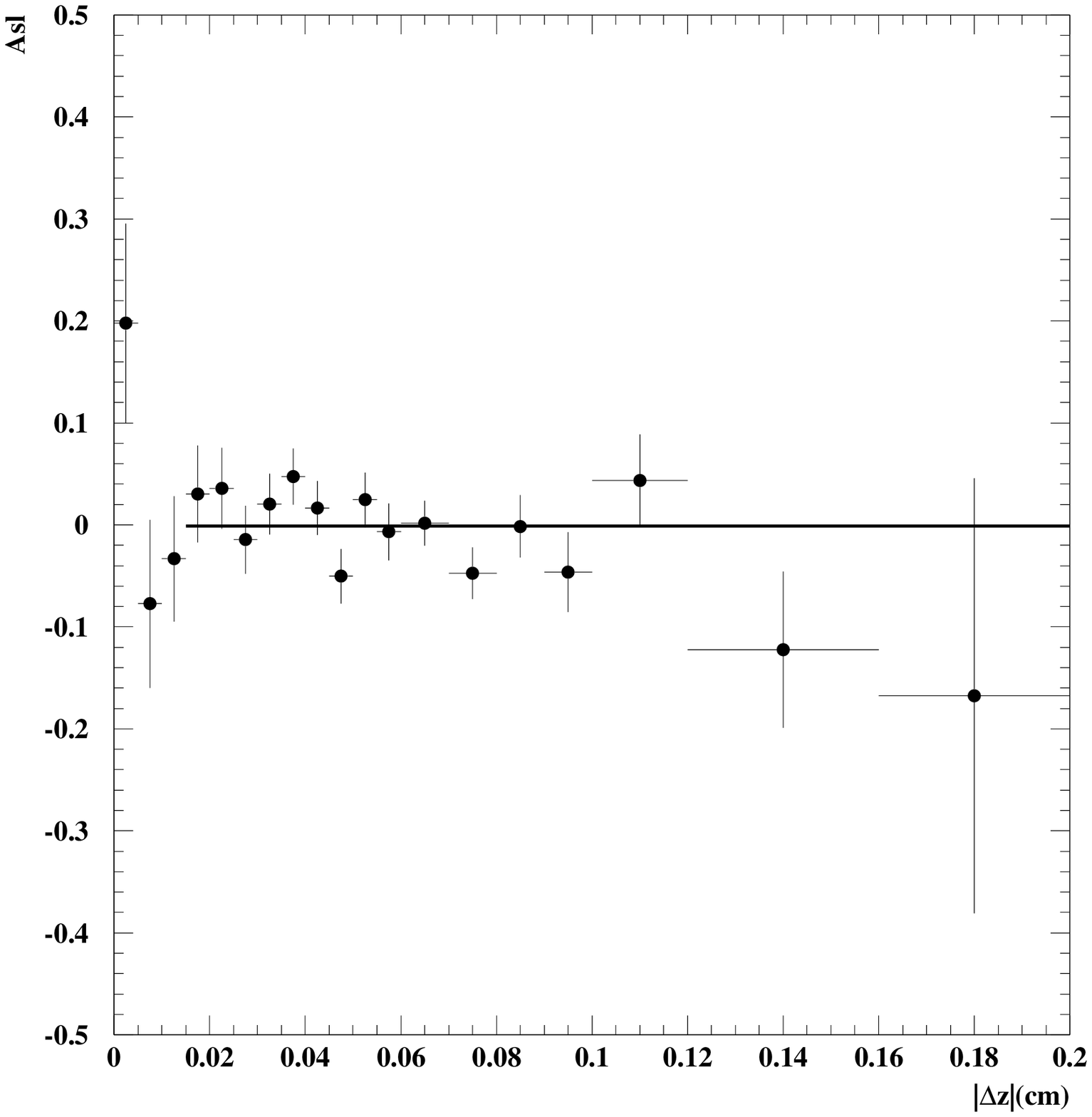}}
} \caption{Reproduction of the experimental results for the $A_{sl}$
asymmetry by Babar \cite{Aubert:2002mn} and Belle
\cite{Nakano:2005jb} collaborations.  The points for the
fitting (the constant solid line in each figure) are taken for $\Delta t > 0.8\G^{-1}$.  As it is seen in
the figures, the time independence of the $A_{sl}$ asymmetry, as well
as its fitting, have not arrived yet to a definite conclusion.
Notice also that the $\w$-effect predicts a second peak at $\Dt \approx 8.2 \G^{-1}$ ($\Delta
z\approx 0.15\, cm,\ \Dt \approx 12.4\, ps$)  which the data
cannot discard; see \fig{moco2} and text therein.}
 \label{asl}
\end{figure}

To close this discussion we find convenient to reproduce here
some experimental results for the $A_{sl}$ asymmetry.  In \fig{asl}
we reproduce Babar and Belle's results on the
asymmetry.  As it can be seen, the fitting to a constant is not
conclusive yet: further measurements are required to give a more
precise answer on the $\w$-effect.

\subsubsection{Statistical analysis}

\begin{figure}[th]
%\begin{center}
%\resizebox{10cm}{6.2cm}{\includegraphics{statistic-analysis.eps}}
%\makebox[\textwidth][c]{
\framebox[1\textwidth]{
\includegraphics[width=.6\textwidth]{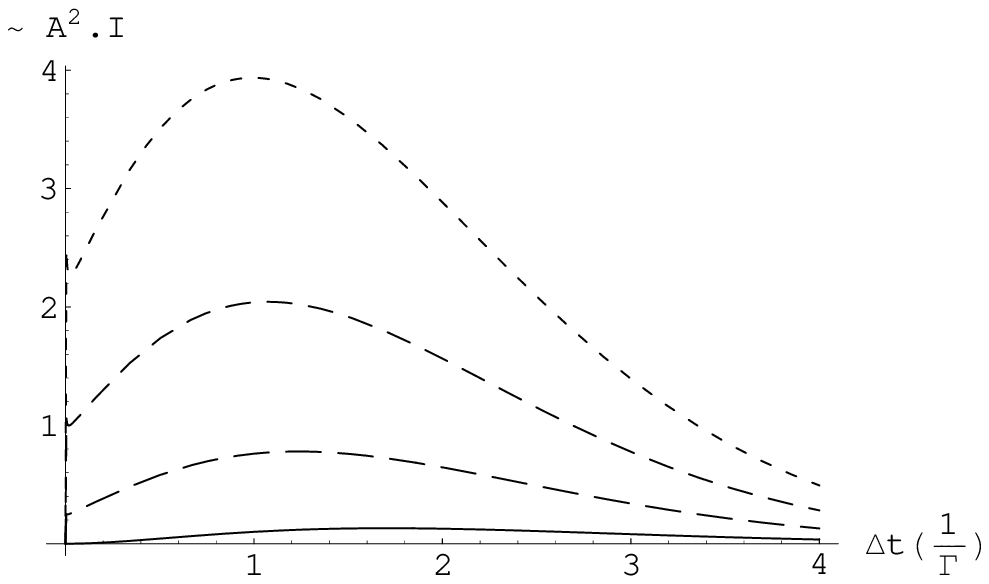}
}%}
%\end{center}
\caption{Figure of merit for measuring the equal-sign dilepton
charge asymmetry; from bottom to top $\w=0$ (solid line),
$\w=0.0003$ (long-dashed), $\w=0.0006$ (medium-dashed) and $\w=0.001$
(short-dashed).  Observe how the peak in the $A_{sl}$ asymmetry for
$\w\neq0$ is reflected here as a growth and a $\Dt$-shift towards
the origin of the maximum. This shows that as $\w\neq0$, its effects
are found to be in the small $\Dt$ region {\it and at the same time}
this region becomes statistically easier to explore.  The figure of merit proves
the high sensibility that exists in the asymmetry to measure $\w$.}
 \label{fig4}
\end{figure}

To complete the analysis of the equal-sign dilepton charge asymmetry, we study how the
statistic requirements are modified due to the presence of $\w$. From
the statistical analysis point of view, the relevant quantity to
observe an asymmetry as a function of $\Dt$ is the figure of merit,
i.e.\ $I(X,X',\Dt)\cdot (A_{sl}(\Dt))^2$.  As it is well known, the minimum
number of events needed to measure the asymmetry is proportional to
its inverse.  The figure of merit is plotted in \fig{fig4} for
different values of $\w$.  In this plot one may see the high
sensibility on $\w$: as $|Re(\w)|$ grows, the maximum also
grows and it is shifted towards smaller $\Dt$'s.  This behaviour
gives to the asymmetry a high sensibility on $\w$: as was
explained above, the effects of $\w$ are found enhanced in the small
$\Dt$ region, whereas the figure of merit tells us that as $|Re(\w)|$
grows, the exploration of this region becomes more effective.

%%%%%%%%%%%%%%%%%%%%%%%%%%%%%%%%%%%%%%%%%%%%%%%%%%%%%%%%%%%%%%%%%%%%%%%%%%%%%%%%%
%%%%%%%%%%%%%%%%%%%%%%%%%%%%%%%%%%%%%%%%%%%%%%%%%%%%%%%%%%%%%%%%%%%%%%%%%%%%%%%%%
%%%%%%%%%%%%%%%%%%%%%%%%%%%%%%%%%%%%%%%%%%%%%%%%%%%%%%%%%%%%%%%%%%%%%%%%%%%%%%%%%
\cleardoublepage\chapter{Conclusions and outlook \hecho{90}}
\label{conclusiones}

In this work we have analyzed and studied the effects of the
violation of the symmetries T and CP, and the possible violation of
CPT, in the correlated B-meson system.  The analysis has been
performed from two different points of view. In Chapter \ref{cp} we
have analyzed how the mixing and decay observables corresponding to
correlated CP and flavour decays depend on these three discrete
symmetries.  In Chapter \ref{w}, instead, we have studied solely the
effects of a novel kind of CPT violation whose effects would arise
in the initial state of the B-factories through the no longer
indistinguishability of $\b$ and $\bb$; producing, hence, a variety
of modification in the usual B-physics observables.  We analyze
separately the conclusions of these two different sets of results.

~

In Chapter \ref{cp} we have shown rigorously how to place a CP-tag.
Although there is no $\cp$ operator which leaves invariant the whole
Hamiltonian including the charged current term, we have shown how a
specific choice of $\ti$ determines a CP$_A$ operator which commutes
with a {\it chosen piece} of this charged current term, under a
$O(\lambda^4)$ approximation.  This piece is the only part of
the charged currents involved in the direct (with no mixing) decay
of the $B$ meson into $J/\psi K_{S,L}$: the Golden Plate decay.
With this determination, the initial state coming from the
$\Upsilon(4S) \to \b\bb$ decay is written in terms of the CP$_A$
eigenstates, and therefore a decay of the first $B$ meson into a Golden
Plate CP+ state places a CP$_A -$ tag in the second $B$ meson. And
vice-versa.

Having obtained a CP$_A$ operator which commutes with the piece of
Hamiltonian involved in the direct Golden Plate decay of the $B$
meson, gives us a powerful tool which makes the analysis of the
adequate correlated decays to get reduced to the $B$-mixing.  In
fact, with this CP-tag and the well-known flavour tag coming from
the flavour specific channels, we can obtain all possible
intensities relating final CP-eigenstate and flavour-specific decays
by studying the one-meson-transition in the $\Dt$-mixing.  The right
choice of the $B$ basis at the beginning and at the end of the
mixing, namely $\{ B_{\pm A} \}$ or $\{ B^0,\, CP_A B^0 \}$ depending on the
decay, reduces the calculation to the mixing of the $B$ meson.
Within any of these basis, we have proved that no phase ambiguities are
found and therefore the $\epsilon$ parameter is rephasing-invariant.

We have completed the analysis of correlated neutral $B$-meson
decays into flavour and CP eigenstates in terms of single $B$-meson
transitions in the time interval $\Delta t$ between both decays. The
action of the discrete symmetries CP, T and CPT is thus described in
terms of initial and final $B$-states.   In Section \ref{intensity
for correlated} a variety of new observables have been proposed to
show the consistency of the entire picture, as well as to explore
the possible existence of physics beyond the \sm.  For instance,
assuming CPT invariance the correlated CP-forbidden $(J/\psi K_S,\,
J/\psi K_S)$ and $(J/\psi K_L,\, J/\psi K_L)$ decays shall have
equal intensities.  As apparent from their CP-forbidden character,
each of them comes out to be proportional to $\sin ^2 (2\beta )$,
with the same $\Delta t$ dependence as the $(\ell^+, \, \ell^+ )$
correlated decay  (see \eq{la milanesa}).  Using the advantages of
the CP-tag, we have also proposed observables for observing direct
T-violation (see \eq{AT}).  At the same time we have remarked the
difference between the T-operation and the $\Dt$-operation
--consisting in the exchange in the order of appearance of the decay
products-- when $\DG\neq 0$. We have also computed the usual
observable which measures $\sin(2\beta)$ through the CP asymmetry
\eq{cp5}.

The genuine symmetries studied have also been combined with the
$\Delta t$ operation.  Although $\Delta t $-exchange and T-reversal
lead to different processes, they must have equal intensities in the
limit $\Delta \Gamma =0$ if CPT is assumed.  We then suggest the
comparison of processes connected by CP$\Delta t$-operation in order
to extract linear terms in $\Delta \Gamma$, as explicitly seen in
Eqs.$\,$(\ref{maite}) and (\ref{maitules}).  In the $B_d$-system,
these CP$\Delta t$-asymmetries are expected to be small.  One could
envisage the use of these observables to determine $\Delta\Gamma_s$
for the $B_s$-system, with expected larger width-difference, by
means of the $B$-factories operating at $\Upsilon (5S)$.

It is worth to note that in studying the double CP decays we have also found that, as in the
case of CP-semileptonic decays, the sign of $\cos (2\beta)$ remains
either hidden or accompanied by the sign of $\Delta \Gamma$.

~

In Chapter \ref{w} we have studied the consequences that CPT
violation could have on the initial $\b\bb$ entangled-state in the
B-factories.  The loss of indistinguishability of
particle-antiparticle yields to the relaxation of Bose-statistics
requirement, which is characterized by the complex parameter $\w$
(\eq{i}).  We have shown that this $\w$-effect would bring out important
conceptual and measurable modifications in the correlated B-meson system.

The most important conceptual change, the demise of flavour tagging,
comes out from the loss of the definite anti-symmetry in the initial
state.  In fact, if $\w\neq0$ then the time evolution of the
initial state will project a component on the $|\b\b\rangle$ and
$|\bb\bb\rangle$ states, which are clearly forbidden in the $\w=0$
case.  The appearance of these {\it forbidden} states produces the
demise of the flavour tag: a flavour-specific $\b$ decay on one side
allows a presence (proportional to $\w$) of the same $\b$
meson on the wave function in the other side {\it at the same time}, and
viceversa.  We show that this effect would be measurable through the
 CP asymmetry consisting of the $\Dt=0$ same sign semi-leptonic
 intensities, $A_{CP}(t)$ and ${\cal A}_{CP}$
 (\eq{eq:Asymmetries:01}).  Notice, however, that although
 conceptually appealing, this observable requires a $|\w|^2$
 precision in the measurement of the intensities.

In order to look for an observable linear in $\w$ we have analyzed
the $\Dt$-dependence of the equal-sign semi-leptonic correlated
decays.  Here the interference terms give rise to the expected
linear in $\w$ modifications.  We have computed the equal-sign
dilepton events in correlated B-decays, $I(\ell^\pm,\ell^\pm,\Dt)$
(see \eq{piropo2} and \fig{fig1}).  We have found corrections linear
in $Re(\w)$ with opposite behaviour depending on the sign of the
leptons, and therefore pointing to their asymmetry  as an optimal
observable where to look for traces of $\w$.

The equal-sign dilepton charge asymmetry, $A_{sl}$, has been already
studied and measured for the case $\w=0$, where its behaviour is
known to be constant and proportional to the CP violating parameter
$Re(\epsilon)$.  The computation of the $A_{sl}$ asymmetry for
$\w\neq0$, on the other hand, shows a strong time-dependence in the
small $\Dt$ region (where it has not been measured yet), and a quasi
time-independent shift in the later (measured) $\Dt\sim\G^{-1}$
region. At small $\Dt\sim|\w| \G^{-1}$ we find a pronounced peak,
whose tail could be detectable for later $\Dt$'s (see Figs.\
(\ref{fig2}-\ref{fig3}).

Current experiments have been performed in the $\Dt \gtrsim \G^{-1}$
region, and are consistent with a possible existence of $\w\neq 0$.
Moreover, the difference between the \sm expectation for
$Re(\epsilon)$ and the measured two-standard-deviation allowed range
for the equal-sign dilepton charge asymmetry, allows us to compute
indirectly the current limits on the value of $Re(\w)$ at a $95\%$ confidence
limit,
 \bea
 -0.084 \leq Re(\w) \leq 0.0100 \qquad 95\%\mbox{ C.L.}
 \eea
 These are the first known limits on $\w$.  This limit can be improved
by the progress in the experimental measure of the
asymmetry or, much more definitive, by the exploration of the small-$\Dt$
region.  This, of course, requires a new treatment of the
background, which complicates the measure of the flavour-specific
channels in that region.

From the statistics point of view, we have shown using the figure of
merit that as $\w\neq0$ its greater effects are found in the small
$\Dt$-region and {\it at the same time} this region becomes easier
to explore.  Hence, the experimental probe of the small-$\Dt$ region
is crucial for determining the possible existence of CPT violation
through {\it distinguishability} of particle-antiparticle.

The results of this thesis advocate and endorse the experimentalists to
explore the low-$\Dt$ region in the equal-sign dilepton events.

The implications due to a non-vanishing value of $\w$ are not only
visible in the dilepton channels.  We plan to explore this effect in
the correlated (flavour,CP) and (CP,flavour) decays, as demanded for
the Golden Plate channels.

%%%%%%%%%%%%%%%%%%%%%%%%%%%%%%%%%%%%%%%%%%%%%%%%%%%%%%%%%%%%%%%%%%%%%%%%%%%%%%%%%
%%%%%%%%%%%%%%%%%%%%%%%%%%%%%%%%%%%%%%%%%%%%%%%%%%%%%%%%%%%%%%%%%%%%%%%%%%%%%%%%%
%%%%%%%%%%%%%%%%%%%%%%%%%%%%%%%%%%%%%%%%%%%%%%%%%%%%%%%%%%%%%%%%%%%%%%%%%%%%%%%%%
\setcounter{chapter}{6} \cleardoublepage
 \chapter{Conclusiones y perspectivas [espa\~nol] \hecho{00}}

\nota{arrgelar la hipenaci\'on cuando est\'e lista} En este trabajo
hemos analizado y estudiado los efectos de la violaci\'on de las
simetr\1as T y CP, y la posible violaci\'on de CPT en el sistema de
mesones B correlacionados.  El an\'alisis ha sido realizado desde
dos puntos de vista diferentes.  En el Cap\1tulo \ref{cp} hemos
analizado c\'omo observables de mixing y decaimiento,
correspondientes a decaimeintos correlacionados de sabor y CP,
dependen de estas tres simetr\1as discretas.  En el Cap\1tulo
\ref{w}, en cambio, hemos estudiado s\'olamente los efectos de una
flamante clase de violaci\'on de CPT cuyos efectos surgir\1an en el
estado inicial de las f\'abricas de B, a trav\'es de la p\'erdida de
indistinguibilidad entre $\b$ y $\bb$; produciendo entonces una
variedad de modificaciones en los observables usuales de la f\1sica
de B's.  Analizamos separadamente las conclusiones de estos dos
conjuntos de resultados diferentes.

~

En el Cap\1tulo \ref{cp} demostramos rigurosamente c\'omo colocar un
r\'otulo de CP (CP-tag).  Aunque no exista un operador $\cp$ que
deje invariante todo el Hamiltoniano incluyendo la parte de
corrientes cargadas, hemos mostrado c\'omo una elecci\'on
espec\1fica de $\Theta$ y $\Theta'$ determina el operador CP$_A$ que
conmuta con una {\it pieza seleccionada} de \'este t\'ermino de
corrientes cargadas, en la aproximaci\'on de ${\cal O}(\lambda^4)$.
 Esta pieza es la \'unica parte de las corrientes cargadas que
 est\'a involucrada en el decaimiento directo (sin mixing) del meson
 B en $J/\psi K_{S,L}$: el decaimiento Golden Plate.  Con esta
 determinaci\'on, el estado inicial proveniente del decaimiento
 $\Upsilon(4S)\to\b\bb$ es escrito en t\'ermino de los autoestados
 de CP$_A$, y entonces un primer decaimiento a un Golden Plate CP+
 coloca un r\'otulo  CP$_A -$  en el segundo mes\'on en vuelo.  Y
 viceversa.

 Haber obtenido un operador CP$_A$ que conmuta con la pieza de
 Hamiltoniano involucrada en el decaimiento Golden Plate, nos ofrece
 una poderosa herramienta que hace que el an\'alisis de los
 decaimientos correlacionados adecuados quede reducido al mixing de
 los B's.  En efecto, con el r\'otulo por CP y el bien conocido
 r\'otulo por sabor que proviene de los canales de
 sabor-espec\1fico, podemos obtener todas las posibles intensidades
 que relacionan decaimientos finales de autoestados de CP y de
 sabor-espec\1fico, estudiando la transici\'on de un mes\'on en el
 mixing durante $\Dt$.  La elecci\'on correcta de la base de B's al
 comienzo y al final del mixing, ll\'amese $\{B_{\pm A} \}$ y
 $\{\b,CP_A \b\}$ seg\'un el decaimiento, reduce todo el c\'alculo
 al mixing de un \'unico mes\'on B.  En cualquiera de estas bases hemos
 probado que no existe ambig\"uedad en las fases, luego $\epsilon$
 es un par\'ametro invariante de fase.

Hemos completado el an\'alisis de los decaimientos de mesones
neutros B correlacionados en canales de sabor y de autoestados de CP
en t\'ermino de transiciones de un \'unico mes\'on B en el intervalo
de tiempo $\Dt$ entre los dos decaimientos.  La acci\'on de las
simetr\1as discretas CP, T y CPT es entonces descrita en t\'ermino
de estados iniciales y finales de mesones B.  En la Secci\'on
\ref{intensity for correlated} una variedad de nuevos observables
han sido propuestos para mostrar la consistencia del modelo, as\1
como explorar la posible existencia de f\1sica mas all\'a del Modelo
Est\'andar.  Por ejemplo, suponiendo in\-va\-rian\-za CPT, los
decaimientos correlacionados CP-prohibidos $(J/\psi K_S,J/\psi K_S)$
y $(J/\psi K_L,J/\psi K_L)$ deben tener la misma intensidad.  Tal
como es aparente de su caracter CP-prohibido, cada uno de ellos
resulta proporcional a $\sin^2(2\beta)$, con la misma dependencia en
$\Dt$ que el decaimiento $(\ell^+,\ \ell^+)$ (ver Ec.~(\ref{la
milanesa})).  Usando las ventajas del r\'otulo por CP, hemos
tambi\'en propuesto observables para medir violaci\'on directa de T
(ver Ec.~(\ref{AT})).  A la vez hemos remarcado la diferencia que
existe entre las operaciones T y $\Dt$ --que consiste en
intercambiar el orden de los decaimientos-- cuando $\DG\neq0$.
 Tambi\'en hemos construido el observable usual para la medici\'on de
$\sin(2\beta)$, la asimetr\1a CP en Ec.~(\ref{cp5}).

Las simetr\1as genuinas estudiadas tambi\'en han sido combinadas con
la operaci\'on $\Dt$.  Aunque el intercambio-$\Dt$ y la operaci\'on
de inversi\'on temporal T conducen a diferentes procesos, estos
deben tener la misma intensidad en el l\1mite $\DG=0$ si CPT es
v\'alido. Teniendo en cuenta esto, sugerimos la comparaci\'on de
procesos conectados por la operaci\'on CP$\Dt$, para as\1 extraer
t\'erminos lineales en $\DG$, como se ve expl\1citamente en las
Ecs.~(\ref{maite}) y (\ref{maitules}).
 En el sistema $B_d$ estas asimetr\1as CP$\Dt$ se esperan que sean
 peque\~nas.  Sin embargo uno puede prever el uso de estos
 observables para determinar $\DG_s$ en el sistema $B_s$, donde se
 esperan mayores anchuras, usando las f\'abricas de B operando en la
 resonancia del $\Upsilon(5S)$.

 Es conveniente notar que al estudiar los decaimientos doble-CP
 hemos nuevamente hallado, como es en el caso de los decaimientos
 CP-se\-mi\-lept\'o\-nicos, que el signo de $\cos(2\beta)$ sigue u
 oculto, o acompa\~nado por el signo de $\DG$.

~

En el Cap\1tulo \ref{w} hemos estudiado las consecuencias que
podr\1a tener vio\-la\-ci\'on de CPT sobre el estado inicial de los
mesones $\b\bb$ correlacionados en las f\'abricas de B.  La
p\'erdida de indistinguibilidad entre part\1cula y antipart\1cula
conduce al relajamiento  del requisito de estad\1stica de Bose,
caracterizado por el par\'ametro $\w$ (Ec.~(\ref{i})). Hemos
mostrado que este efecto-$\w$ podr\1a implicar importantes
modificaciones conceptuales y medibles en el sistema de mesones B
correlacionados.

El cambio conceptual m\'as importante, el 'fin' del r\'otulo por
sabor, se debe a la p\'erdida de la anti-simetr\1a definida en el
estado inicial.  En efecto, si $\w\neq0$ entonces la evoluci\'on
temporal del estado inicial tendr\'a una proyecci\'on sobre los
estados $|\b\b\rangle$ y $|\bb\bb\rangle$ no nula, la cual est\'a
claramente prohibida en el caso de $\w=0$.   La aparici\'on de estos
{\it estados prohibidos} produce el 'fin' del r\'otulo por sabor: un
decaimiento de sabor-espec\1fico $\b$ en un lado permite una
presencia (proporcional a $\w$) del mismo mes\'on $\b$ en el otro
lado {\it al mismo tiempo},  y viceversa.  Hemos demostrado que este
efecto ser\1a medible a trav\'es de una asimetr\1a CP que consista
de las intensidades dilept\'onicas de mismo signo con $\Dt=0$,
$A_{CP}(t)$ y ${\cal A}_{CP}$ (Eq.~(\ref{eq:Asymmetries:01})).
N\'otese, de todos modos, que aunque conceptualmente importante,
este observable requiere una precisi\'on de $|\w|^2$ en la
medici\'on de las intensidades.

Con el objeto de buscar un observable lineal en $\w$ hemos analizado
la dependencia en $\Dt$ de los decaimientos correlacionados
semilept\'onicos de mismo signo.  Aqu\1 los t\'erminos de
interferencia dan lugar a las buscadas modificaciones lineales en
$\w$.  Hemos computado los eventos de dos leptones del mismo signo
en los decaimientos correlacionados de mesones B,
$I(\ell^\pm,\ell^\pm,\Dt)$ (ver Ec.~(\ref{piropo2}) y \fig{fig1}).
 Hemos hallado correcciones lineales en $\w$ con comportamientos
 opuestos dependiendo en los signos de los leptones, apuntando de
 este modo a su asimetr\1a como un \'optimo observable donde buscar
 rastros del efecto-$\w$.

La asimetr\1a de carga de eventos dilept\'onicos del mismo signo,
$A_{sl}$, ha sido ya estudiada y medida para el caso $\w=0$, donde
su comportamiento se predice constante y proporcional al par\'ametro
de violaci\'on de CP $Re(\epsilon)$.  El c\'alculo de la asimetr\1a
$A_{sl}$ con $\w\neq0$, sin embargo, muestra una fuerte dependencia
temporal en la regi\'on de $\Dt$ peque\~nos (donde no ha sido medida
aun), y un corrimiento cuasi-independiente del tiempo para tiempos
posteriores $\Dt\sim\G^{-1}$ (donde si se ha medido).  Para
peque\~nos $\Dt\sim|\w|\G^{-1}$ hallamos un pico pronunciado, cuya
cola podr\1a ser detectable en $\Dt$ posteriores (ver
Figs.~(\ref{fig2}-\ref{fig3})).

Los experimentos actuales han sido realizados en la regi\'on
$\Dt\gtrsim\G^{-1}$, y son consistentes con la posible existencia de
$\w\neq0$.  Es m\'as, la diferencia entre el valor que predice el
Modelo Est\'andar para $Re(\epsilon)$ y el rango permitido por las
dos desviaciones est\'andares de la medici\'on de la asimetr\1a
$A_{sl}$, nos permite establecer indirectamente los l\1mites
actuales para el valor de $Re(\w)$ a un nivel de $95\%$ de l\1mite
de confianza, \bea -0.084 \leq Re(\w) \leq 0.0100 \qquad 95\%\mbox{
L.C.} \eea Estos son los primeros l\1mites conocidos para $\w$. Este
l\1mite puede ser mejorado a trav\'es de una mejor y m\'as precisa
medici\'on de la asimetr\1a o, en forma m\'as definitiva, explorando
la regi\'on de $\Dt$ peque\~nos.  Esto, claro est\'a, requiere un
nuevo tratamiento del fondo, ya que \'este complica la medici\'on de
canales espec\1ficos de sabor en esa regi\'on.

Desde el punto de vista de la estad\1stica, hemos demostrado usando
la figura de m\'erito que si $\w\neq0$ entonces sus efectos se
hallan en la regi\'on de $\Dt$'s peque\~nos y {\it a la misma vez}
esta regi\'on es m\'as f\'acil de explorar.  Luego, el an\'alisis
experimental de la regi\'on de peque\~nos $\Dt$'s es crucial para
determinar la posible existencia de violaci\'on de CPT a trav\'es de
{\it distinguibilidad} de part\1cula y antipart\1cula.

Los resultados de esta tesis ponderan y respaldan a los experimentales
a explorar la regi\'on de peque\~nos $\Dt's$ en los eventos
dilept\'onicos del mismo signo.

Las implicaciones de un posible $\w$ diferente de cero son visibles no s\'olo en los
canales dilept\'onicos.  Nuestro plan es explorar este efecto en los
decaimientos correlacionados (sabor,CP) y (CP,sabor), para observar
las correcciones que producir\1a en los canales Golden Plate.

%%%%%%%%%%%%%%%%%%%%%%%%%%%%%%%%%%%%%%%%%%%%%%%%%%%%%%%%%%%%%%%%%%%%%%%%%%%%%%%%%
%%%%%%%%%%%%%%%%%%%%%%%%%%%%%%%%%%%%%%%%%%%%%%%%%%%%%%%%%%%%%%%%%%%%%%%%%%%%%%%%%
%%%%%%%%%%%%%%%%%%%%%%%%%%%%%%%%%%%%%%%%%%%%%%%%%%%%%%%%%%%%%%%%%%%%%%%%%%%%%%%%%

%zx
\newpage
~
\cleardoublepage
\addcontentsline{toc}{chapter}{Agradecimientos}
\vskip 2cm
\begin{center}
{\bf AGRADECIMIENTOS}
\end{center}

All\'a, por los comienzos de 2001, habr\1an pasado no m\'as de
algunas semanas de mi licenciatura, no ten\1a siquiera un proyecto
de qu\'e hacer en el futuro, pero ya me encontraba ofreciendo
(>canjeando?) agradecimientos para mi tesis doctoral. Hoy, 15 de
noviembre de 2005, como todo hombre que se digna de s\1 mismo, me
debo a mi Palabra.

\begin{itemize}
\item[I] Gracias a la Universidad de Valencia por el apoyo econ\'omico durante estos cuatro
a\~nos a trav\'es de la Beca {\it Cinc Segles}. Gracias a
Fundaci\'on Antorchas y Fundaci\'on K\'onex por haber dep\'ositado
su confianza en m\1 y, en los tiempos m\'as duros de Argentina,
haberme adjudicado sendas becas para que pudiese venir a realizar mi
doctorado en la Universidad de Valencia. Gracias a Sof\1a Rawson y a
Luis Ovsejevich por su grandeza y dignidad, sus desinteresados
respaldo y apoyo, y por su confianza en m\1. Gracias a los
Departamentos de F\1sica de la Universidad de Valencia y de la
Universita Degli Studi di Roma La Sapienza, cuyos recursos fueron
indispensables para realizar los trabajos contenidos en esta tesis.

\item[II]
Debo agradecer fuerte y enormemente a Pepe Bernab\'eu su sabia gu\1a
y ayuda, y su cons\-tan\-te respaldo y apoyo durante estos cuatro
a\~nos, as\1 como tambi\'en le agradezco por haberme aceptado para
realizar este doctorado bajo su supervisi\'on.

Gracias a Joannis Papavassiliou por su ayuda y gu\1a, tambi\'en a Miguel Nebot y Nick Mavromatos por las, finalmente ricas, tertulias e intercambio de ideas.

Gracias a Alejandro Szynkman, Luis Epele, Carlos Garc\1a-Canal y
Daniel G\'omez-Dumm por su apoyo para trabajar con ellos durante mi
estancia en la Universidad de La Plata.

Gracias a C.~Garc\1a-Canal, N.~Mavromatos y L.~Silvestrini por sus \'utiles comentarios y aportes en la lectura del manuscrito.

Gracias a Vicente Vento y a Miguel Angel Governa, quienes llevar\'an aqu\1 el r\'otulo de part\1cipes necesarios.

\item[III] Vaya un gracias may\'usculo para Los Viejos, Ingrid y Hugo,
por su constante apoyo en todas y cada una de las situaciones que me
ha tocado en suerte vivir. Otro gracias especial a Agust\'in, por su
interminable apoyo y camarader\1a.

Gracias a Carla, por su fiel apoyo, por su compa\~n\1a, por su colaboraci\'on.  Y tambi\'en gracias, por su eterna comprensi\'on.  

Gracias a los gom\1as, Federico y Mart\1n, gracias a la Eli; gracias
a Pablo Eggarter; gracias a Carmen y a Jorgelina, gracias a su familia; gracias a Mirta y familia; gracias a Gonzalo, camarada de mar agitado;
gracias a Asunci\'on por su implacable ayuda y apoyo, y tambi\'en gracias a Alicia, Cristina y Pilar;
gracias a Ada, por su cari\~no y colaboraci\'on; gracias a Marek y Minou, gracias a la familia
Kaczorowski; gracias a Ale Szynkman, el {\it sensei} del Sur; gracias a Diego Mazzitelli;
gracias al Sr.~Jose Fernando Ben\1tez-Villena, al Sr.~Jorge Cabero, al Sr.~Fabio Lanci, al Sr.~Fausto Troilo y al Sr.~Massimo Milano; gracias a Carlo, Adriana y Sergio, gracias a la
familia Zilli en toda su extensi\'on; gracias a Inma, gracias a Natalia, Angel y a la familia Basante-L\'opez; 
gracias a Sara; gracias a las vecinas del primero, las Navarro-Cayuela; gracias a Cristian Batista, a Rolando
Somma, a Ge\-rar\-do Ort\1z; gracias a Raquel Antelo; gracias a mis dos Yamahitas, compa\~neras de aventuras; gracias al Bar Zakate y los s\'olitos
compa\~neros que all\1 paran, Jaqueline, Bea, Julia, Jorge, Claudia,
etc.; Gracias a Julieta Ugartemend\1a, gracias a sus vecinos
venezolanos y compa\~neras de piso; gracias a Joannis, Catalina, Teresa, David y a El Limbo; gracias a
Cinthia; gracias a Carmen y a Elena; gracias a Caro y Guilleaume, gracias a Jul Bidu; gracias a Andrea Trenes; gracias a Carolina Toledo; gracias a Miguel Socolovsky, esp\1ritu de la revoluci\'on; gracias a Warhol y a la
extensi\'on de caracteres que all\1 habitan, gracias a Diego, Gato, Juancho y
sus compa\~neras; gracias a Susana Asencio y a Sakura, develadoras de horizontes; gracias al Gordo Tato y al Antro por esos infinitos asados; gracias a la Tierra, por habernos dado la m\'usica; gracias a la vida, que me ha dado tanto.
\end{itemize}

%%%%%%%%%%%%%%%%%%%%%%%%%%%%%%%%%%%%%%%%%%%%%%%%%%%%%%%%%%%%%%%%
\newpage 
~ \thispagestyle{empty} \newpage
~ \thispagestyle{empty}

\addons{

\newpage
\addcontentsline{toc}{chapter}{Ep\1logo} \vskip 2cm
\begin{center}
{\bf Ep\1logo}
\end{center}
\vskip 4cm

 \noindent
 Desde aquella ya casi p\'erdida {\it \'ultima l\1nea}, \newline
 convergen en m\1 y en estas cansadas palabras \newline
 un abanico de a\~nos que hoy ignoro si se han ido. \newline
 Amplia y vasta, el agua que deb\1a ver correr, \newline
 esa otra agua, la que se siente bajar del monte, \newline
 es clara, es rica, es prometedora.  Tambi\'en desafiante. \newline
 Desde esta pieza, perdida en un Invierno boreal, \newline
 sent\1 el fugaz d\1a irse, detr\'as del ventanal: \newline
 navidad y a\~no nuevo s\'olo los s\'e festejar en verano. \newline
 Esta pelea la luchamos del lado de los m\'as d\'ebiles.

} % termina el "addons"

\parami{
~\newpage
\appendix
\subsubsection{Belle}
The Kek-Belle features are found essentially in hep-ex/0409012,
hep-ex/0505017 (new!) and PRD67, 052004.  Summarizing:
\bea
\langle\beta\gamma\rangle &=& 0.425 \nn\\
0.015cm < &\Delta z& < 0.2cm \nn\\
\frac{1}{\Gamma} &\leftrightarrow& 0.0186 cm\nn\\
0.80\frac{1}{\G} < &\Dt& < 10.7 \frac{1}{\G}
\eea

\subsubsection{Babar}
The SLAC-Babar features are found essentially in hep-ex/0202041.
Summarizing:
\bea
\langle\beta\gamma\rangle &=& 0.55 \nn\\
0.02cm < &\Delta z& < ?=0.2cm \nn\\
\frac{1}{\Gamma} &\leftrightarrow& 0.0240 cm\nn\\
0.83\frac{1}{\G} < &\Dt& < ?=8.3 \frac{1}{\G}
\eea

\subsubsection{B-features}
we have
\bea
\frac{1}{\Gamma} &=& 1.53 \times 10^{-12} s \nn\\
\Dm &=& 0.77 \G
\eea
}%termina el parami

\newpage

\end{document}